\documentclass[a4paper,11pt]{article}

\pdfoutput=1

\usepackage{jheppub}
\usepackage{amsmath}
\usepackage{xspace}
\usepackage{hyperref}
\usepackage{mdwlist}
\usepackage[dvipsnames]{xcolor}
\usepackage{subcaption}
\usepackage[normalem]{ulem}
\usepackage{multirow}

\newcommand{\eq}[1]{eq.~\eqref{eq:#1}}
\newcommand{\eqs}[2]{eqs.~\eqref{eq:#1} and \eqref{eq:#2}}
\renewcommand{\sec}[1]{section~\ref{sec:#1}}

\newcommand{\app}[1]{appendix~\ref{app:#1}}
\newcommand{\fig}[1]{figure~\ref{fig:#1}}
\newcommand{\figs}[2]{figures~\ref{fig:#1} and \ref{fig:#2}}
\newcommand{\tab}[1]{table~\ref{tab:#1}}
\newcommand{\tabs}[2]{tables~\ref{tab:#1} and \ref{tab:#2}}

\newcommand{\nn}{\nonumber}

\newcommand{\refcite}[1]{ref.~\cite{#1}}
\newcommand{\refscite}[1]{refs.~\cite{#1}}

\newcommand{\Pythia}{\textsc{Pythia}\xspace}

\allowdisplaybreaks[2]

\title{Power Counting Energy Flow Polynomials}

\author[a,b,c]{Pedro Cal,}
\author[d,e]{Jesse Thaler,}
\author[b,c]{Wouter J. Waalewijn}

\affiliation[a]{Deutsches Elektronen-Synchrotron DESY, Notkestr. 85, 22607 Hamburg, Germany}

\affiliation[b]{Institute for Theoretical Physics Amsterdam and Delta Institute for Theoretical Physics, University of Amsterdam, Science Park 904, 1098 XH Amsterdam, The Netherlands}

\affiliation[c]{Nikhef, Theory Group, Science Park 105, 1098 XG, Amsterdam, The Netherlands}

\affiliation[d]{Center for Theoretical Physics, Massachusetts Institute of Technology, Cambridge, MA 02139, USA}

\affiliation[e]{The NSF AI Institute for Artificial Intelligence and Fundamental Interactions}

\emailAdd{pedro.cal@desy.de}
\emailAdd{jthaler@mit.edu}
\emailAdd{w.j.waalewijn@uva.nl}

\abstract{
Power counting is a systematic strategy for organizing collider observables and their associated theoretical calculations.
In this paper, we use power counting to characterize a class of jet substructure observables called energy flow polynomials (EFPs).
EFPs provide an overcomplete linear basis for infrared-and-collinear safe jet observables, but it is known that in practice, a small subset of EFPs is often sufficient for specific jet analysis tasks.
By applying power counting arguments, we obtain linear relationships between EFPs that hold for quark and gluon jets to a specific order in the power counting.
We test these relations in the parton shower generator \Pythia, finding excellent agreement.
Power counting allows us to truncate the basis of EFPs without affecting performance, which we corroborate through a study of quark-gluon tagging and regression.
}

\preprint{\vbox{%
\hbox{DESY-22-077}
\hbox{MIT-CTP 5434}
\hbox{Nikhef 22-005}
}}

\begin{document}
\maketitle

%%%%%%%%%%%%%%%%%%%%%%%%%%%%%%%%%%%%%%%%%%%%%
\section{Introduction}
\label{sec:intro}
%%%%%%%%%%%%%%%%%%%%%%%%%%%%%%%%%%%%%%%%%%%%%

Most collisions at the Large Hadron Collider (LHC) involve jets.
The patterns exhibited by these collimated sprays of hadrons---known as a jet's \emph{substructure}---reveal important information about the partons initiating jet formation and about the underlying hard scattering process.
In the early days of jet substructure, tailored observables were developed to probe jets in specific theoretically-motivated ways~\cite{Seymour:1991cb,Seymour:1993mx,Butterworth:2002tt,Butterworth:2007ke,Butterworth:2008iy,Kaplan:2008ie,Thaler:2008ju,Almeida:2008yp,Thaler:2010tr}.
In recent years, the focus has shifted towards using machine learning strategies, with the goal of exploiting all information available inside of jets~\cite{Gallicchio:2011xq,Cogan:2014oua,Almeida:2015jua,Baldi:2016fql,Komiske:2016rsd,Guest:2016iqz,Louppe:2017ipp,Dreyer:2018nbf,Andreassen:2018apy,Komiske:2018cqr}.
This in turn raises the question about how jet information should be represented to ensure robust jet analyses, e.g.~whether the inputs to a neural network should be a jet image, a sequence/set of particles, a clustering tree, a graph, or a collection of observables.

In this paper, we use the technique of power counting to systematically organize jet observables from first principles.
Power counting has already seen successes in identifying specific jet observables~\cite{Larkoski:2014gra,Larkoski:2014zma,Larkoski:2015uaa}, and it forms the basis of precision calculations in soft-collinear effective theory~\cite{Bauer:2000yr,Bauer:2001ct,Bauer:2001yt,Bauer:2002nz,Beneke:2002ph}.
Here, we apply power counting to study energy flow polynomials (EFPs)~\cite{Komiske:2017aww}, which are an overcomplete linear basis for infrared-and-collinear safe jet substructure.
EFPs have been used in a variety of machine learning tasks~\cite{Komiske:2018vkc,Kasieczka:2019dbj,Faucett:2020vbu,Collado:2020fwm,Collado:2020ehf,Dillon:2021gag,Lu:2022cxg,Bradshaw:2022qev}, so it is a natural context to study how best to represent jet information.
By exploiting power counting,
we show how to simplify the EFP basis for analysis tasks involving quark and gluon jets.

Each EFP is an $N$-point correlator on jets with angular degree $d$, which can be represented by a graph with $N$ nodes and $d$ edges.
The set of all multigraphs corresponds to the set of all EFPs.
While a complete basis of jet substructure observables is conceptually important, any practical application has to limit the basis in some way.
For the related case of $N$-subjettiness~\cite{Thaler:2010tr,Thaler:2011gf,Stewart:2010tn}, the convergence as a function of $N$ was explored in \refscite{Datta:2017rhs,Moore:2018lsr} in the context of machine learning. 
For EFPs, a natural way to truncate the basis is to restrict the angular degree $d$ of the EFPs, which corresponds to the limiting the total number of graph edges.

 \begin{table}[t]
\centering
\begin{tabular}{|cc||c|c|c|c|c|c|c|}
\hline
\multicolumn{2}{|c||}{Degree }                                  & 0 & 1 & 2 & 3 & 4 & 5 & 6 \\ \hline\hline
\multicolumn{1}{|c|}{\multirow{2}{*}{All EFPs (\tab{All-Prime-EFPs})}}                     & by degree    & 1  &  1 & 3  & 8  & 23  & 66  & 212   \\ \cline{2-9} 
\multicolumn{1}{|c|}{}                                                            & cumulative      & 1  &  2 & 5  & 13  & 36  & 102  & 314  \\ \hline\hline

\multicolumn{1}{|c|}{\multirow{2}{*}{Strongly-ordered basis (\tab{LLbasis})}}                       & by degree         & 1 &   1&   2 &  4 & 7  & 12  & 22  \\ \cline{2-9} 
\multicolumn{1}{|c|}{}                                                             & cumulative     & 1 &   2 &  4 & 8  & 15  & 27  & 49  \\ \hline \hline

\multicolumn{1}{|c|}{\multirow{2}{*}{2-collinear basis (\tab{EFP_basis_6})}}          & by degree        & 1  &  1&  2 & 4  & 8  &17   & 37  \\ \cline{2-9} 
\multicolumn{1}{|c|}{}                                                             &cumulative   & 1  &   2 & 4  & 8  & 16  &  33 & 70  \\ \hline \hline

\multicolumn{1}{|c|}{\multirow{2}{*}{$z^2$-truncated basis (\tab{z2_basis})}} & by degree      & 1 &   1&   3&  5 & 9  & 13  & 20  \\ \cline{2-9} 
\multicolumn{1}{|c|}{}                                                             & cumulative     & 1 &   2 &  5 & 10  & 19  & 32  & 52 \\ \hline \hline

\multicolumn{1}{|c|}{\multirow{2}{*}{Color-reduced basis (\tab{ColorReducedBasis})}} & by degree      & 1 &   1&   2&  4 & 8  & 16  & 34  \\ \cline{2-9} 
\multicolumn{1}{|c|}{}                                                             & cumulative     & 1 &   2 &  4 & 8  & 16  & 32  & 66  \\ \hline

\end{tabular}
\caption{The number of all EFPs of/up to degree $d$, as well as the number of basis elements needed in the strongly ordered, 2-collinear, $z^2$-truncated, and color-reduced (1-collinear) expansion.
For our studies, we restrict our attention to $d > 0$ EFPs.}
\label{tab:counting}
\end{table}

As we will show, even when restricting to fixed degree $d$, there is substantial redundancy between the EFPs when 
working to a certain level of approximation.
To obtain these redundancy relations, we employ power counting, inspired by the pioneering work in \refcite{Larkoski:2014gra}.
We consider several different schemes for performing the power counting, as summarized in \tab{counting}.
These range from strongly-ordered emissions (typical of  calculations at leading-logarithmic order) to an expansion in the number of collinear (energetic) emissions in which all angular correlations are kept.
The EFP relations we obtain are valid for single prong jets, i.e.~those initiated by a light quark or gluon.
We test these EFP relations using the parton shower generator \Pythia~\cite{Sjostrand:2014zea}, finding reasonable to very good agreement, depending on the choice of power counting scheme.

For any application of power counting, one has to decide the meaning of ``leading power.''
This in turn corresponds to making an assumption about what form the optimal observable should take for a given task.
We explore two different power counting assumptions in the body of the text: 
\begin{itemize}
    \item \textbf{Strongly-ordered expansion}:  The emissions in the jet are assumed to be strongly ordered in both energy and angle. 
    \item \textbf{Collinear expansion}:  The jet is assumed to consist of collinear and collinear-soft emissions, and we expand in the number of collinear emissions, keeping all angular information.  
\end{itemize}
 We consider both the 1-collinear and 2-collinear expansion.
 The strongly-ordered expansion is a further expansion of the 1-collinear case. 
 The expansions we consider are not directly related to the logarithmic accuracy of a calculation (which depends on the details of the observable).
 In general, leading-logarithmic (LL) accuracy lies between the strongly-ordered and 1-collinear expansion, while the 2-collinear expansion holds at next-to-leading logarithmic (NLL) order.%
\footnote{\label{footnote:LLdef} Here we count logarithms in the cross section, $\int^{\mathcal{O}} {\rm d} \mathcal{O'}\, {\rm d} \sigma/{\rm d} \mathcal{O'} = \sum_{n,k}c_{n,k} \alpha_s^n L^k + \dots$, where $L$ are logarithms of the observables $\mathcal{O}$. In this counting, $k=2n$ corresponds to LL, and $k=2n-1$ corresponds to NLL.}

  In both the collinear and strongly-ordered expansions, we assume that the optimal observable for a given jet substructure task is well approximated by a single EFP, or by a sum of EFPs with no fine-tuned cancellations between terms.
  In practice, this means that we start from the full set of EFPs, using the redundancy relations to reduce the basis.
 Like with any linearly redundant system, the choice of EFP basis elements is not unique, with differences appearing beyond the chosen level of accuracy.  
 We also explore an alternate scheme in \app{energy-expansion}:
  \begin{itemize}
    \item \textbf{Energy truncation}:  The optimal observable for a given jet substructure task is assumed to have an expansion as a series in the momentum fractions $z$.
\end{itemize}
The energy expansion yields reasonable performance, but not much conceptual insight.

Power counting not only allows us to conceptually simplify the EFP basis, but in the 1-collinear case, it also reduces their computational cost.
Naively, the complexity to compute an $N$-point EFP on $M$ particles scales like $\mathcal{O}(M^N)$.
Using variable elimination, this can be reduced to $\mathcal{O}(M^t)$, where $t$ is the tree-width of the graph representing the EFP~\cite{Komiske:2017aww}.%
\footnote{For the special case where the angular distance can be expressed through an inner product, this can be further reduced to $\mathcal{O}(v^3 M)$~\cite{Komiske:2019asc}, where $v$ is the maximum number of lines connecting to a single node.}
In the limit of just one collinear emission, power counting allows us to further ``cut open'' the highest tree-width graphs and express them in terms of lower tree-width graphs.
As discussed in \app{color-reduced}, this yields computational gains with a modest decrease in machine learning performance.

To demonstrate that our reduced bases of EFPs perform as well as using all EFPs on single-prong jets, we carry out quark/gluon tagging and regression studies.
While the collinear and strongly-ordered expansions differ substantially in the numerical accuracy of the  relations derived by power counting, interestingly, their performance in the regression study is nearly identical.
This suggests that alternative approaches to obtaining the relations between EFPs should be possible to get the correct coefficients at LL accuracy.
Indeed, we demonstrate that much better expressions for EFPs in terms of the strongly-ordered basis can be obtained via linear regression.

The rest of this paper is organized as follows.
In \sec{EFPreview}, we discuss the power counting for EFPs and obtain relations between them that hold in the strongly-ordered expansion, as well as for 1 or 2 collinear particles.
Using these relationships, we identify a reduced basis of EFPs in \sec{bases} using the strongly-ordered, 1- and 2-collinear expansions.
The energy truncation and corresponding basis is discussed in \app{energy-expansion}, with complete results archived at \refcite{cal_pedro_2022_6542205}.
An alternative, computationally advantageous, basis for the 1-collinear expansion is given in \app{color-reduced}.
We test the accuracy of these relations between EFPs in \sec{results} and explore the machine learning performance of the reduced EFP bases in \sec{tagging}.
We conclude in \sec{conc} with a summary and outlook.

%%%%%%%%%%%%%%%%%%%%%%%%%%%%%%%%%%%%%%%%%%%%%
\section{Power counting of energy flow polynomials ~\label{sec:EFPreview}}
%%%%%%%%%%%%%%%%%%%%%%%%%%%%%%%%%%%%%%%%%%%%%

%-----------------------------------------------------------------------
\subsection{Review of energy flow polynomials}
%-----------------------------------------------------------------------

EFPs are $N$-point correlators on jets~\cite{Komiske:2017aww}, and they are represented by a graph with $N$ nodes.
For each node, the energy fractions $z_i$ of all particles $i$ in a jet are summed over.
The terms in these nested sums are weighted by the angles between each of the particles whose momentum fractions appear, where the exponent of $\theta_{ij}$ is equal to the number of lines between nodes contributing $z_i$ and $z_j$:
%
%%%
\begin{align}
 \includegraphics[width=0.98\textwidth]{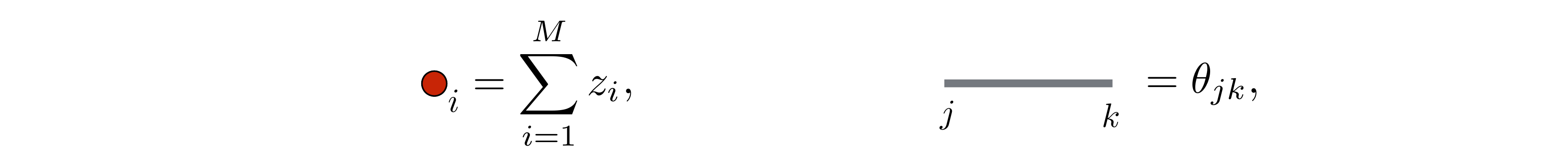}  \raisetag{2.0\baselineskip} 
\end{align}
%%%
where $M$ is the number of particles.
The precise definition of $z_i$ and $\theta_{jk}$ depends on the application of interest.
For hadron colliders, it is typical to use:
\begin{equation}
    z_i = \frac{p_{Ti}}{\sum_{j=1}^M p_{Tj}}, \qquad \theta_{jk} = \big(\Delta R_{jk} \big)^\beta,
\end{equation}
where $p_{Ti}$ is the transverse momentum of particle $i$, $\Delta R_{jk}$ is the angular distance between particles $j$ and $k$ on the rapidity-azimuth cylinder, and $\beta > 0$ is an angular weighting exponent.
For the numerical studies in this paper, we use $\beta=2$, though the power counting arguments are independent of $\beta$.
Unlike for the angles, the energy fractions must have an exponent equal to one to ensure collinear safety.

\begin{table}[t]\centering
 \includegraphics[width=0.83\textwidth]{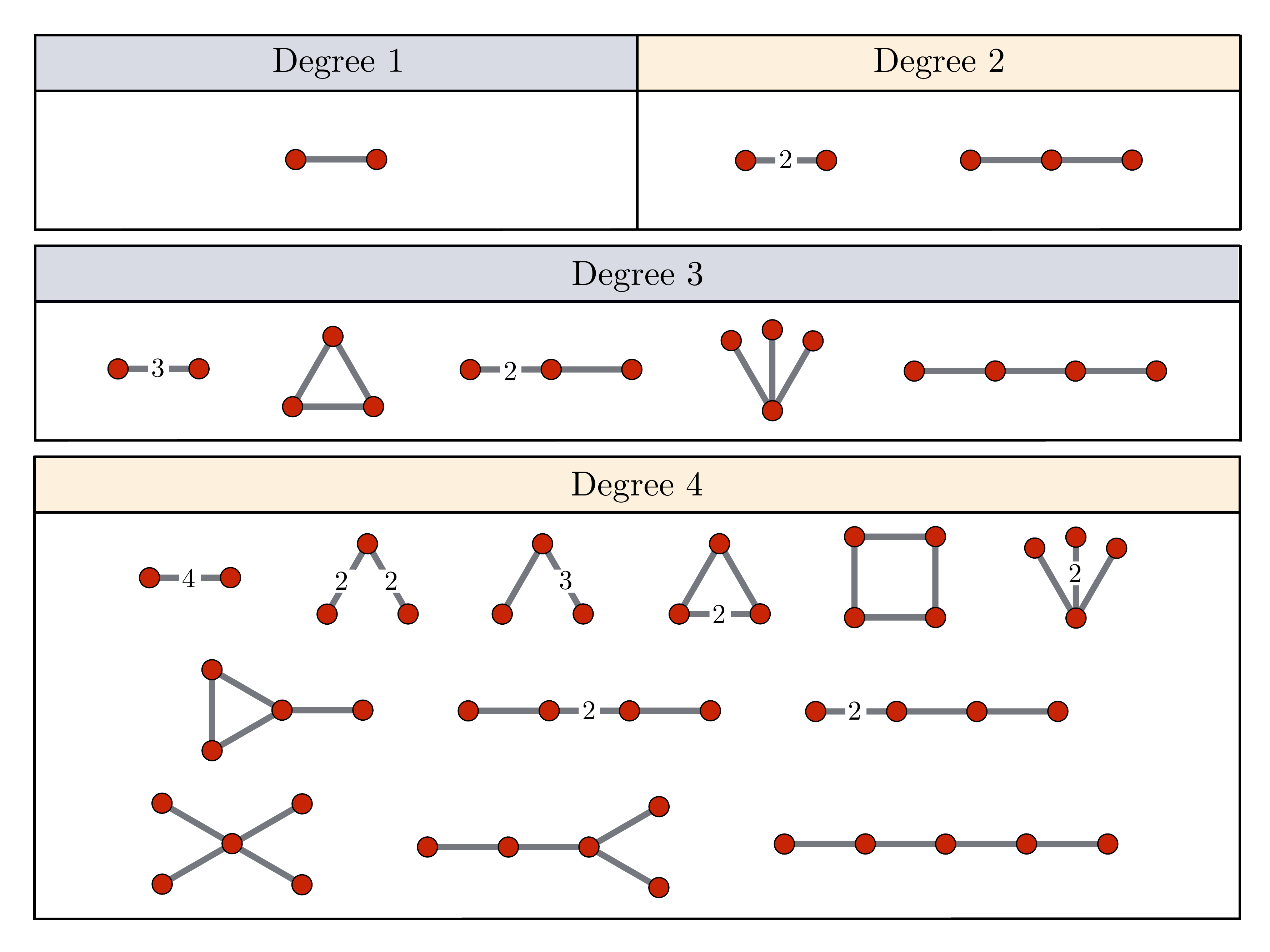} \hfill 
 \caption{
 All prime (connected) EFPs up to degree 4.
 For legibility, the numbers indicate the multiplicity of lines connecting the corresponding nodes.
 To form composite EFPs, one multiplies together these prime elements.
 \label{tab:All-Prime-EFPs}}
 \end{table}

As an example, the two-point correlator with three lines between the two nodes is given by:
%%%
\begin{align}
 \includegraphics[width=0.98\textwidth]{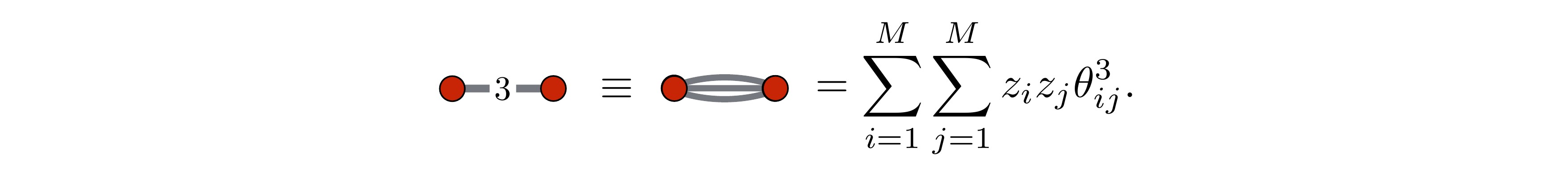}  \raisetag{2.0\baselineskip} 
 \label{eq:full-dumbell}
\end{align}
%%%
To simplify the graphs, we write the number of lines between nodes as a number, since drawing multiple lines between nodes becomes less legible for complicated EFPs.
Two more non-trivial examples are: 
%%%
\begin{align}
 \includegraphics[width=0.98\textwidth]{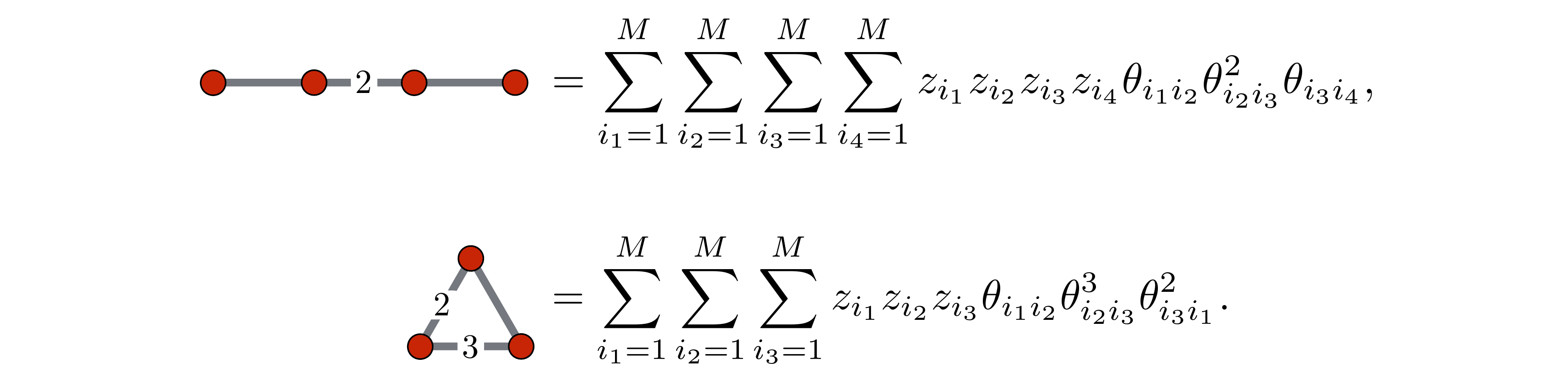}  \raisetag{1.9\baselineskip}  
\end{align}
%%%
The prime (i.e.~connected) EFPs up to degree 4 are shown in \tab{All-Prime-EFPs}.
Composite EFPs can be written as the product of prime EFPs.
The chromatic number of a graph is the number of colors needed to paint the nodes such that no two connected nodes have the same color.
A single dot corresponds to the constant 1, which we omit in our EFP regression studies.

%-----------------------------------------------------------------------
\subsection{Power counting in the strongly-ordered expansion}
%-----------------------------------------------------------------------

\begin{figure}[t]\centering
 \includegraphics[width=0.98\textwidth]{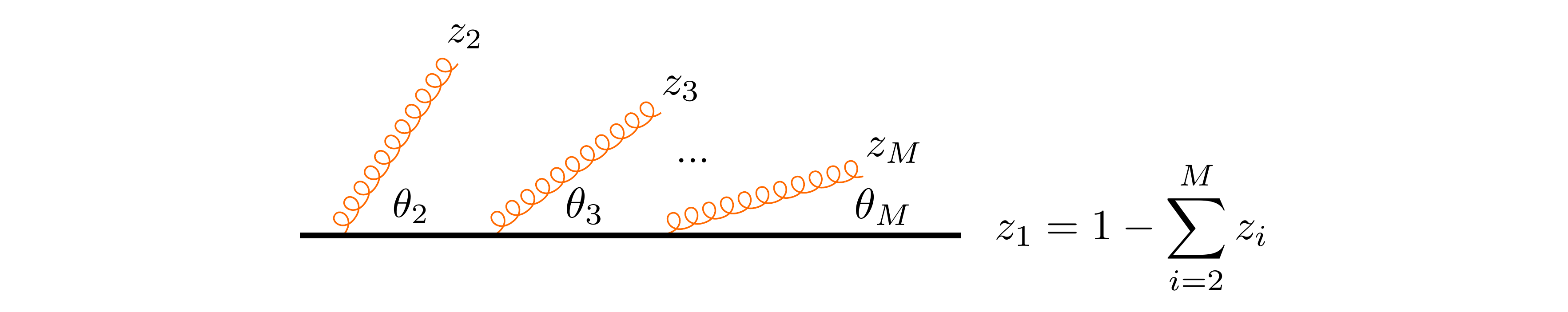} \hfill 
 \caption{At strongly-ordered (SO) accuracy, all emissions (orange) in a jet are assumed to be collinear \emph{and} soft, as well as strongly ordered.
 The parton initiating the jet (black) has momentum fraction $z_1 = 1 - \sum_{i=2}^M z_i \approx 1$ after the emissions. \label{fig:so}}
 \end{figure}

In the strongly-ordered (SO) expansion, we consider emissions that are both collinear and soft, as depicted in \fig{so}.
The jet thus consists of a hard parton ($i=1$) and $M\!-\!1$ collinear-soft gluons ($i=2,\dots, M$), which are strongly ordered in energy and angle:
%%%
\begin{align}\label{eq:strongordering}
\textbf{Strongly-ordered expansion:} \quad z_{i+1} \ll z_i\,, \qquad \theta_{1,i+1} \ll \theta_{1,i} \text{ for } i>1\,.
\end{align}
%%%
In this case, the squared matrix element, describing the probability of producing a certain jet, can be factorized into a product of matrix elements for $1 \to 2$ processes.

The measurement can also be simplified in the SO expansion.
For example, for EFPs with bipartite graphs (i.e.~with chromatic number 2), the observable will be dominated by the leading collinear-soft emission, so one can simplify the EFPs up to power corrections.
This is closely related to LL accuracy in calculations, though there one only has strong ordering in a single variable; e.g.~for the LL resummation of jet mass it is convenient to order emissions in their contribution to the mass.
In the SO limit, we are effectively assuming simultaneous strong ordering for multiple variables.

We start by applying the power counting in \eq{strongordering} to the simplest dumbbell EFP from \eq{full-dumbell}: 
%%%
\begin{align}\label{eq:LLdumbbell}
 \includegraphics[width=0.98\textwidth]{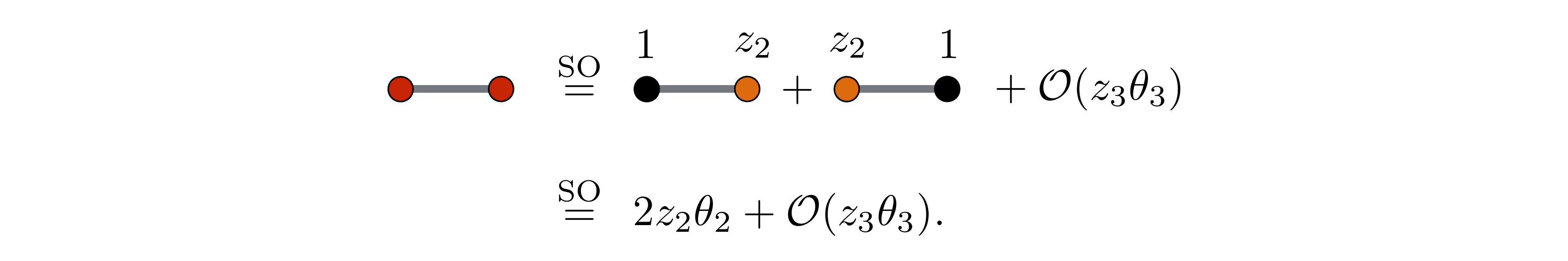} \raisetag{2.0\baselineskip}   
\end{align}
%%%
Since the momentum fraction of the hard parton is larger than that of the collinear-soft, we want to focus on the corresponding terms in the sums arising from the nodes.
However, the term in which both sums involve the hard parton vanishes, because it is weighted by a power of $\theta_{11} = 0$.
The leading contribution thus arises from the hard parton (black) contribution in the sum of one of the nodes and a collinear-soft (orange) contribution from the other.
There are two permutations, resulting in an overall factor of 2.

By using the expansion in \eq{strongordering}, more complicated EFPs can be written in terms of simpler building blocks.
As an example, consider the four-dot EFP:
%%%
\begin{align}
 \includegraphics[width=0.98\textwidth]{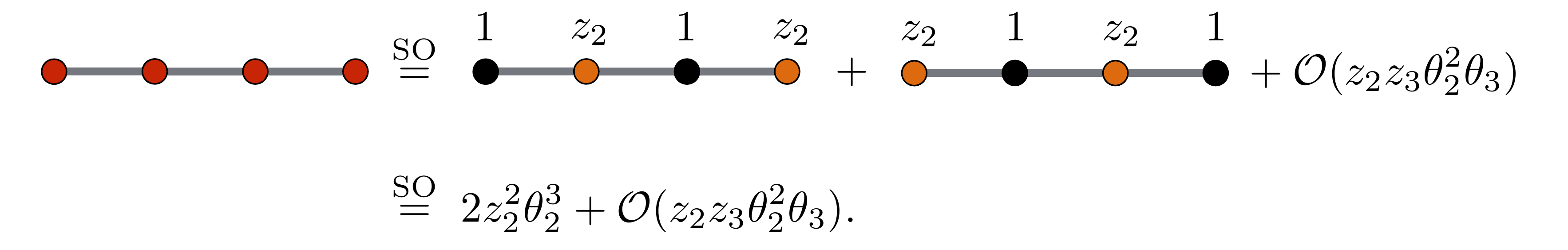} \raisetag{2.0\baselineskip}   
\end{align}
%%%
By comparing to \eq{LLdumbbell}, we see that the 4-dots EFP can be identified as a product of dumbbells.
Therefore, we conclude that in the SO expansion:
%%%
\begin{align} \label{eq:4dotsSO}
 \includegraphics[width=0.98\textwidth]{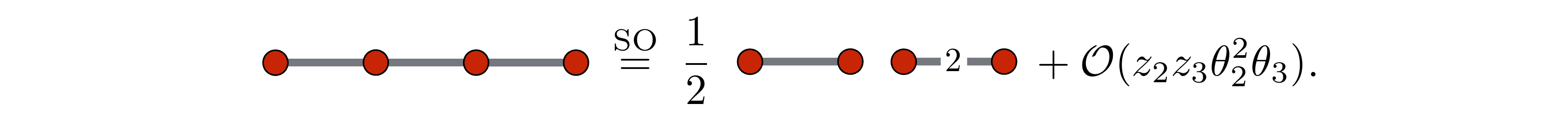} \raisetag{2.0\baselineskip}   
\end{align}
%%%
As another example, consider the crocodile EFP:
%%%
\begin{align}
 \includegraphics[width=0.98\textwidth]{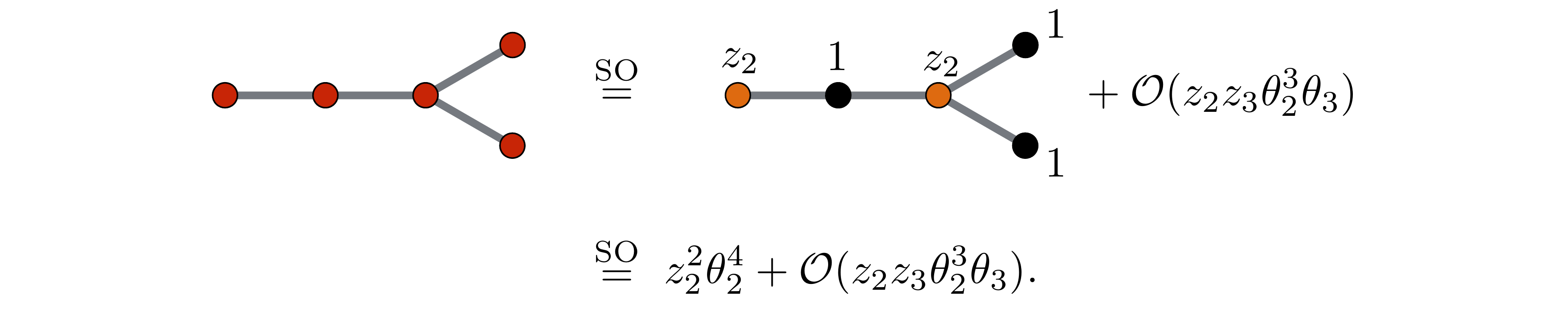} \raisetag{2.0\baselineskip}   
\end{align}
%%%
This leads to the relationship:
\begin{align} \label{eq:smallcrocSO}
 \includegraphics[width=0.98\textwidth]{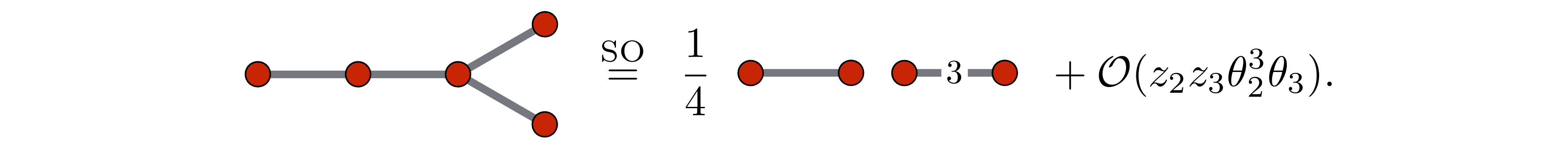} \raisetag{2.0\baselineskip}   
\end{align}
Using these results, we will derive the strongly-ordered EFP basis in \sec{SO_basis_present}.

%-----------------------------------------------------------------------
\subsection{Power counting in the 1-collinear expansion}
%-----------------------------------------------------------------------

\begin{figure}[t]\centering
 \includegraphics[width=0.98\textwidth]{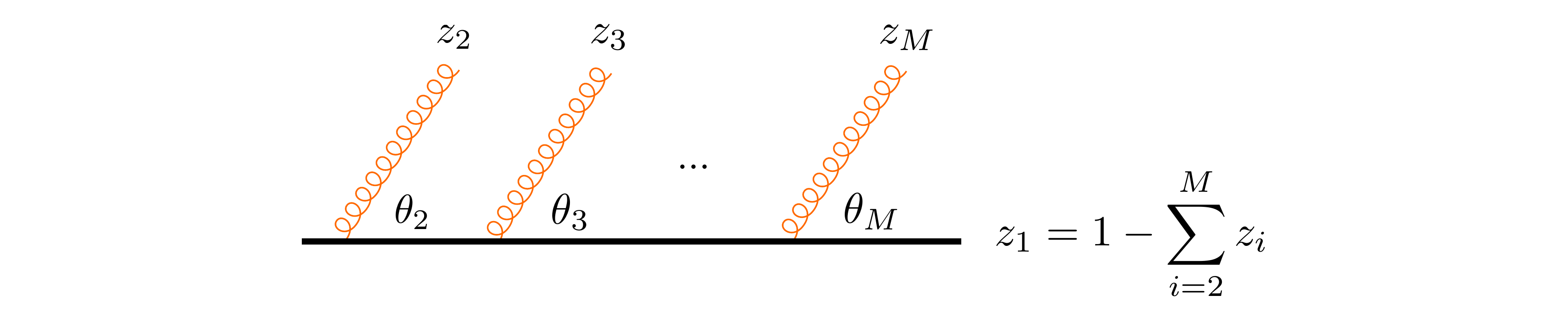} \hfill 
 \caption{At 1-collinear (1c) accuracy, all emissions (orange) are assumed to be collinear and soft.
 Unlike the SO expansion, though, here the emissions are \emph{not} strongly ordered, i.e.\ no assumptions are made on their relative energy and angular scalings. \label{fig:csoft}}
 \end{figure}

The relationships between EFPs obtained when assuming strong ordering in \eq{strongordering} are not numerically accurate, as will be shown in \sec{SO-basis}.
We therefore explore a different power counting scheme that yields much more robust relationships.

In the $n$-collinear approximation, a jet consists of $n \geq 1$ collinear parton(s) and $M-n$ collinear-soft gluons, as shown in \fig{csoft}.
In this case, we do not assume a hierarchy in angles, and we treat the energies of all collinear-soft gluons as being parametrically of the same size.
For the 1-collinear approximation, we thus assume: 
%%%
\begin{align} \label{eq:pc}
 \textbf{1-collinear expansion:} \quad z_1 = 1 + \mathcal{O}(z)\,, \qquad z_i \sim z \ll 1 \text{ for } i>1\,, \qquad \theta_{ij} \sim \theta \ll 1\,,
\end{align}
%%%

The strongly-ordered expansion in \eq{strongordering} is therefore a further expansion of the 1-collinear case.
Whereas the strongly-ordered expansion is worse than LL accurate, the 1-collinear expansion is better than LL accurate, since it includes no strong ordering of the collinear-soft emissions.

We start again by examining the dumbbell EFP, but this time applying the power counting in \eq{pc}:
%%%
\begin{align} \label{eq:dumbbell}
 \includegraphics[width=0.98\textwidth]{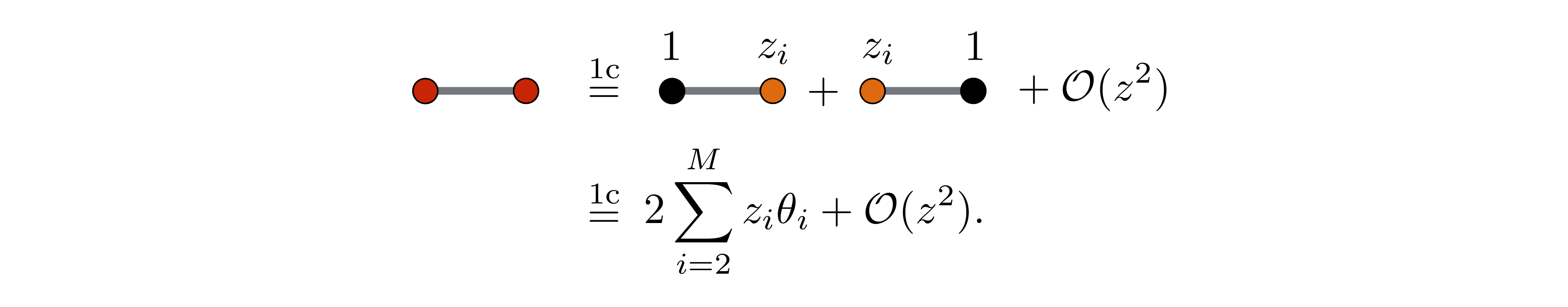} \raisetag{2.0\baselineskip}
\end{align}
%%%
In contrast to \eq{LLdumbbell}, the sum over all collinear-soft emissions is kept.
Similarly, for the triangle EFP:
%%%
\begin{align} \label{eq:triangle}
 \includegraphics[width=0.98\textwidth]{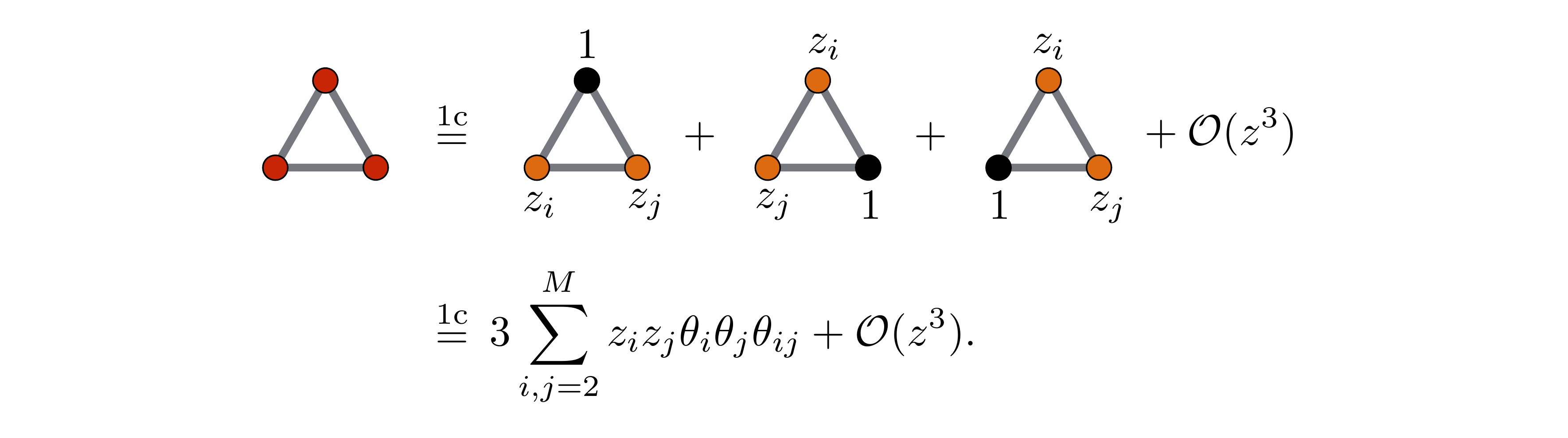} \raisetag{2.0\baselineskip}    
\end{align}
%%%
Since the graph is fully connected, the hard parton can only contribute in the sum of one of the nodes.

By using the expansion in \eq{pc}, more complicated EFPs can be written in terms of simpler building blocks.
As an example, consider the four-dot EFP:
%%%
\begin{align}\centering
 \includegraphics[width=0.98\textwidth]{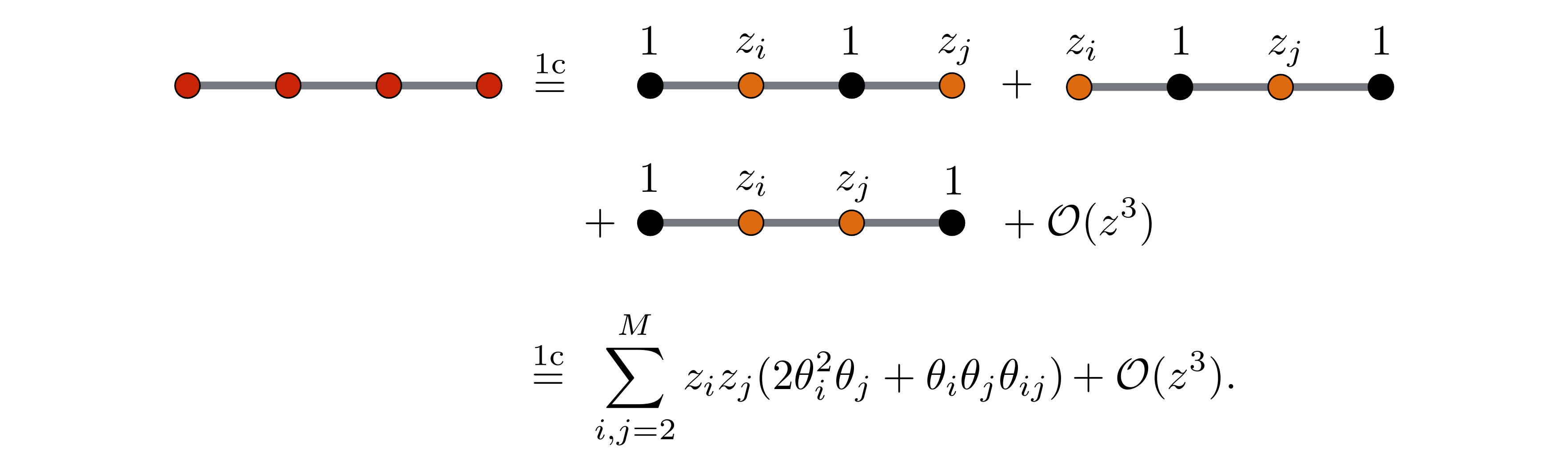} \raisetag{2.4\baselineskip}    
\end{align}
%%%
The sums in the first term of the expansion factorize and can be represented as a product of two EFPs.
By comparing to \eq{dumbbell}, these can be identified as dumbbells. The remaining term can be identified as the triangle EFP in \eq{triangle}.
Therefore we conclude that at 1-collinear accuracy:
%%%
\begin{align} \label{eq:4dotsLL}
 \includegraphics[width=0.98\textwidth]{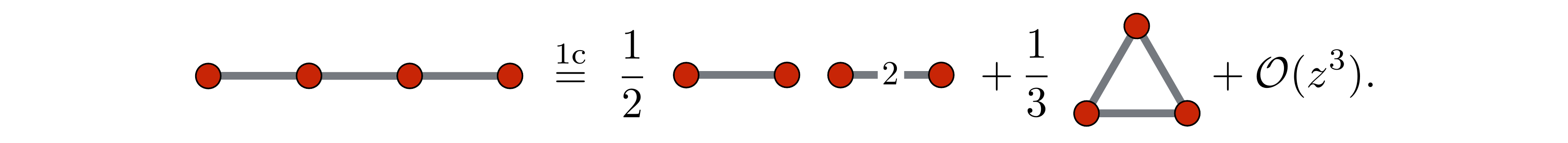} \raisetag{1.6\baselineskip}    
\end{align}
%%%

As another example, consider the crocodile EFP:
%%%
\begin{align} \label{eq:smallcroc}
 \includegraphics[width=0.98\textwidth]{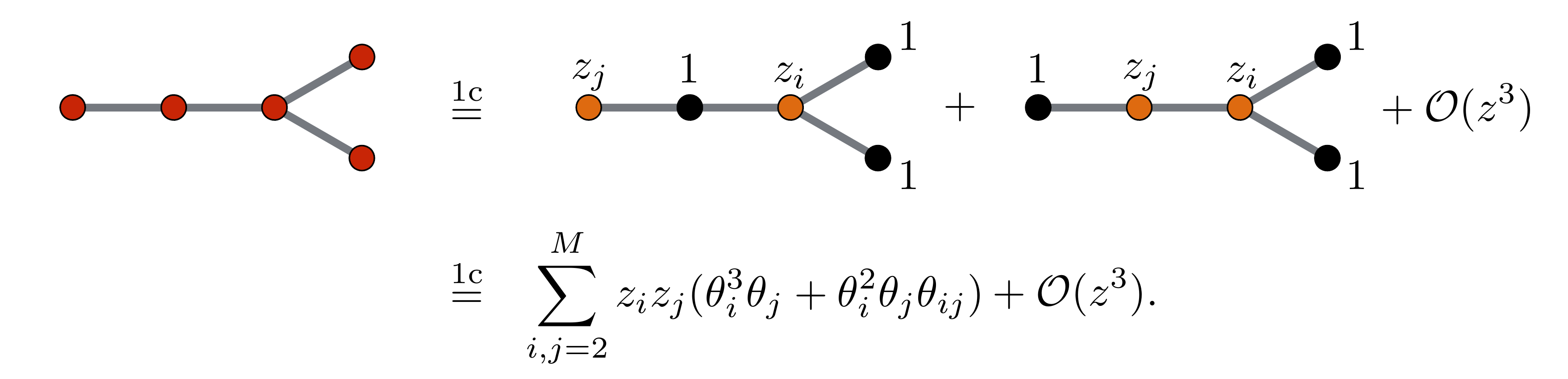} \raisetag{2.0\baselineskip}    
\end{align}
%%%
The first term can directly be identified as a product of dumbbells.
The second term looks like the triangle in \eq{triangle} except that $\theta_i \to \theta_i^2$.
This suggests considering a triangle with two lines on one side:
%%%
\begin{align}\label{eq:triangle2}
 \includegraphics[width=0.98\textwidth]{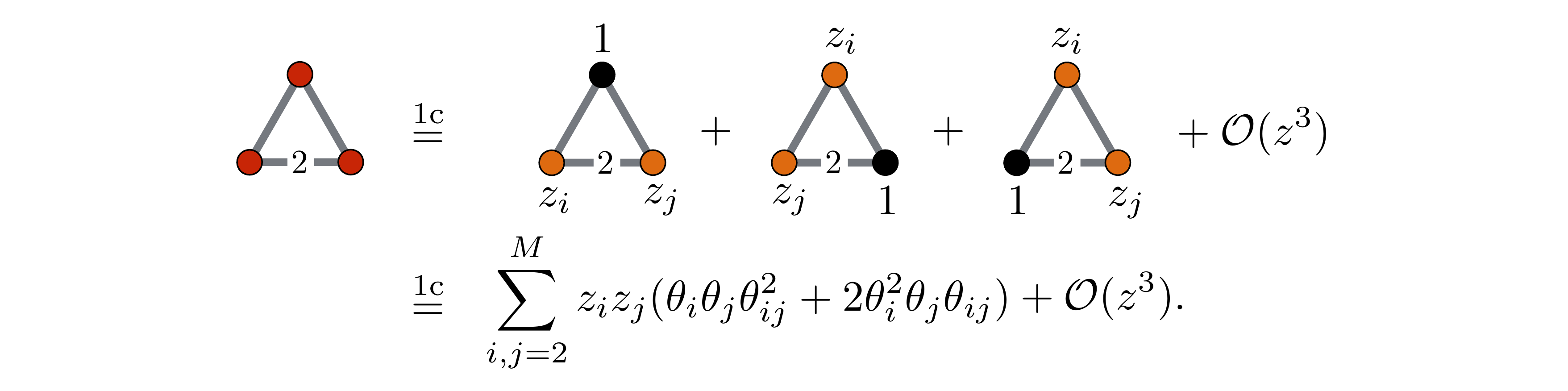} \raisetag{2.0\baselineskip}    
\end{align}
%%%
This, however, also produces an unwanted $\theta_i \theta_j \theta_{ij}^2$ term.
It turns out that the desired angular structure is instead produced by the martini glass EFP:
%%%
\begin{align} \label{eq:martini}
 \includegraphics[width=0.98\textwidth]{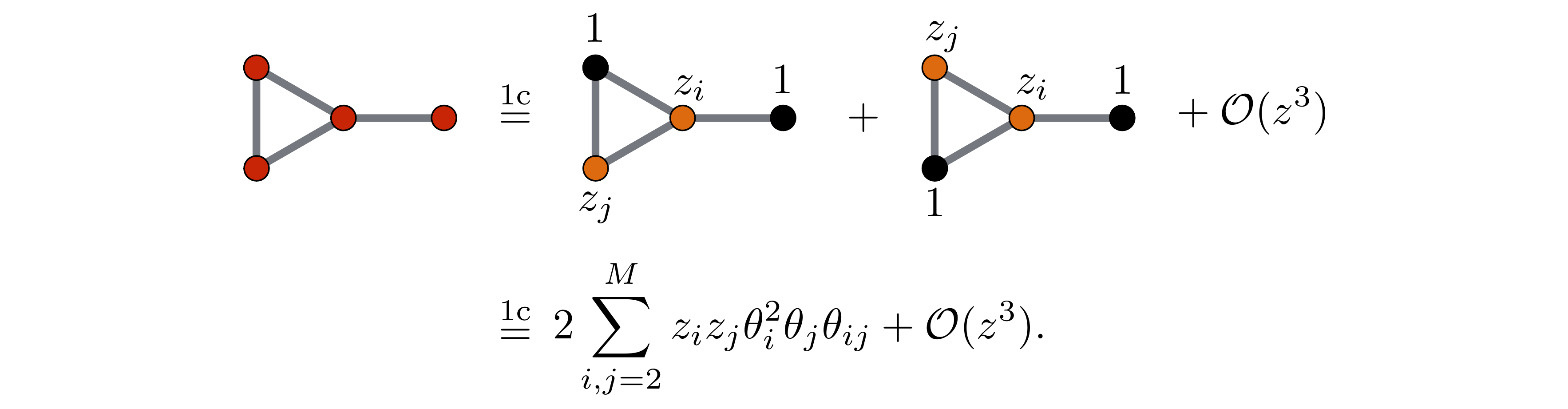} \raisetag{2.0\baselineskip}    
\end{align}
%%%
This allows us to write the crocodile EFP in \eq{smallcroc} as:
%%%
\begin{align} \label{eq:smallcrocLL}
 \includegraphics[width=0.98\textwidth]{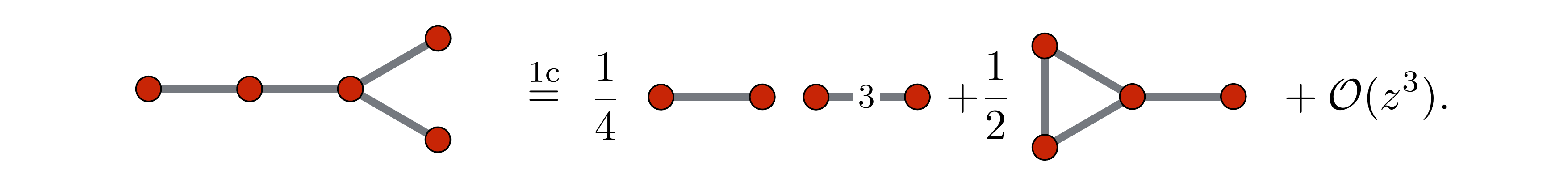} \raisetag{2.4\baselineskip}    
\end{align}
%%%
In our basis of EFPs in the 1-collinear expansion, described in \sec{1c_basis_present}, we will also need the $\theta_i \theta_j \theta_{ij}^2$ structure, which we obtain by taking the difference of \eqs{triangle2}{martini}.

%-----------------------------------------------------------------------
\subsection{Power counting in the 2-collinear expansion}
%-----------------------------------------------------------------------

\begin{figure}[t]
\centering
 \includegraphics[width=0.98\textwidth]{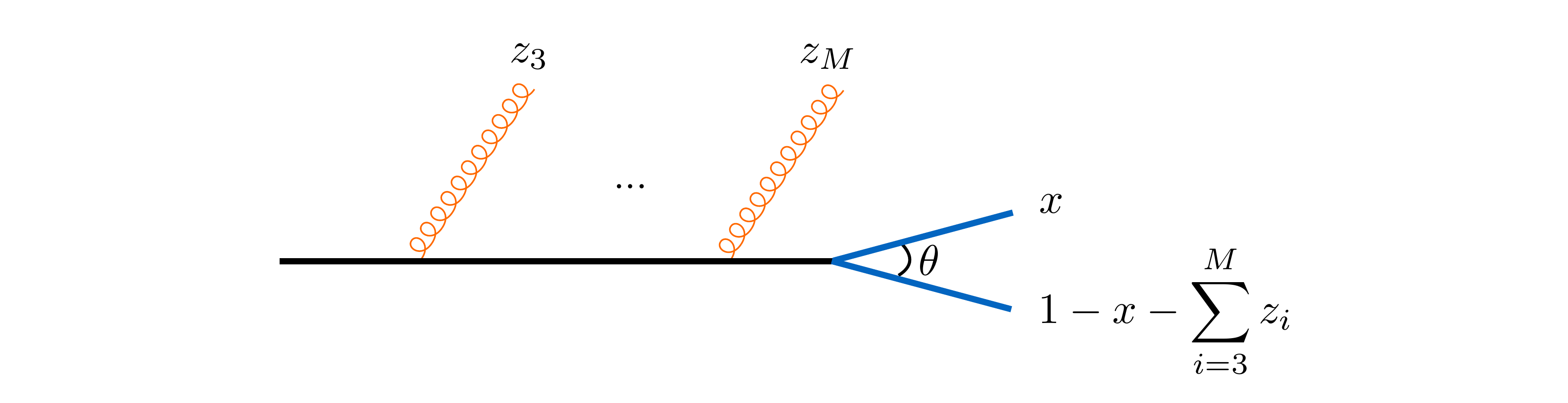} 
 \caption{In the 2-collinear (2c) approximation, one can have one collinear emission with momentum fraction $x$, in addition to collinear-soft emissions, such that the momentum fraction of the initial parton is $1-x-\sum_i z_i \approx 1-x$.
 The angle between the collinear partons is denoted by $\theta\equiv \theta_{12}$.}
  \label{fig:coll}
 \end{figure}
 
In the 2-collinear approximation, we have two collinear partons in the final state, in addition to collinear-soft emissions, as pictured in \fig{coll}. This corresponds to:
%%%
\begin{align} \label{eq:pc2}
 \textbf{2-collinear expansion:} \quad z_1 = x\,, \quad z_2 = 1-x\,, \quad z_i \sim z \ll 1 \text{ for } i>2\,, \quad \theta_{ij} \sim \theta \ll 1 \,,
\end{align}
%%%
%
where $z_2$ has $\mathcal{O}(z)$ corrections, and $x$ is $\mathcal{O}(1)$.
This expansion goes beyond NLL accuracy, for which one of the emissions from the initial parton is soft or collinear, because the collinear-soft emissions are not strongly ordered.%
\footnote{This statement is true when we count logarithms in the cross section, see footnote \ref{footnote:LLdef}. When counting logarithms in the exponent of the cross section (i.e.~in $\ln {\rm d} \sigma$), NLL requires the exponentiation of the one-loop non-cusp anomalous dimension, and it is not obvious that this would be reproduced in the 2-collinear expansion.} 

We now show some examples of the 2-collinear approximation. Let us start by expanding the dumbbell EFP:
%%%
\begin{align}
 \includegraphics[width=0.98\textwidth]{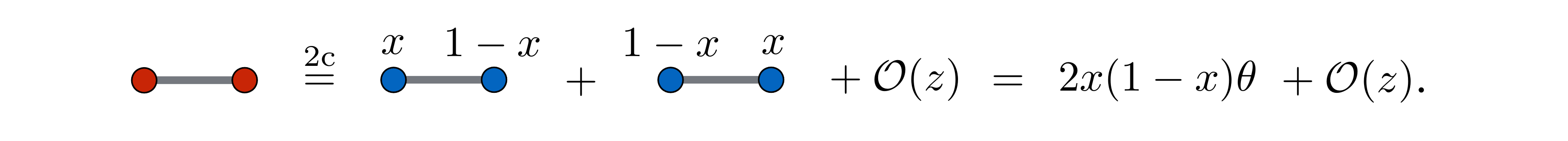} \raisetag{1.4\baselineskip}    
 \label{eq:dumbbell-2coll}
\end{align}
%%%
As another example, we look at the ``H'' EFP.  In the 1-collinear approximation it reads:
%%%
\begin{align} \label{eq:HLL}
 \includegraphics[width=0.98\textwidth]{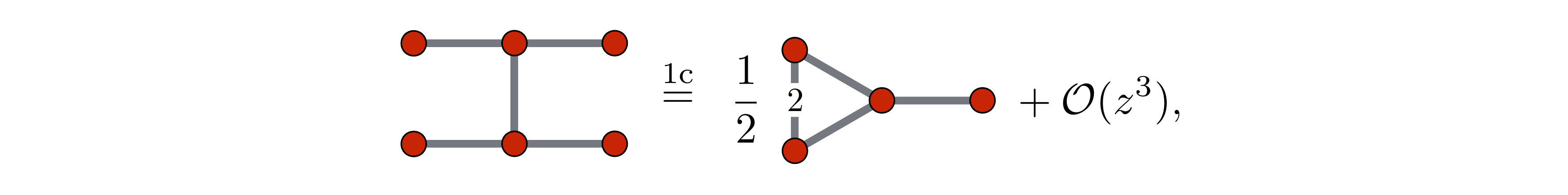} \raisetag{1.9\baselineskip}    
\end{align}
%%%
which also holds when one of the emissions is soft.
In the 2-collinear approximation, the EFP is dominated by the terms involving the two collinear particles (blue nodes):
%%%
\begin{align}
 \includegraphics[width=0.98\textwidth]{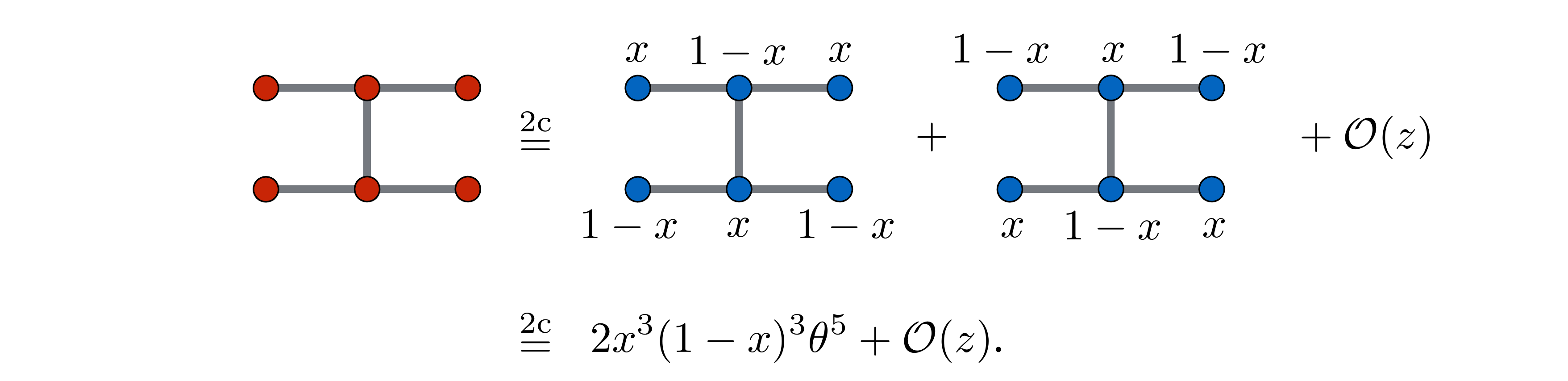} \raisetag{1.2\baselineskip}    
\end{align}
%%%
This means that in the 2-collinear approximation we can use \eq{dumbbell-2coll} to write the H EFP as
%%%
\begin{align} \label{eq:HCR}
 \includegraphics[width=0.98\textwidth]{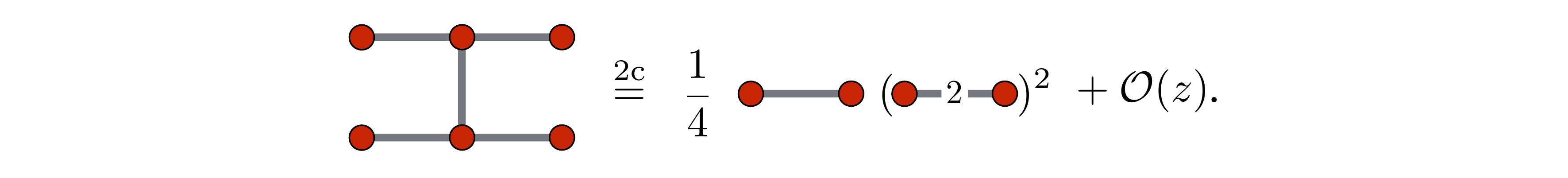} \raisetag{2.0\baselineskip}    
\end{align}
%%%
Interestingly, the general solution (denoted by $\uparrow$2c for ``up to 2-collinear'') is the sum of \eq{HLL} and \eq{HCR}, because the martini glass is suppressed in the 2-collinear approximation and the product of dumbbells is suppressed in the 1-collinear approximation:
%%%
\begin{align} \label{eq:HNLL}
 \includegraphics[width=0.98\textwidth]{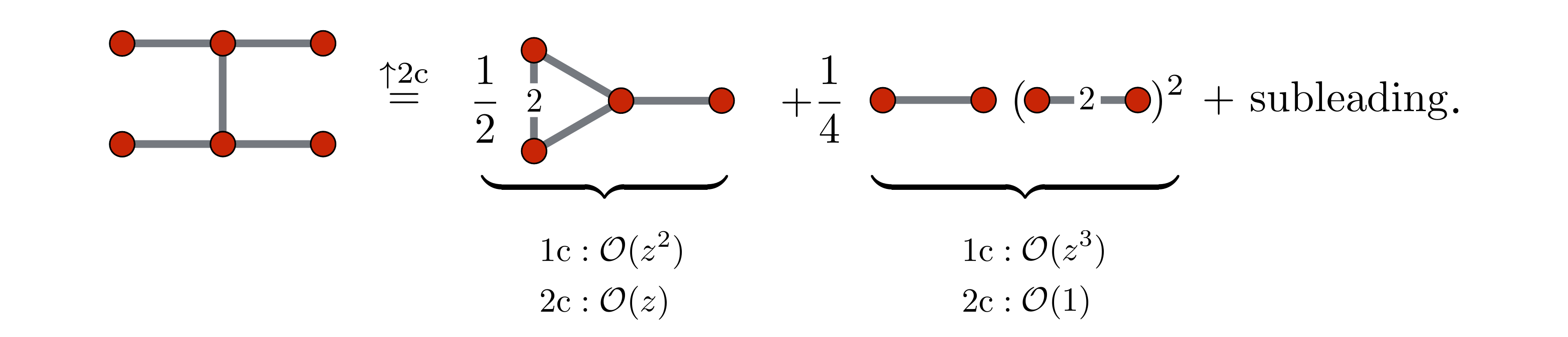} \raisetag{2.0\baselineskip}    
\end{align}
%%%

As a less trivial example, consider the ``A" EFP.
Because this is a 3-color graph and there are only two collinear particles, one needs to include collinear-soft emissions.
The leading contributions are shown below, where the assignment of the nodes is indicated by blue nodes for collinear and orange nodes for collinear-soft emissions:
%%%
\begin{align}
 \includegraphics[width=0.98\textwidth]{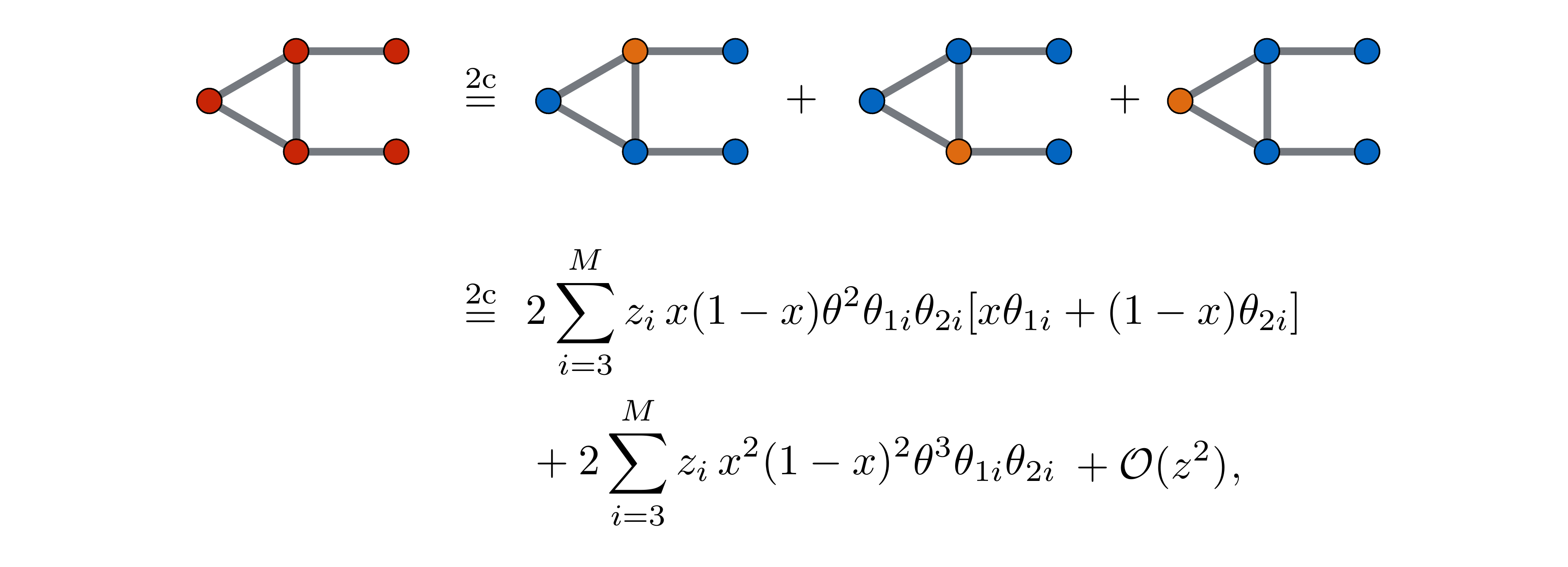} \raisetag{1.8\baselineskip}    
\end{align}
%%%
where we simplified the first term using $x (1-x)^2 + x^2(1-x) =x(1-x)$.
The momentum fractions of the blue nodes can be $x$ or $1-x$, must differ for connected nodes, and all possibilities are summed over.
In the 2-collinear approximation, we also obtain:
%%%
\begin{align}
 \includegraphics[width=0.98\textwidth]{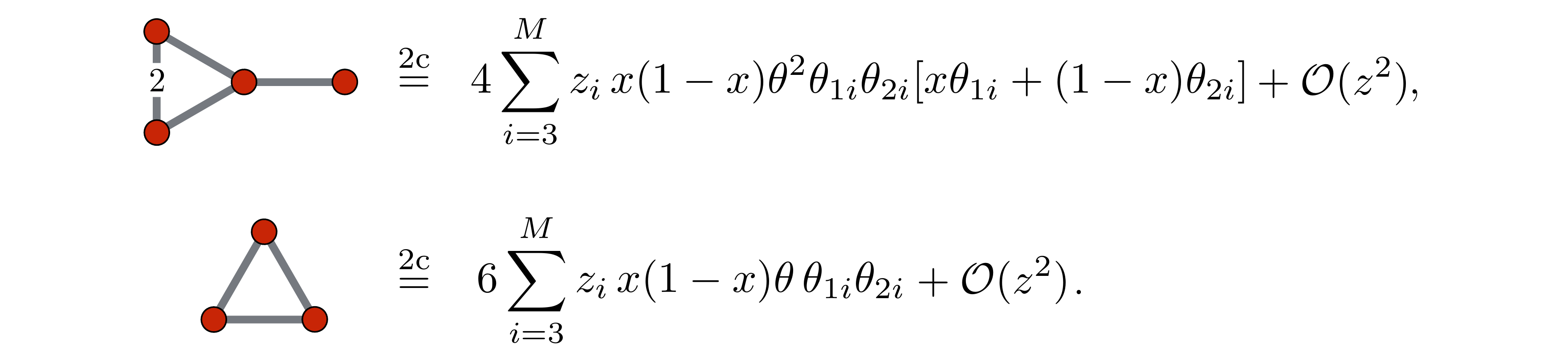} \raisetag{1.7\baselineskip}    
\end{align}
From this we obtain the 2-collinear solution, which in this case also holds in the 1-collinear approximation:
%%%
\begin{align} \label{eq:ANLL}
 \includegraphics[width=0.98\textwidth]{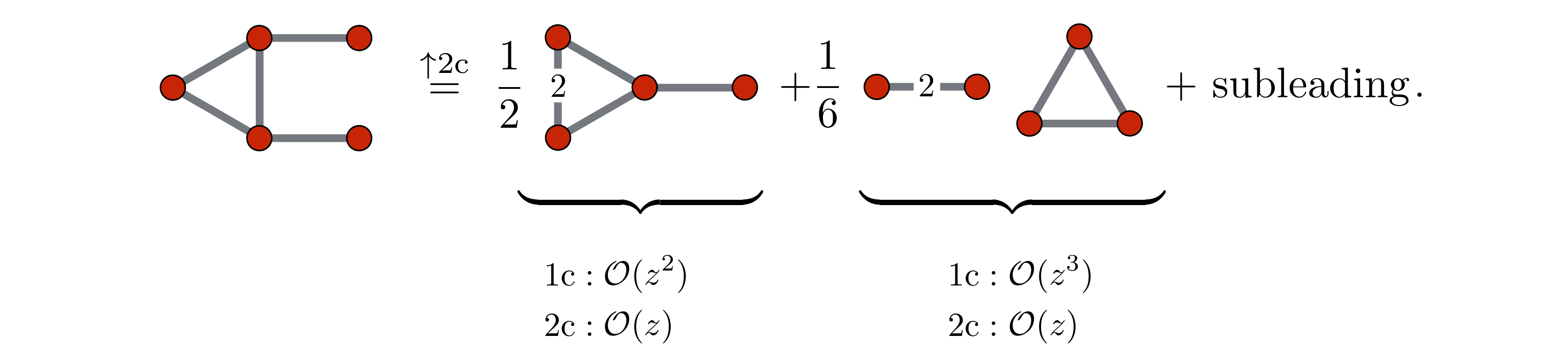} \raisetag{4.7\baselineskip}   
\end{align}
%%%
Specifically, the first term in the above equation is the solution in the 1-collinear approximation and the second term is suppressed in this case.

The above example shows that, in general, the up-to-2-collinear solution is not obtained by simply adding the individual 1- and 2-collinear solutions.
The key to systematically obtaining the full solution lies in considering the degeneracy of the individual solutions, i.e.~including all EFPs terms that are higher order in $z$ as part of the solution, with unspecified coefficients.
The up-to-2-collinear solution is then obtained by taking the intersection of the individual 1- and 2-collinear solutions, which is by definition a solution to both approximations.
We pursue this strategy in \sec{2c_basis_present}.

%%%%%%%%%%%%%%%%%%%%%%%%%%%%%%%%%%%%%%%%%%%%%
\section{Constructing reduced EFP bases}
\label{sec:bases}
%%%%%%%%%%%%%%%%%%%%%%%%%%%%%%%%%%%%%%%%%%%%%

In the previous section, we discussed specific relations between EFPs in the SO, 1- and 2-collinear approximations.
We can now put these to use systematically to reduce the full basis of EFPs.
We consider two approaches:
\begin{itemize}
\item 
 \textbf{Strongly-ordered basis:} Starting from all EFPs, we use relationships obtained in the SO expansion to eliminate as many elements as possible. 
\item
 \textbf{$n$-collinear basis:} Similar to how we obtain the SO basis, except that we only use relationships obtained in the up-to-$n$-collinear expansion.
\end{itemize}
A third approach based on a $z$ expansion is presented in \app{energy-expansion}.
The linear relations obtained in these limits are presented in \refcite{cal_pedro_2022_6542205} up to degree 6. As shown in \sec{tagging}, quark/gluon discrimination starts to saturate at around degree 6, which motivates truncating our analysis at this degree; expanding to higher degrees is a straightforward but tedious exercise.

There is of course a certain arbitrariness when using relationships to reduce a basis.
Though the difference is formally beyond the order that we are working at, the choice may matter.
In the 1-collinear approximation, in particular, it is possible to use a color-reduced basis in which only graphs with chromatic number up to $c-1$ are needed as basis elements for graphs of chromatic number $c$, as discussed in \app{color-reduced}.
This color-reduced basis is not as useful when considering the 2-collinear approximation, however, as it would require a substantial extension.
In the body of this paper, we choose a 1-collinear basis that requires a minimal extension to lift to the 2-collinear case.

%-----------------------------------------------------------------------
\subsection{The strongly-ordered basis}
\label{sec:SO_basis_present}
%-----------------------------------------------------------------------

\begin{table}[t]
  \includegraphics[width=.96\textwidth]{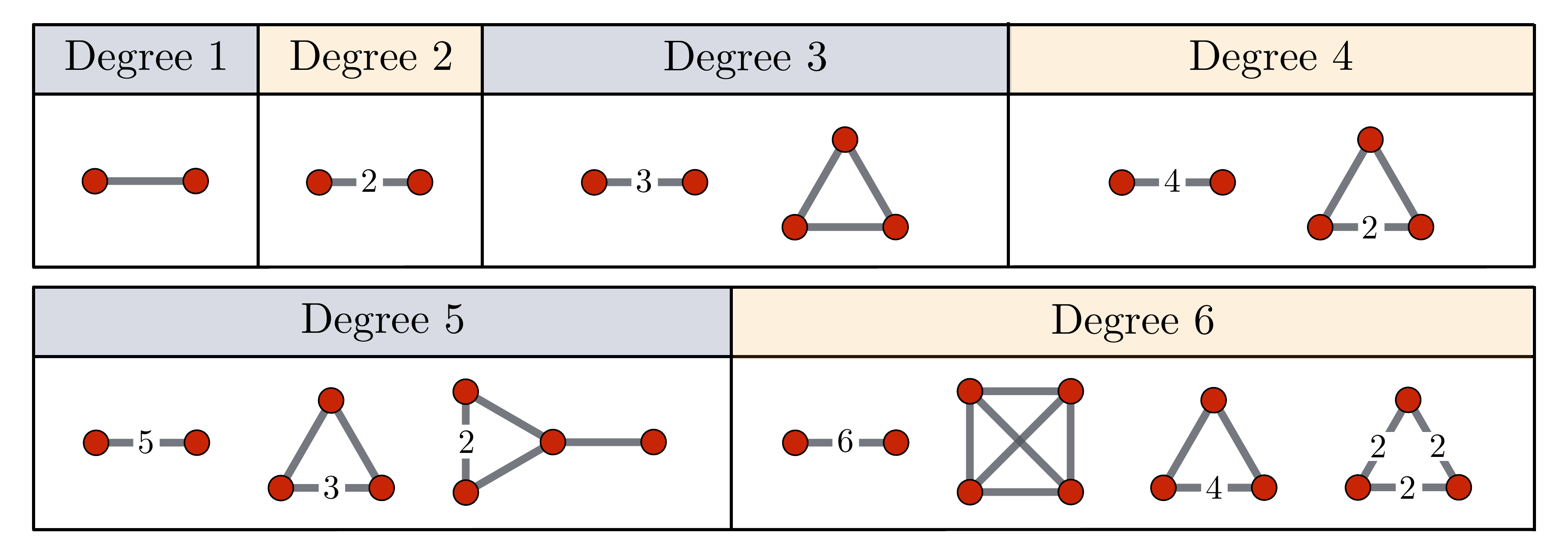}
  \caption{Basis of prime EFPs up to degree 6 in the strongly-ordered (SO) expansion.
  \label{tab:LLbasis}}
\end{table}

In the SO approximation from \eq{strongordering}, any EFP is reduced to a polynomial of terms of the form:
\begin{equation}
    \prod^{c}_{i=2} z_i^{n_i} \theta_i^{m_i},
\end{equation}
with $n_2\geq ... \geq n_c$  and $m_2> ... > m_c$, where $c$ is the chromatic number of the EFP in question.
In fact, up to degree 6 all except for one EFP consist of a single such term (with the exception being the martini glass with a double line on the rim at degree 5).
The construction of the SO basis is then performed by considering all terms of this form that can appear at a given degree:
\begin{align}
\text{Degree 1: }\hspace{0.5cm}& z_2 \theta_2; \nn \\
\text{Degree 2: }\hspace{0.5cm}& z_2 \theta_2^2;\nn \\
\text{Degree 3: }\hspace{0.5cm}& z_2 \theta_2^3,\,\,\, z_2 z_3 \theta_2^2 \theta_3; \nn \\
\text{Degree 4: }\hspace{0.5cm}& z_2 \theta_2^4,\,\,\, z_2 z_3 \theta_2^3 \theta_3; \nn \\
\text{Degree 5: }\hspace{0.5cm}& z_2 \theta_2^5,\,\,\, z_2 z_3 \theta_2^4 \theta_3, \,\,\, z_2 z_3 \theta_2^3 \theta_3^2; \nn \\
\text{Degree 6: }\hspace{0.5cm}& z_2 \theta_2^6,\,\,\, z_2 z_3 \theta_2^5 \theta_3, \,\,\, z_2 z_3 \theta_2^4 \theta_3^2,\,\,\, z_2 z_3 z_4 \theta_2^4 \theta_3 \theta_4 . 
\end{align}
Once we select representative EFPs that correspond to each of these terms in the SO approximation, all others can be written in terms of the chosen elements.
This yields the SO basis, which can be seen in \tab{LLbasis}.

Note that we could not have chosen a basis for which all of the graphs are fully connected.%
\footnote{In \refcite{Moult:2016cvt} by one of the authors, an erroneous statement was made implying that fully connected graphs form a complete basis for infrared-and-collinear-safe observables.  This statement was corrected in \refcite{Komiske:2017aww}.  Here, we see that this statement is not even true in the SO limit.  Mea culpa.}
This would require replacing the martini glass at degree 5 by a triangle.
However, the triangles with sides 2-2-1 and 3-1-1 are equivalent, since there are two large angles $\theta_{12} \sim \theta_{23}$ and one small angle $\theta_{23}$.
Thus, a partially connected graph is needed to capture the angular scaling of the martini glass.

%-----------------------------------------------------------------------
\subsection{The 1-collinear basis}
\label{sec:1c_basis_present}
%-----------------------------------------------------------------------

        \begin{table}[p]
        \centering
      \includegraphics[width=0.9\textwidth]{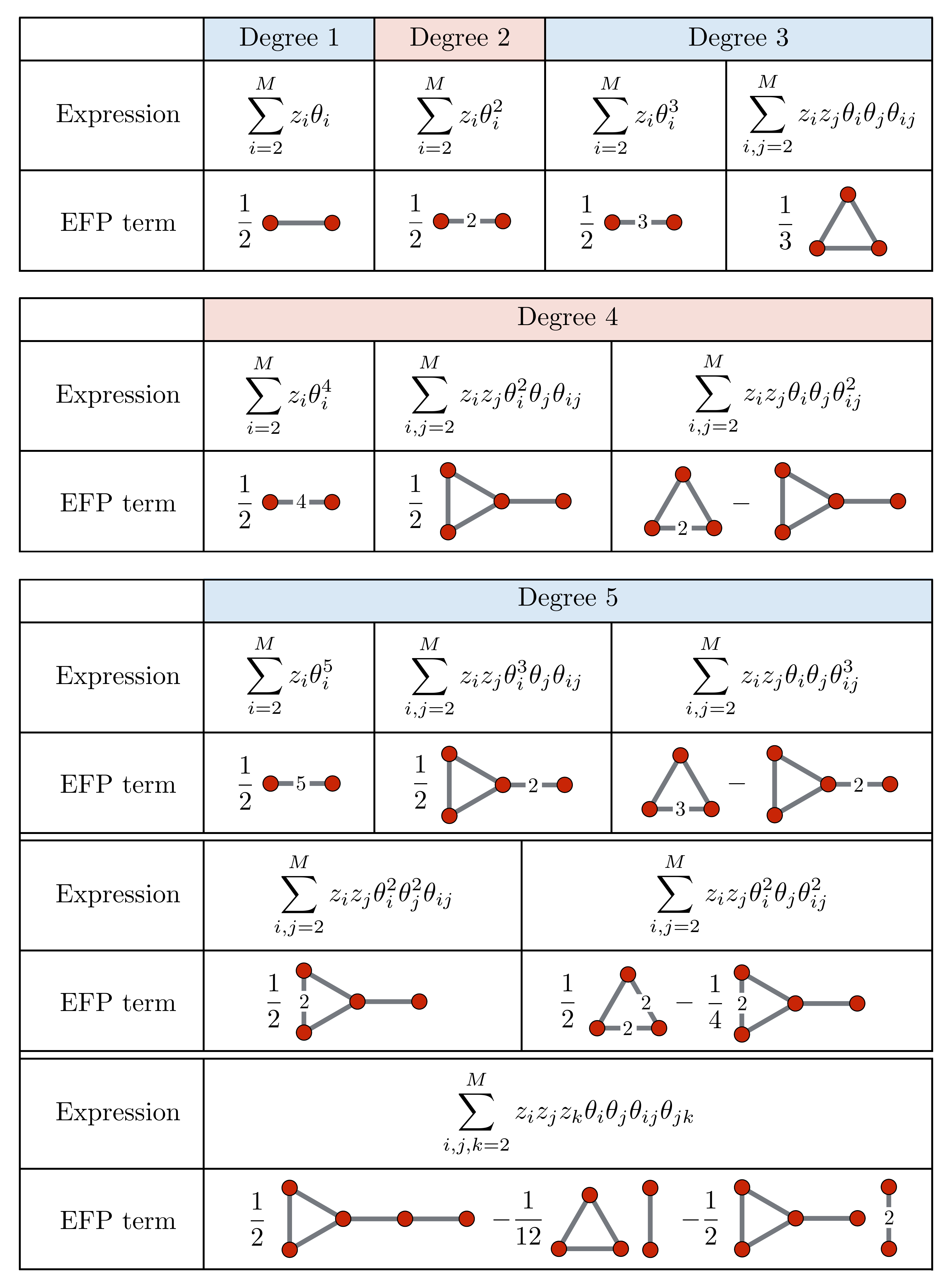}  
      \caption{Possible monomial structures that can appear in an EFP in the 1-collinear approximation up to degree 5, and the corresponding EFPs in terms of which they can be expressed. \label{tab:EFP_basis_elements}}
     \end{table}

        \begin{table}[t]\centering
      \includegraphics[width=0.92\textwidth]{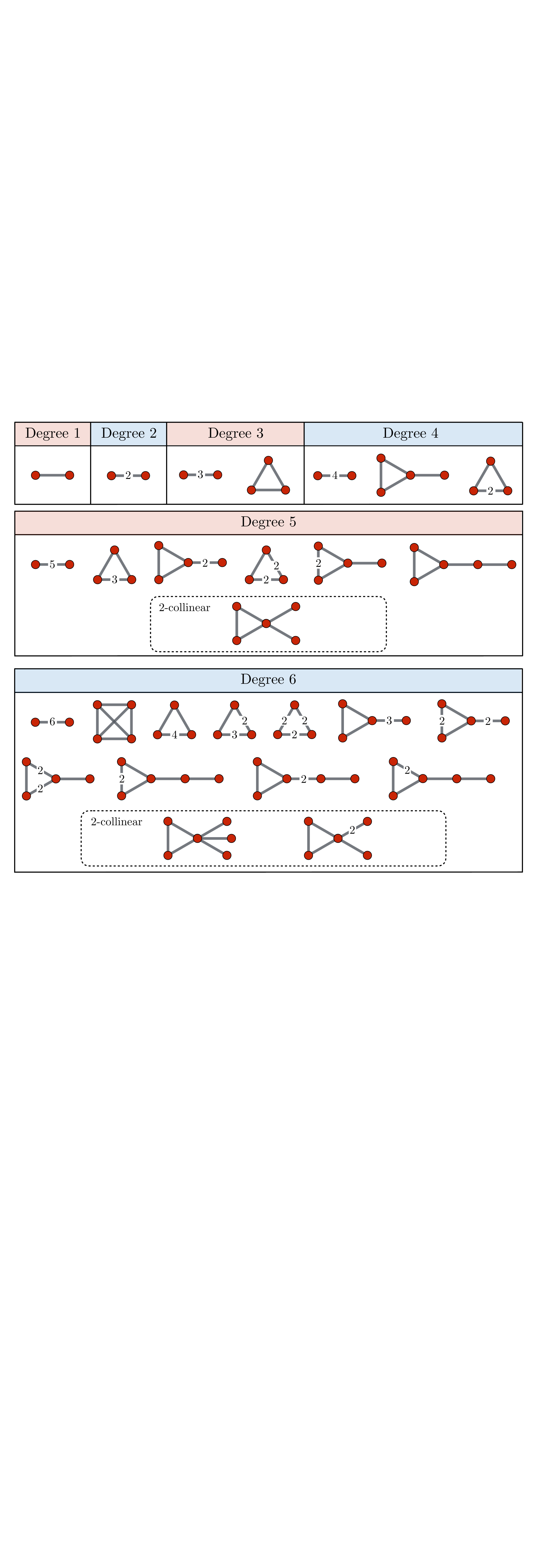}  
      \caption{Basis of prime EFPs up to degree 6 in the 1-collinear expansion. A generic EFP up to degree 6 can be expressed as a polynomial in terms of these bases elements. To extend this basis to the 2-collinear approximation requires one new basis element at degree 5 and two at degree 6. \label{tab:EFP_basis_6}}
     \end{table}

To systematically construct the $1$-collinear basis, we consider all possible monomial structures in the sum over momentum fractions and angles that can appear in the 1-collinear approximation from \eq{pc}.
These structures and the corresponding EFPs are summarized in \tab{EFP_basis_elements} up to degree 5.
Generic EFPs can be obtained as polynomials of these ingredients.
The full basis of prime EFPs up to degree 6 is given in \tab{EFP_basis_6}.
We have validated this basis by expressing all EFPs with $d \le 6$ as polynomials of these basis elements.

In the 1-collinear approximation, we could have alternatively used a color-reduced basis, discussed further in \app{color-reduced}.
In the 1-collinear approximation, there is a single collinear parton with momentum fraction 1.
This collinear parton that has be inserted somewhere on the EFP, allowing us to ``cut" the EFP there.
Letting a black dot indicating the collinear parton, we can cut open fully connected graphs via: 
\begin{align}
\label{eq:color-reduced}
 \includegraphics[width=0.98\textwidth]{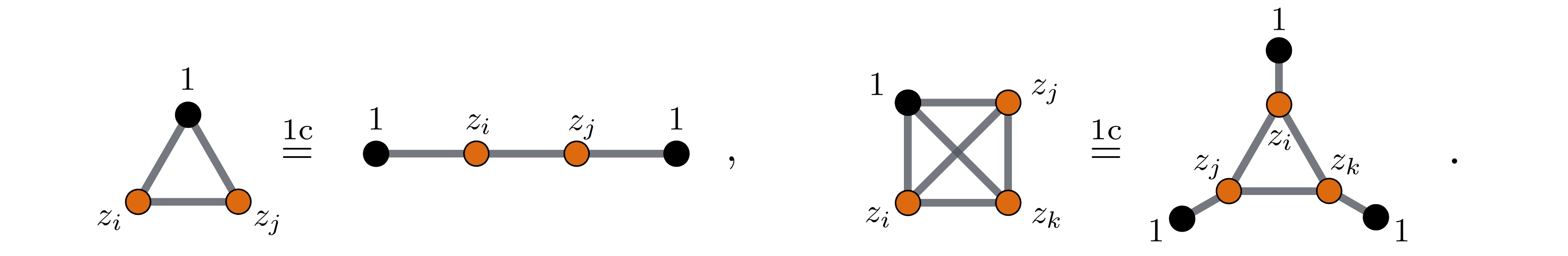}
 \raisetag{2.3\baselineskip}    
\end{align}
This leads to a computational gain, since cutting open a graph generally decreases the computational cost by a factor of $\mathcal{O}(M)$.
That said, the above relationships do not hold in 2-collinear approximation, and one cannot, for example, write the 3-color triangle as some combination of 2-color graphs.

%-----------------------------------------------------------------------
\subsection{The 2-collinear basis}
\label{sec:2c_basis_present}
%-----------------------------------------------------------------------

While there is considerable freedom in choosing basis elements in the 1-collinear case, the specific elements in \tab{EFP_basis_6} were selected to minimize the difference between the 1-collinear and 2-collinear bases.
In particular, \tab{EFP_basis_6} was constructed to ensure that essentially the same basis elements could be used in the up-to-2-collinear approximation.

To verify this, we expressed all EFPs up to degree 6 in terms of our 1-collinear basis, using the relationships that hold in the up-to 2-collinear approximation from \eqs{pc}{pc2}.
We found that this was possible for all but one EFP at degree 5 and three at degree 6, requiring the new basis elements shown in the dashed boxes of \tab{EFP_basis_6}.
The structures of these new basis elements in the 2-collinear approximation do not match that of any of the other EFPs.
For example,  
%%%
\begin{align}
 \includegraphics[width=0.98\textwidth]{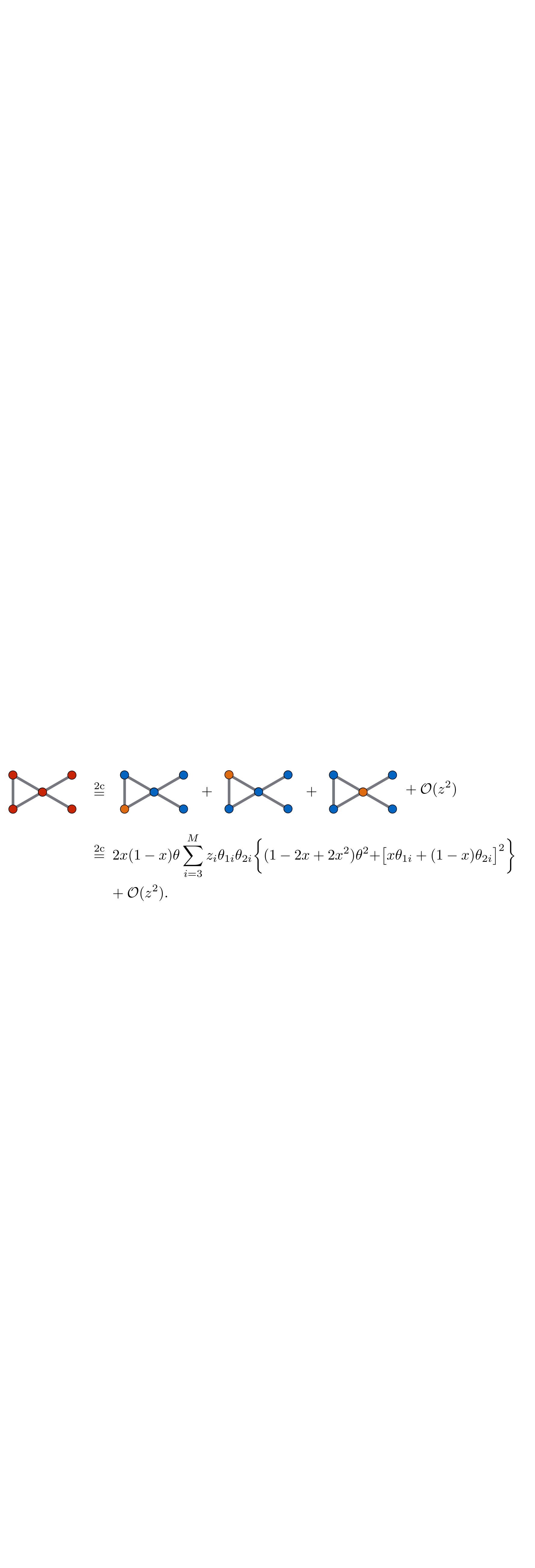} \raisetag{1.5\baselineskip}    
\end{align}
%%%
The last contribution on the first line is the challenging one, as it involves  four angles $\theta_{1i}, \theta_{2i}$ with a complicated $x$ dependence.
Since there is only a single momentum fraction $z_i$, it cannot be written as a product of two EFPs.
Similarly, removing one of the ``antennae" to change one of the lines to a double line does not capture the $x$ dependence.
We therefore conclude that this is the minimal extension needed to go from the 1-collinear to the 2-collinear basis.

%%%%%%%%%%%%%%%%%%%%%%%%%%%%%%%%%%%%%%%%%%%%%
\section{Testing linear relations}
\label{sec:results}
%%%%%%%%%%%%%%%%%%%%%%%%%%%%%%%%%%%%%%%%%%%%%

We now test the linear relations that express EFPs in terms of basis elements in various approximations.
The dataset used for this study was retrieved from \refscite{Zenodo:EnergyFlow:Pythia8QGs, Komiske:2018cqr}, in which gluon jets are obtained from $Z (\to \nu \bar{\nu})+g$ and quark jets from $ Z(\to \nu \bar{\nu}) + (u, d, s) $.
These events are generated in \Pythia 8.226~\cite{Sjostrand:2014zea} with $\sqrt{s} = 14$ TeV, hadronization and underlying event turned on, using the \texttt{WeakBosonAndParton:qqbar2gmZg} and \texttt{WeakBosonAndParton:qg2gmZq} processes.
The jets are identified using the anti-$k_T$ jet algorithm~\cite{Cacciari:2008gp,Cacciari:2011ma} with radius parameter $R=0.4$, and we select for jets with transverse momentum $p_T^{{\rm jet}} \in [500,550]$ GeV and rapidity $|y^{\rm jet}| < 1.7$.

%-----------------------------------------------------------------------
\subsection{Results with the strongly-ordered basis}
\label{sec:SO-basis}
%-----------------------------------------------------------------------

\begin{figure}[p]
\subfloat[]{
  \includegraphics[width=.48\textwidth]{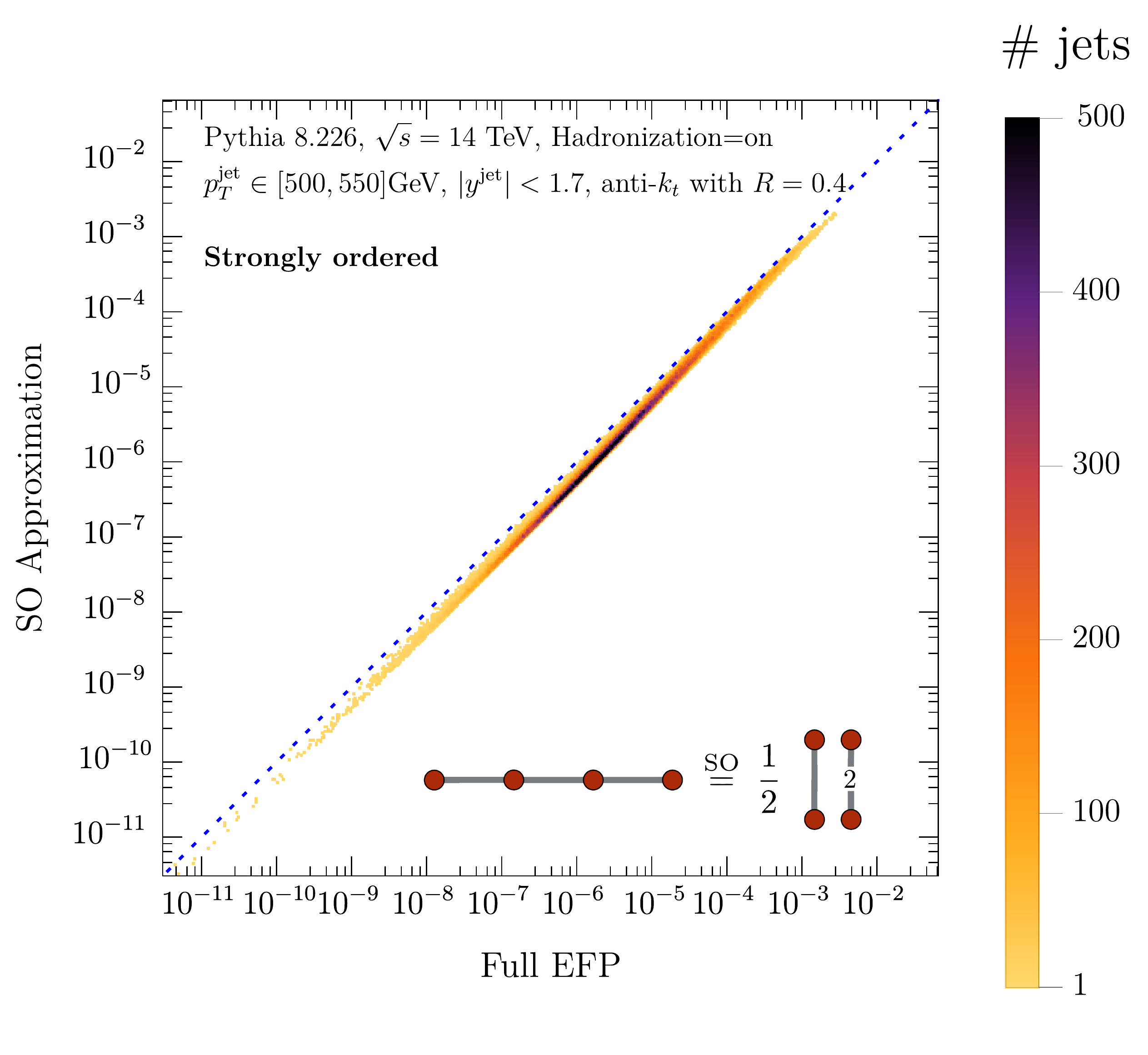}
  }
  \hfill 
  \subfloat[]{
    \includegraphics[width=.48\textwidth]{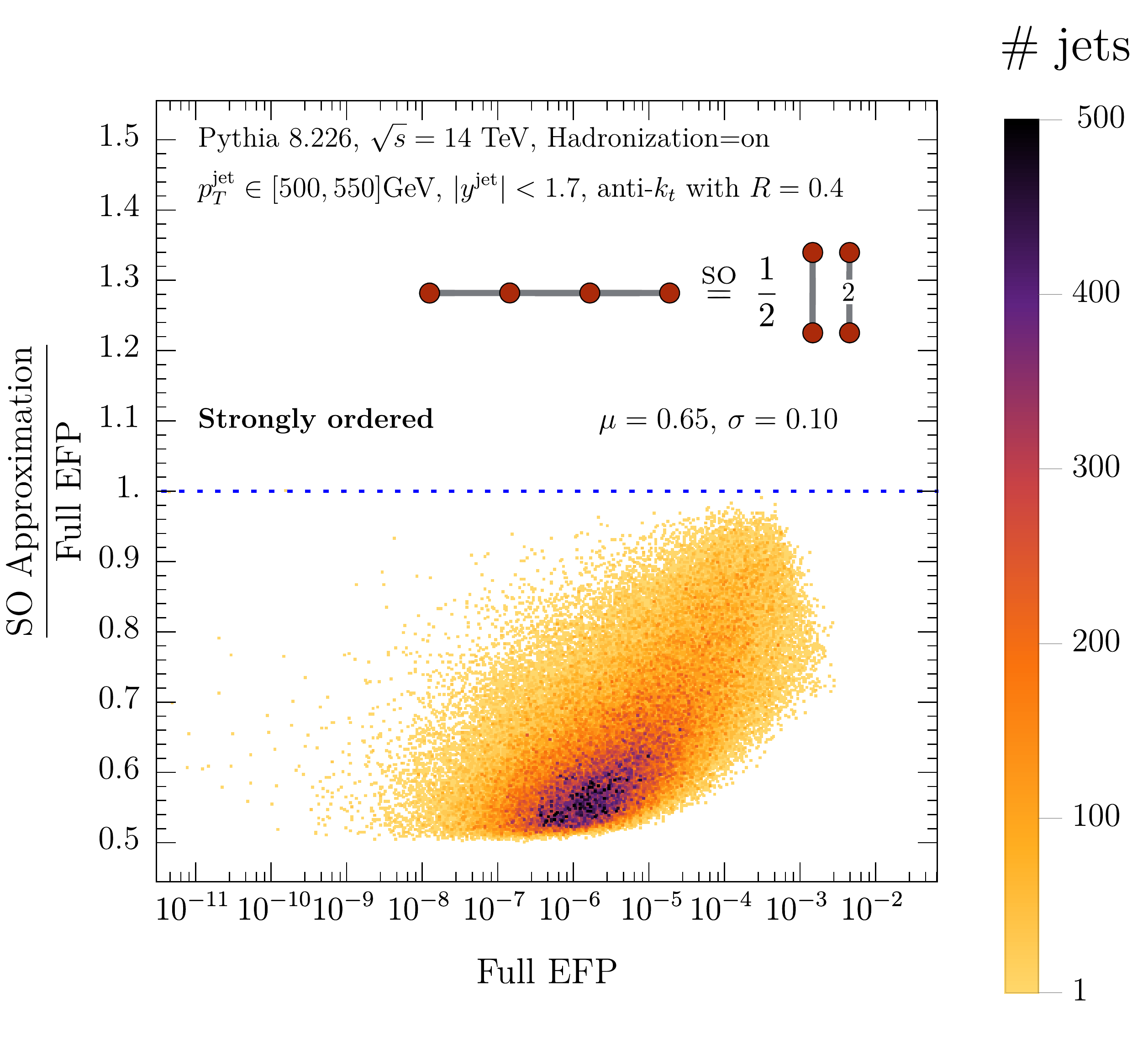}
    }\\ 
    \subfloat[]{
      \includegraphics[width=.48\textwidth]{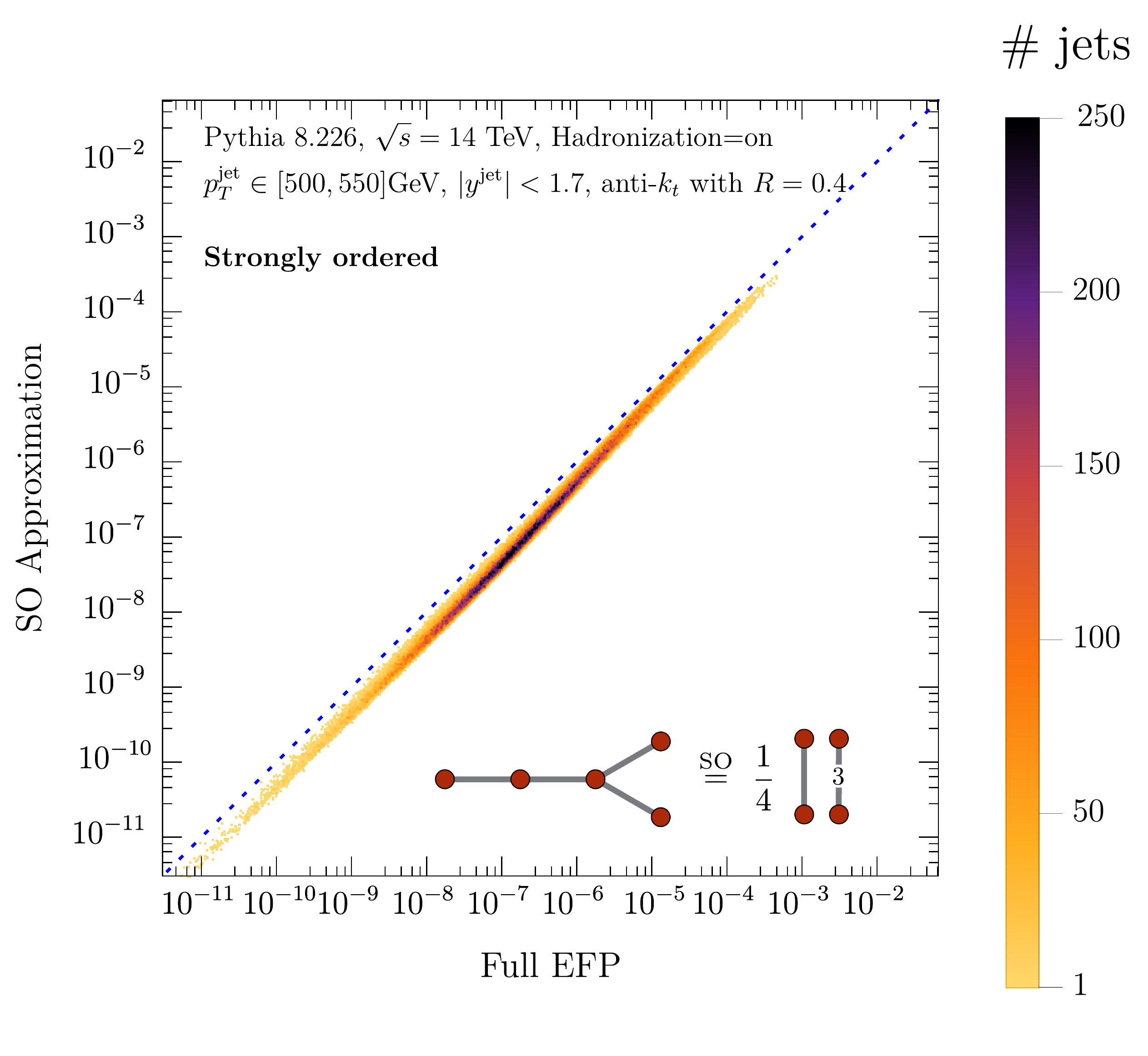}
      }
    \hfill 
    \subfloat[]{
    \includegraphics[width=.48\textwidth]{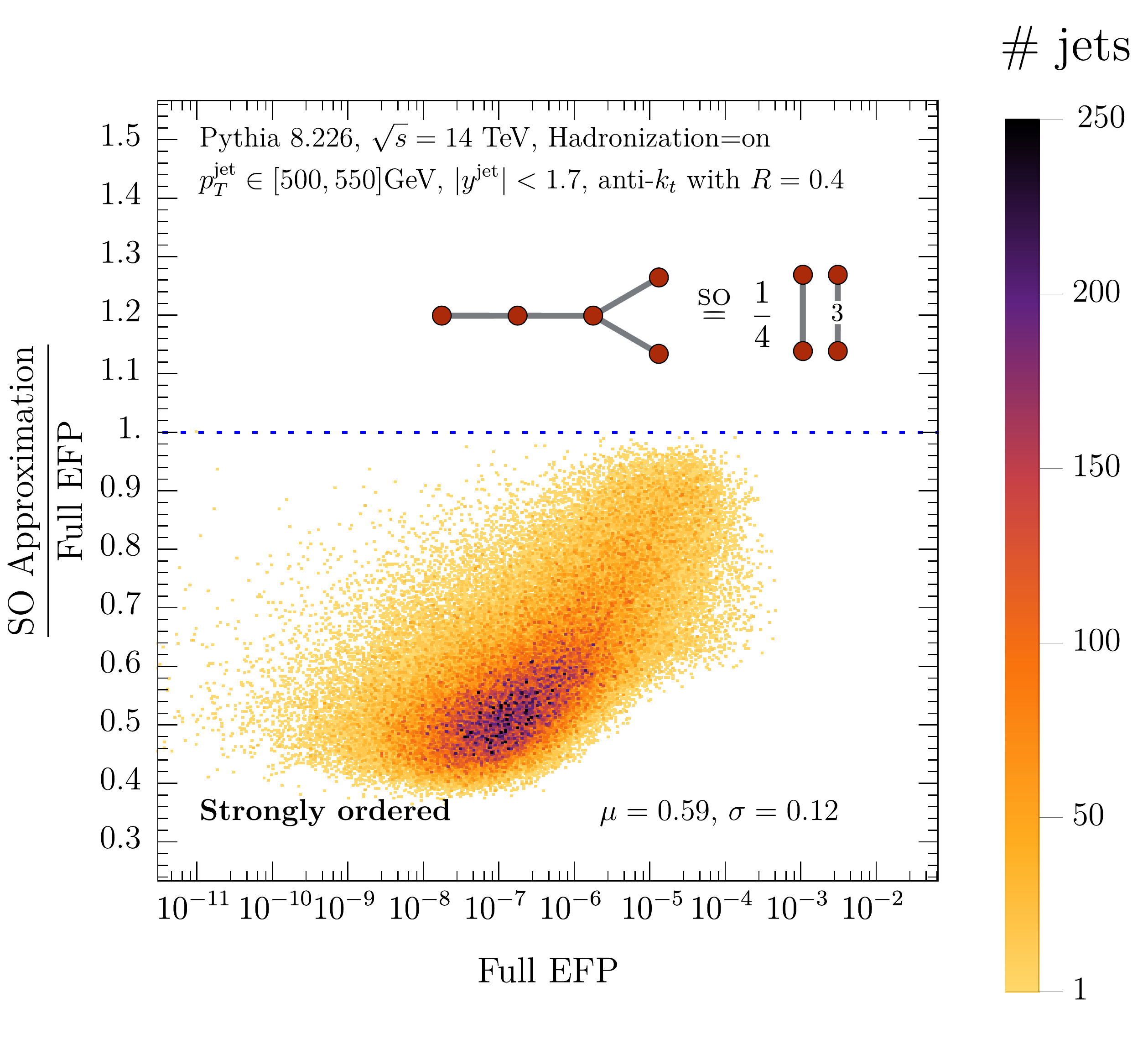}
    }
      \caption{
      Testing the strongly-ordered relationship for the 4-dot EFP in \eq{4dotsSO} (top row) and the crocodile EFP in \eq{smallcrocSO} (bottom row).
      Results are shown as a correlation plot (left column) and ratio (right column).
      The average $\mu$ and standard deviation $\sigma$ of the ratio is also shown.}
       \label{fig:LL_corr}
\end{figure}

We start with some of the examples we discussed explicitly before: the four dots in \eq{4dotsSO} and the crocodile in \eq{smallcrocSO}.
In the left panels of \fig{LL_corr}, we compute the left-hand and right-hand side of the equation on each jet, and plot the values as the horizontal and vertical coordinates.
In the right panels of \fig{LL_corr}, we show the ratio between these values.
While there is a clear correlation, the average ratio $\mu$ differs from 1 by a factor of 0.6, and the spread $\sigma$ in the correlation is at the 10\% level.
This motivated us to consider the more accurate $n$-collinear expansions, which we discuss in \sec{n-coll-basis}.

   \begin{figure}[p]
   \centering
    \begin{subfigure}{0.48\textwidth}
     \includegraphics[width=\textwidth]{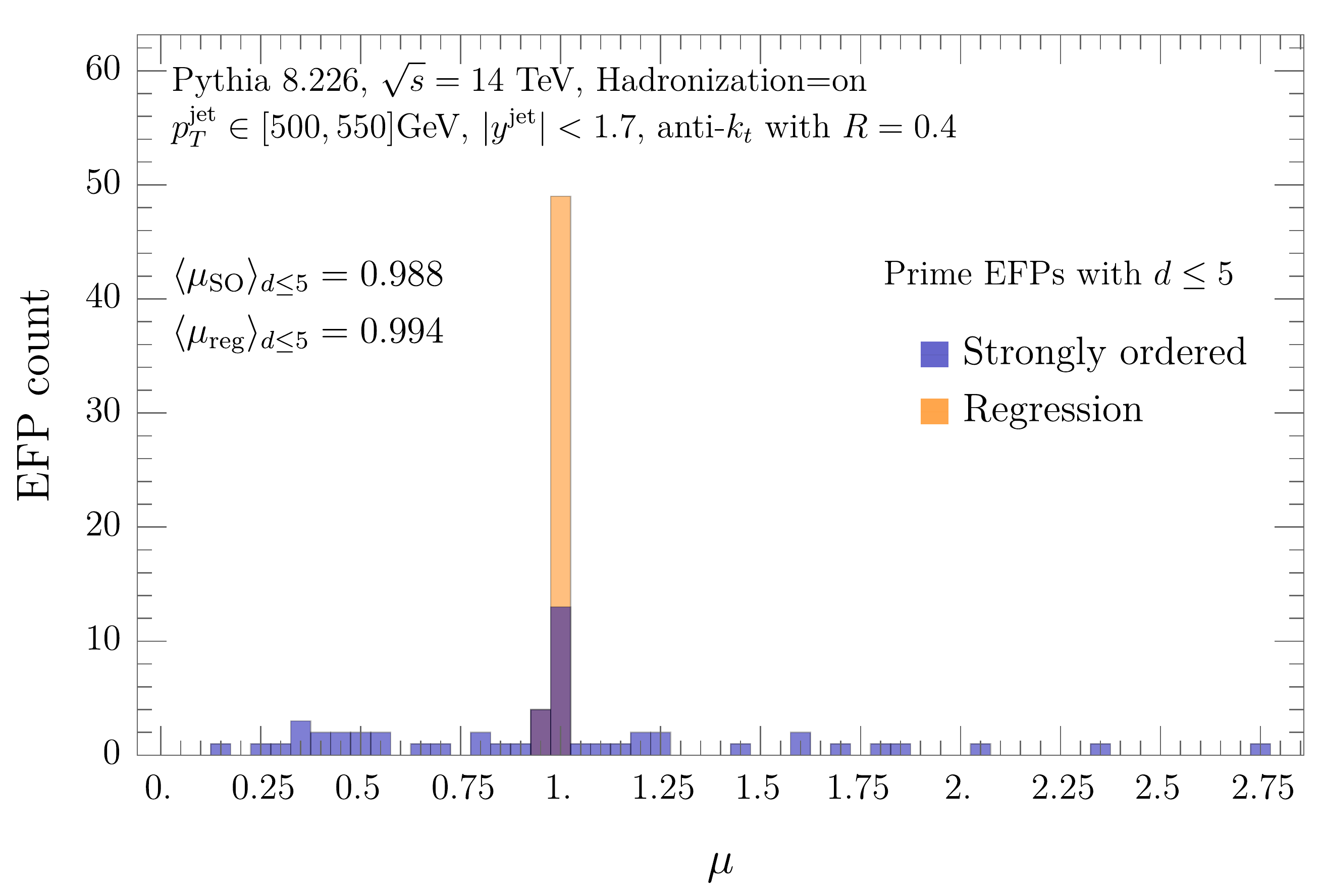}
     \caption{}
     \label{fig:SO_hist_a}
     \end{subfigure}     
     \hfill
    \begin{subfigure}{0.48\textwidth}     
     \includegraphics[width=\textwidth]{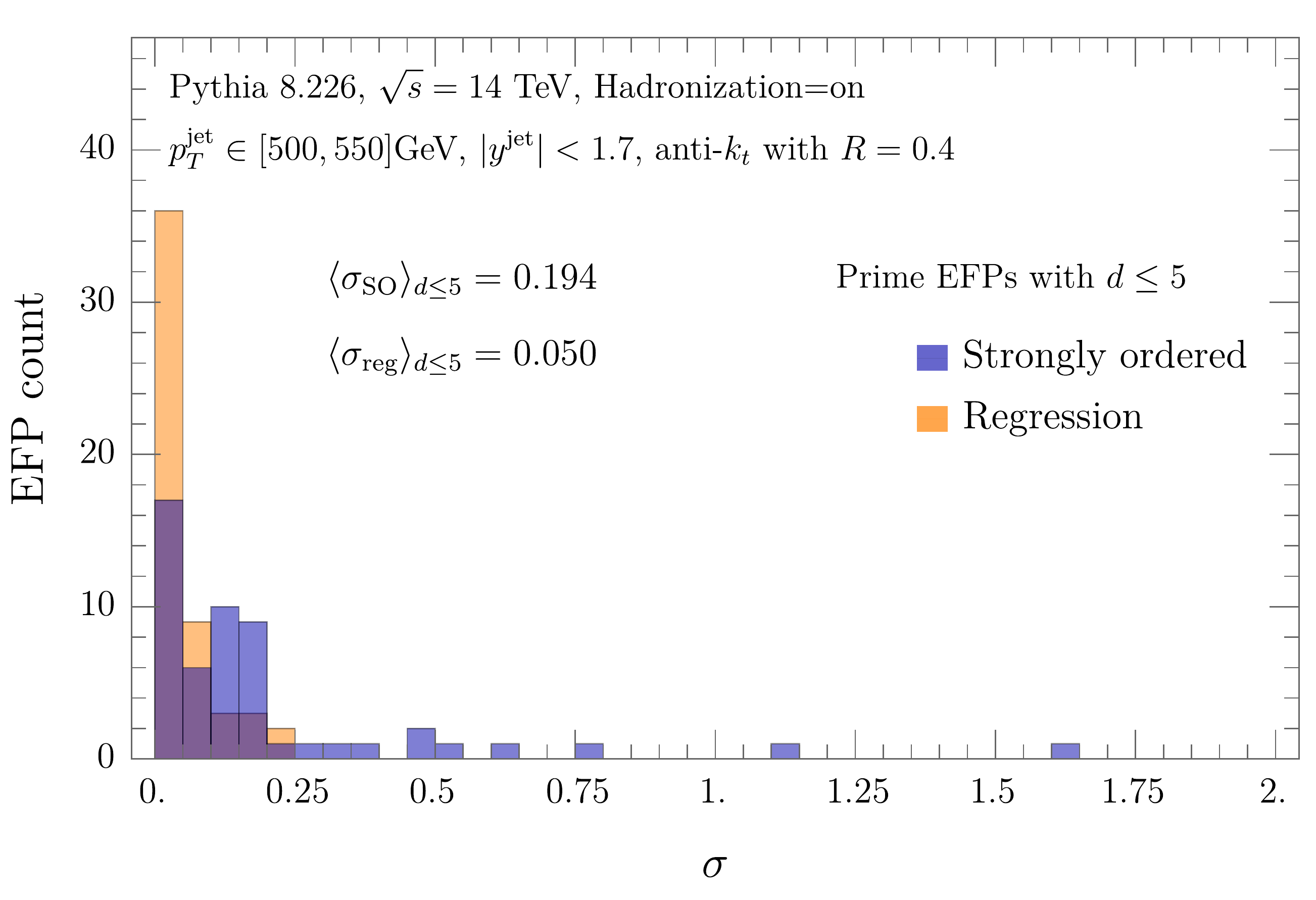} 
     \caption{}
     \label{fig:SO_hist_b}
     \end{subfigure} 
     \\
     \vspace{0.5cm}
     \begin{subfigure}{0.6\textwidth}
      \hspace{-0.45cm}\includegraphics[width=\textwidth]{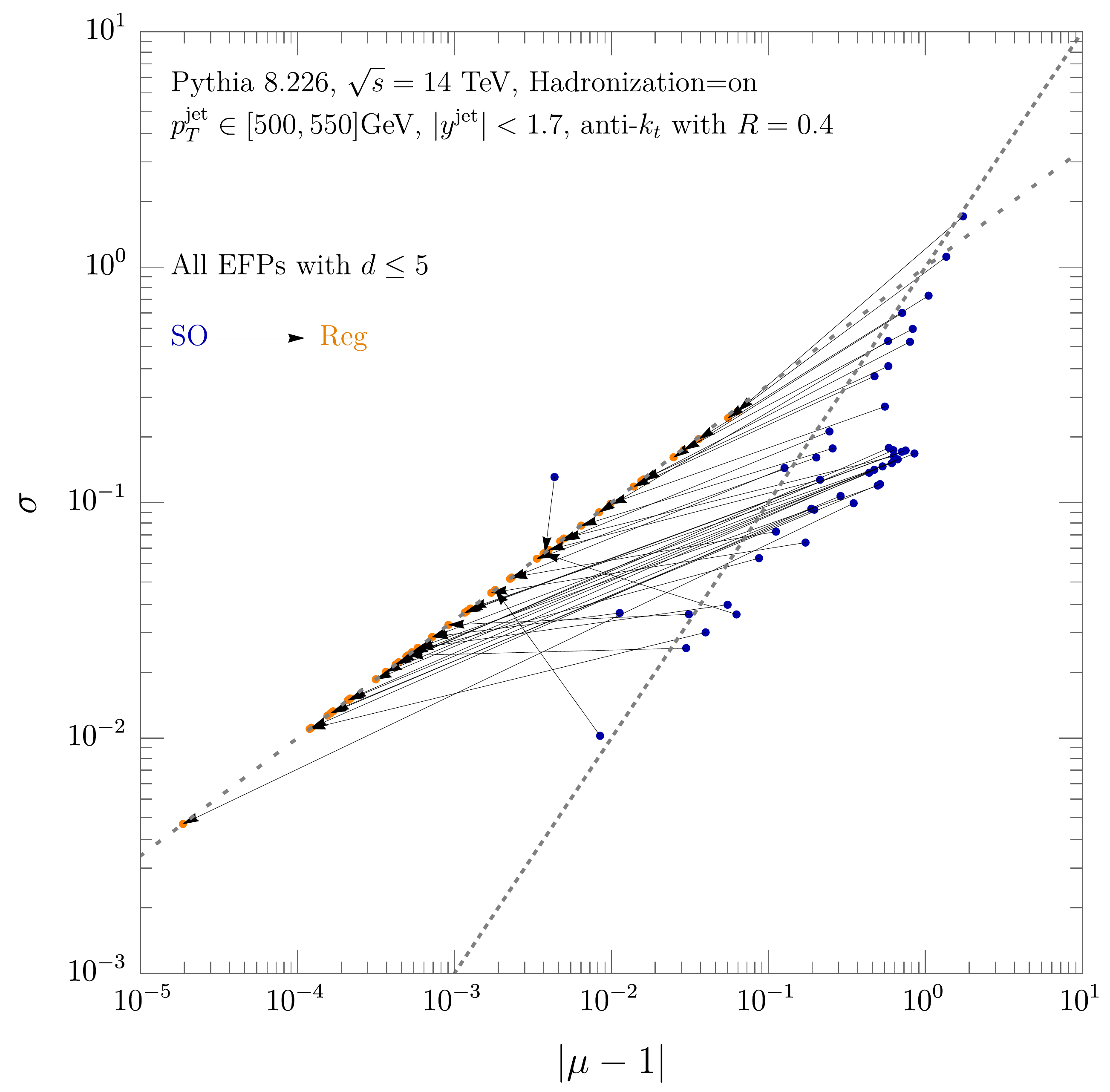} 
     \caption{}
     \label{fig:SO_hist_c}
     \end{subfigure}
\caption{
Histogram for the average $\mu$ (a) and standard deviation $\sigma$ (b) of the ratio between prime EFPs up to degree 5 and their expression in terms of SO basis elements.
Here, the relationship was found using power counting (blue) and regression (orange).
The improvement from using regression is visualized with arrows in the scatter plot (c) for $|\mu-1|$ and $\sigma$.
The dashed line corresponds to $\sigma = |\mu-1|$, and the dotted line corresponds to $\sigma^2 = |\mu-1|$.
}
\label{fig:SO_hist}
     \end{figure}

Moving beyond individual cases, we tested the expression of all EFPs up to degree 6 in terms of the SO basis.
In the blue histograms of \figs{SO_hist_a}{SO_hist_b}, we summarize the average $\mu$ and spread $\sigma$ of the ratio for prime EFPs (i.e.~EFPs represented by connected graphs) that have a non-trivial power counting relation.
As expected, $\mu$ is peaked around $1$ and $\sigma$ is peaked around 0.     
While there are some relationships that are very accurate, there is quite a spread, in line with \fig{LL_corr}.
In \fig{SO_hist_c}, we show the same results as a scatter plot with $|\mu-1|$ and $\sigma$ on the axes, allowing one to see the correlation between them.
Empirically, we find that most relations satisfy $\sigma \simeq |\mu-1|$, which one might anticipate because the average and spread are controlled by the same subleading power corrections.

Interestingly, even though the individual relations in the SO approximation are not very accurate, any EFP is still very well approximated by a linear combination of SO basis elements.
This is shown in the orange histograms of \figs{SO_hist_a}{SO_hist_b}, where we perform a linear regression for all prime EFPs in terms of the elements of the SO basis.
This helps explain the curious fact we will encounter in \sec{tagging}, where the tagging performance of the SO basis matches that of the 2-collinear basis, even though the individual linear relations are considerably poorer.
Note that this regression does not include a constant term (i.e.~single dot EFP), so $\mu$ does not automatically equal 1, since $d>0$ EFPs go to zero in the collinear and soft limits.
Empirically in \fig{SO_hist_c}, we see that the regression results satisfy $\sigma^2 \simeq |\mu-1|$, which arises because with regression, the average can be fine-tuned to be smaller than the spread.

%-----------------------------------------------------------------------
\subsection{Results with the $n$-collinear basis}
\label{sec:n-coll-basis}
%-----------------------------------------------------------------------

     \begin{figure}[p]
     \subfloat[]{
     \includegraphics[width=0.48\textwidth]{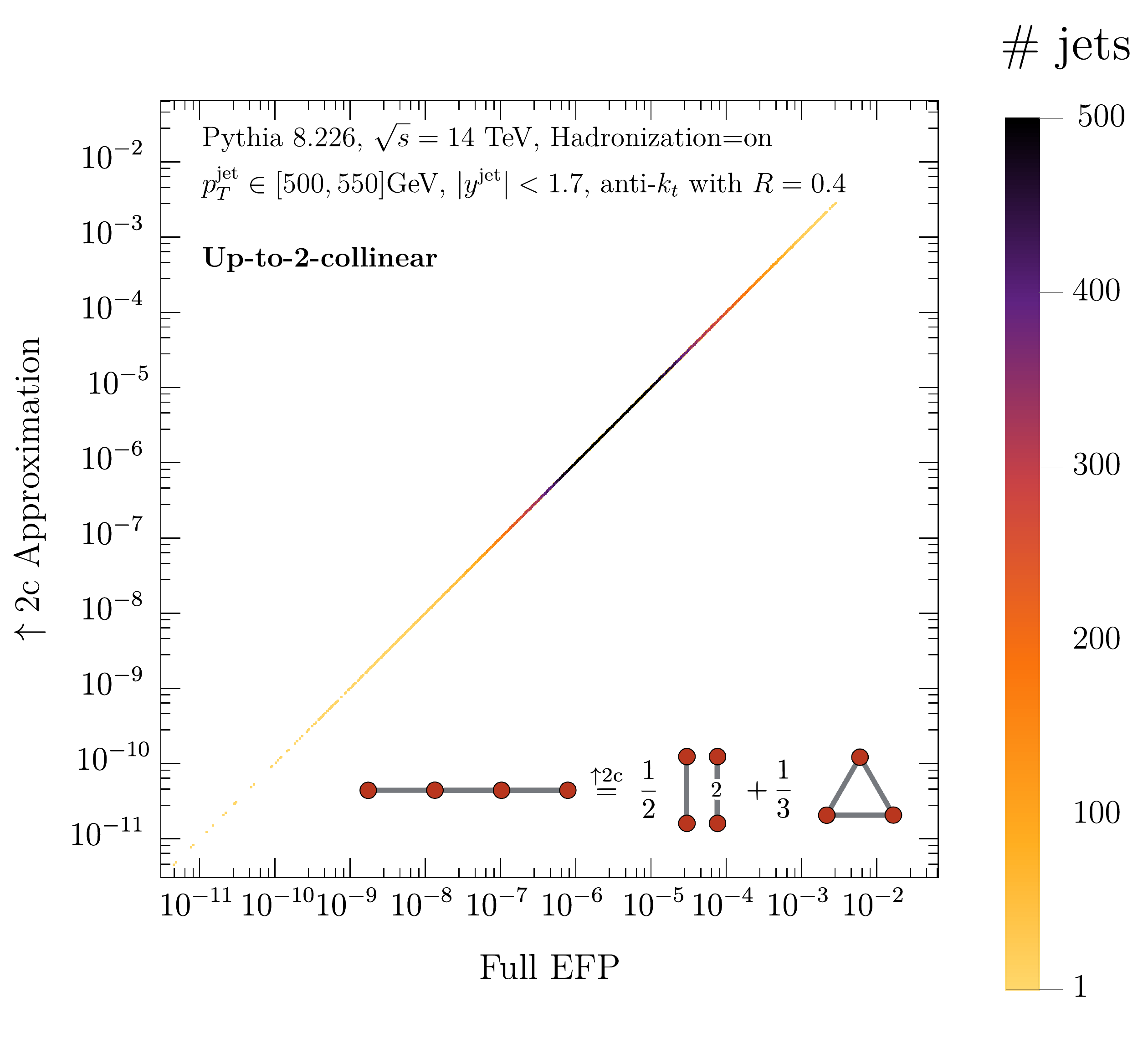}
     }
     \hfill 
     \subfloat[]{
     \includegraphics[width=0.48\textwidth]{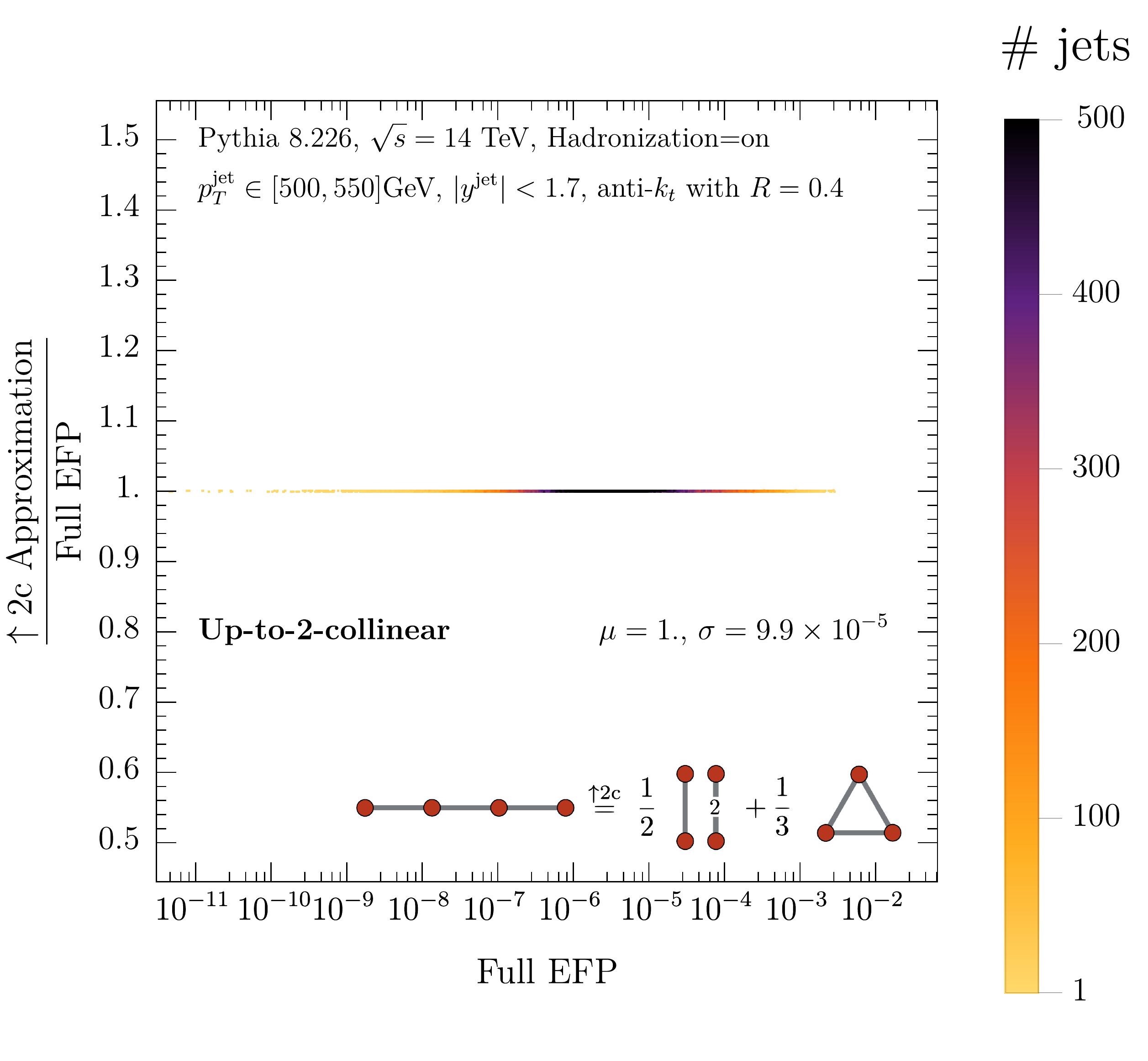}
     }\\ 
     \subfloat[]{
     \includegraphics[width=0.48\textwidth]{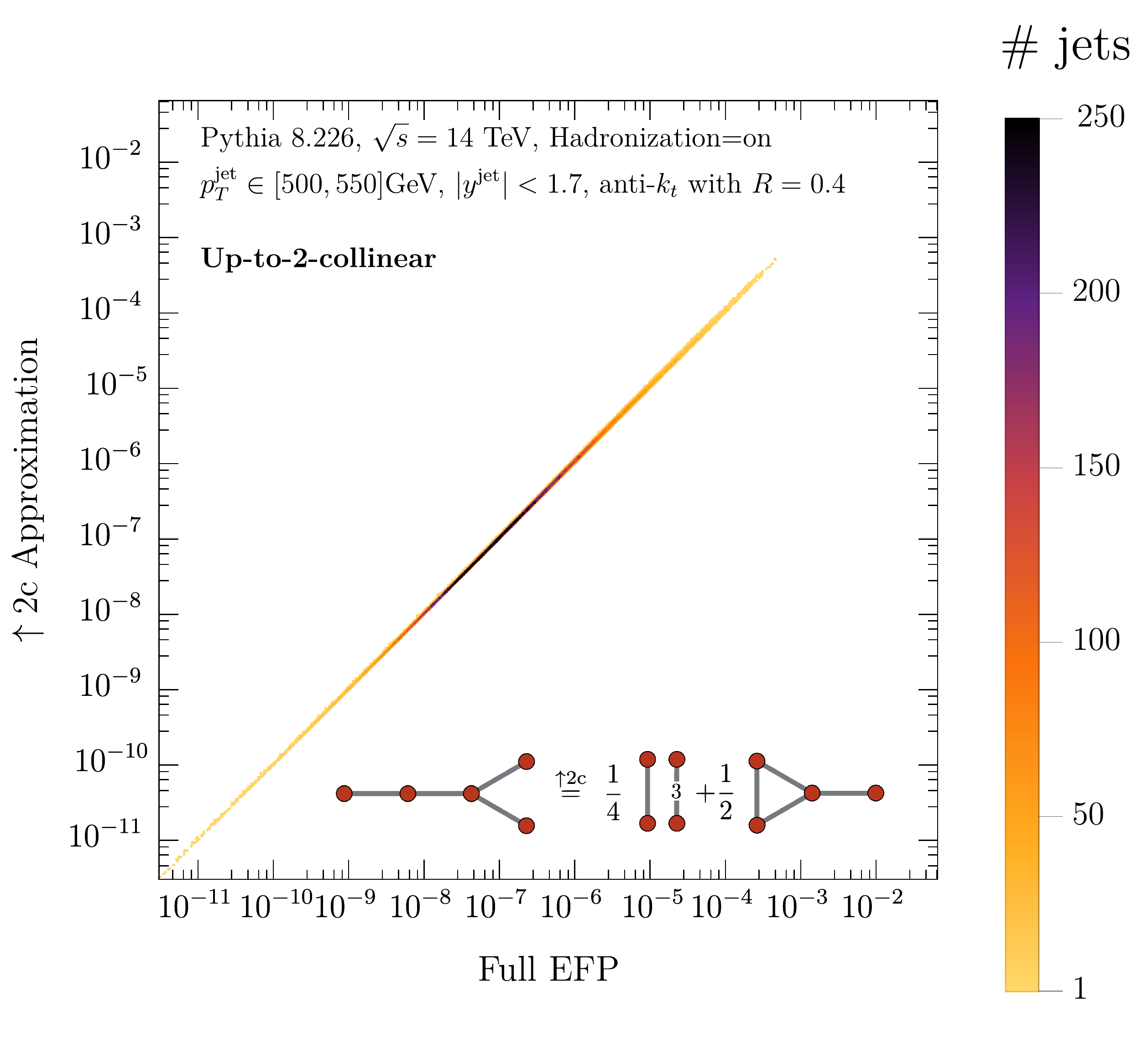}
     }
      \hfill
      \subfloat[]{\includegraphics[width=0.48\textwidth]{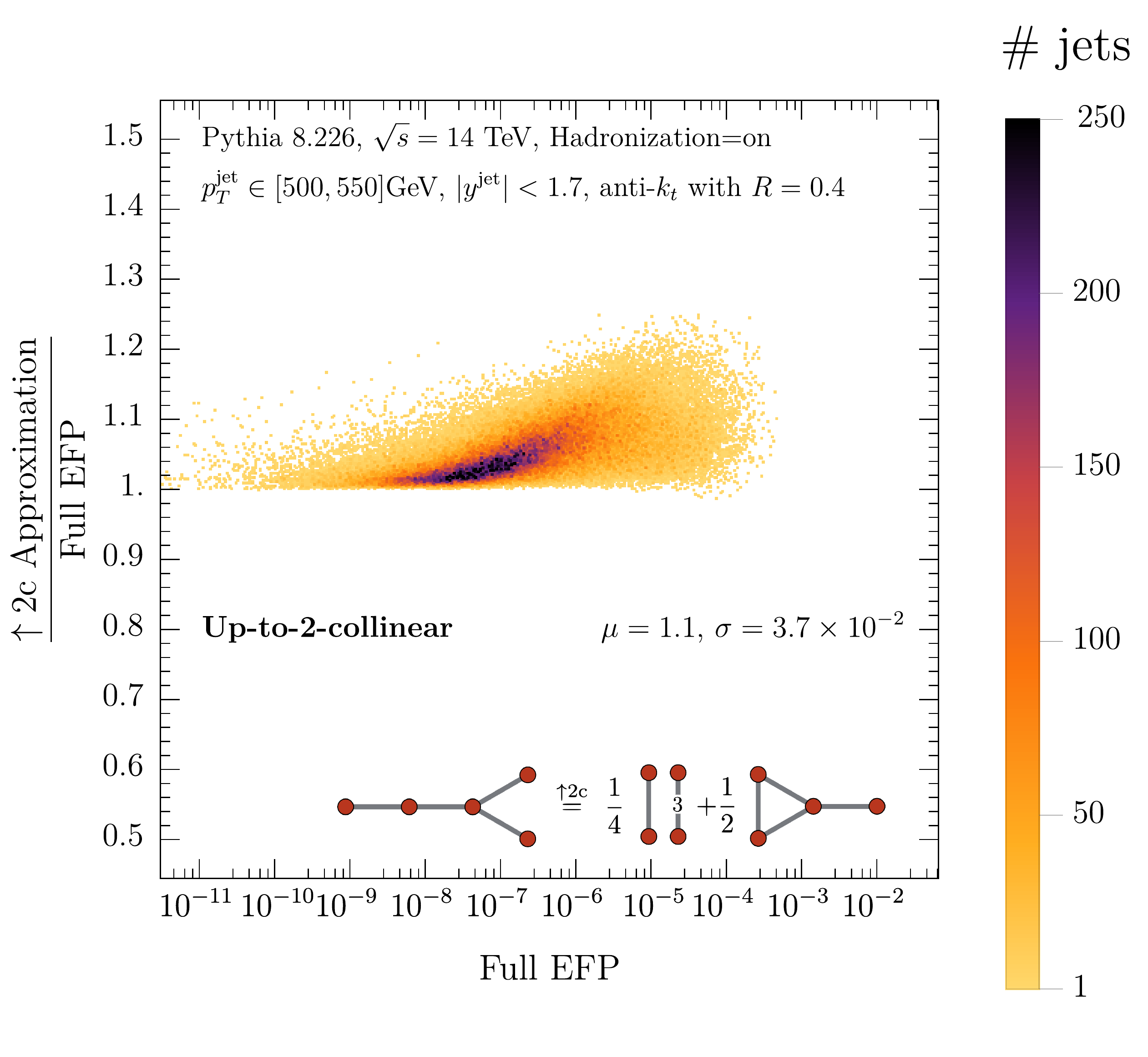}}
     \caption{Same as \fig{LL_corr}, but using the relationships obtained from the up-to-2-collinear expansion, which are much more accurate. \label{fig:corr1}}
     \end{figure}

We now test linear relations in the collinear expansion.
We start by considering the same examples as in \fig{LL_corr}, given for four dots in \eq{4dotsLL} and the crocodile in \eq{smallcrocLL}.
While these relationships were derived in the 1-collinear approximation, they in fact also hold for the 2-collinear approximation.
The correlation holds extremely well for four dots, while for the crocodile the mean of the ratio differs by about 10\% from 1 and the spread is a few percent.
In both cases, the improvement over the strongly ordered expansion is substantial.

   \begin{figure}[p]
     \subfloat[]{
     \includegraphics[width=0.48\textwidth]{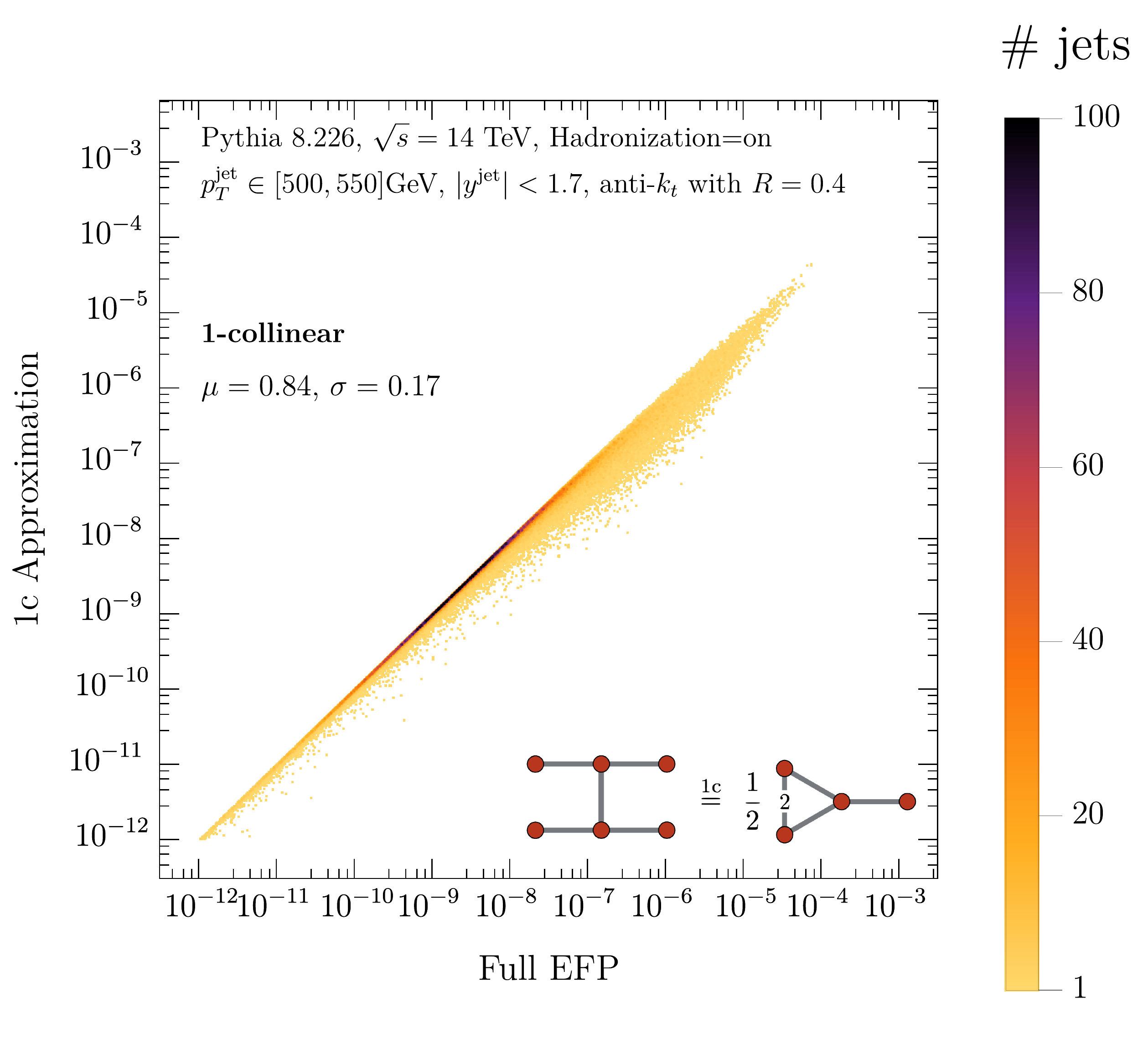}
     }
     \hfill 
     \subfloat[]{
     \includegraphics[width=0.48\textwidth]{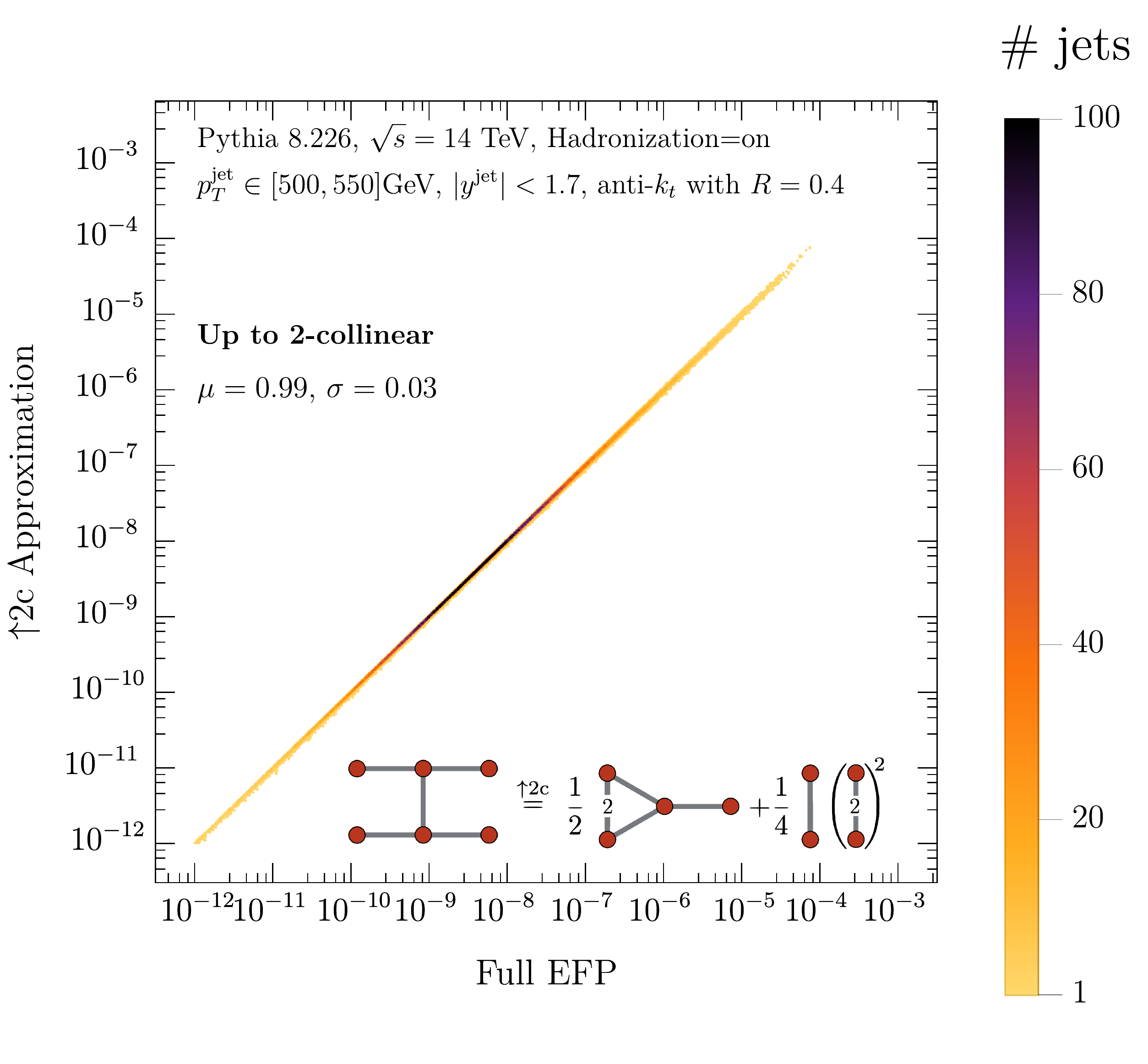}
     }\\ 
     \subfloat[]{\includegraphics[width=0.48\textwidth]{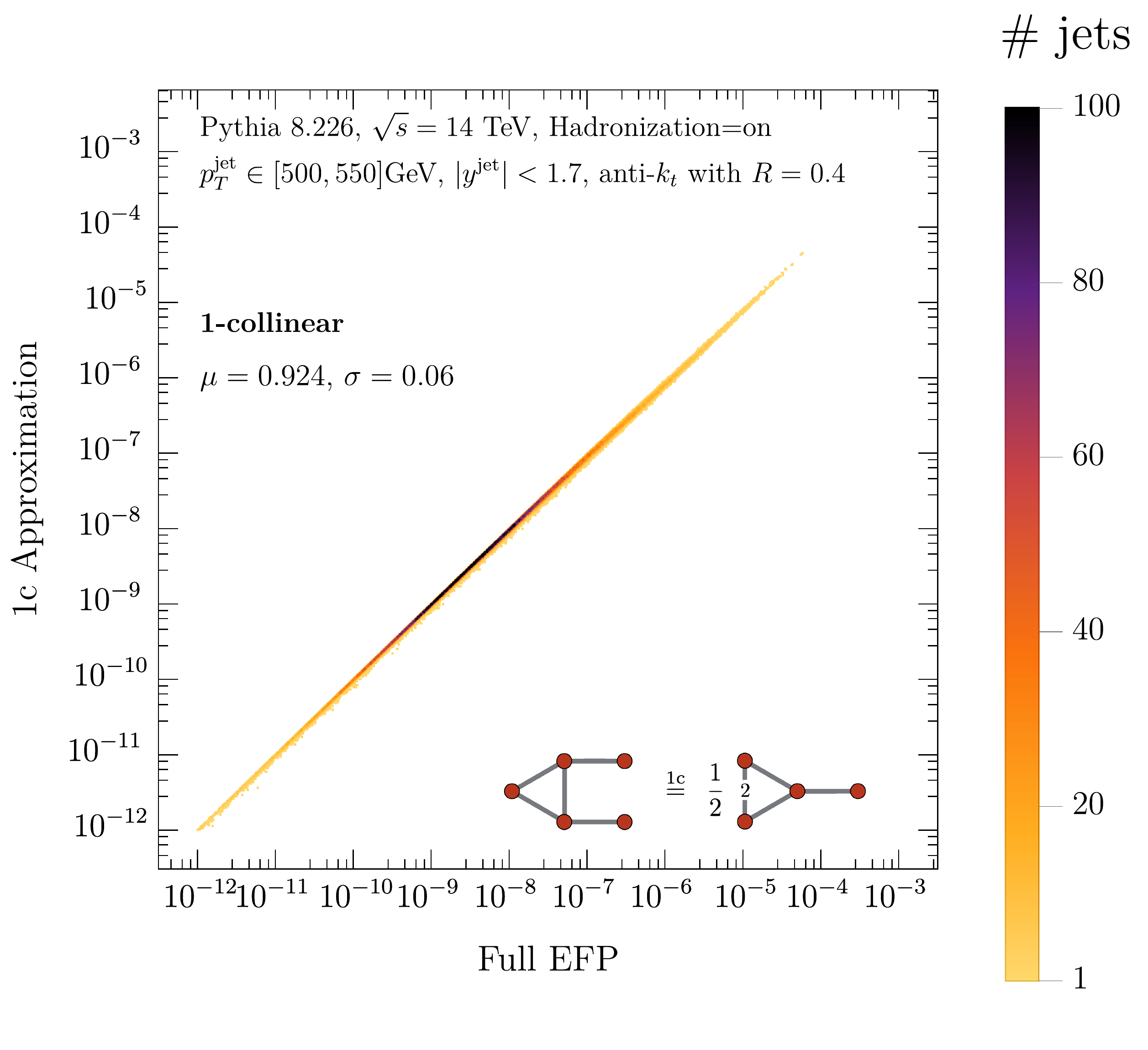}
     }
      \hfill
      \subfloat[]{
      \includegraphics[width=0.48\textwidth]{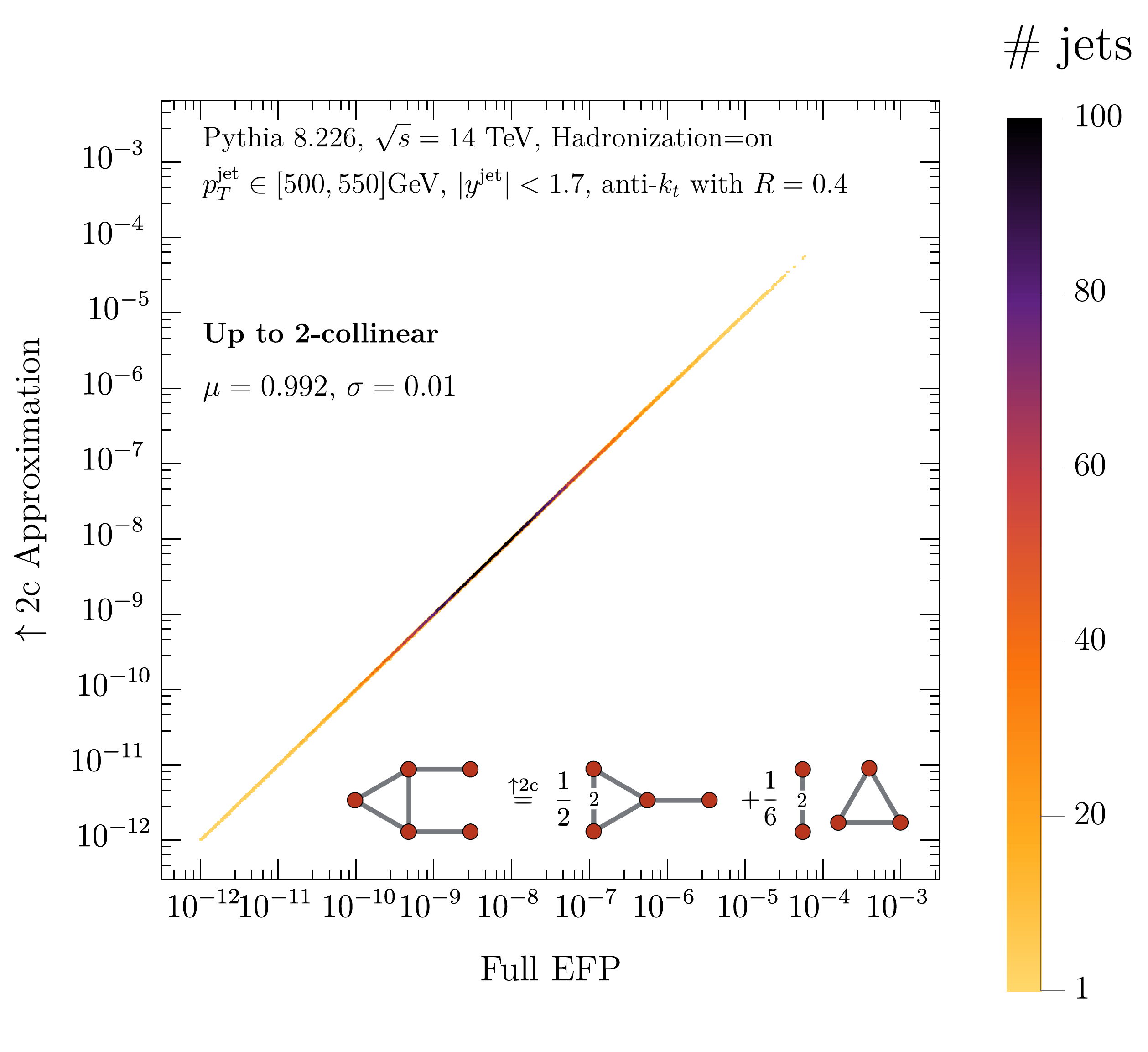}
      }
      \caption{Testing the relationship for the ``H" EFP in \eq{HNLL} (top row) and ``A" EFP in \eq{ANLL} (bottom row) in the 1-collinear approximation (left column) and including the 2-collinear approximation (right column), which leads to a substantial improvement. The average $\mu$ and standard deviation $\sigma$ of the ratio is shown. \label{fig:corr2}}
     \end{figure}
     
Next, we consider two EFPs where the 2-collinear approximation differs from the 1-collinear approximation, namely the ``H" and ``A" EFPs from \eqs{HNLL}{ANLL}.
In \fig{corr2}, we show the correlation plot for the 1-collinear approximation on the left and including the 2-collinear approximation on the right.
While the ratio is not plotted, the average $\mu$ and standard deviation $\sigma$ of the ratio are indicated in the figure.
As is clear from their values, and from the correlation plots, there is a substantial improvement from including the 2-collinear approximation.
    
   \begin{figure}[p]
     \subfloat[]{
     \includegraphics[width=0.48\textwidth]{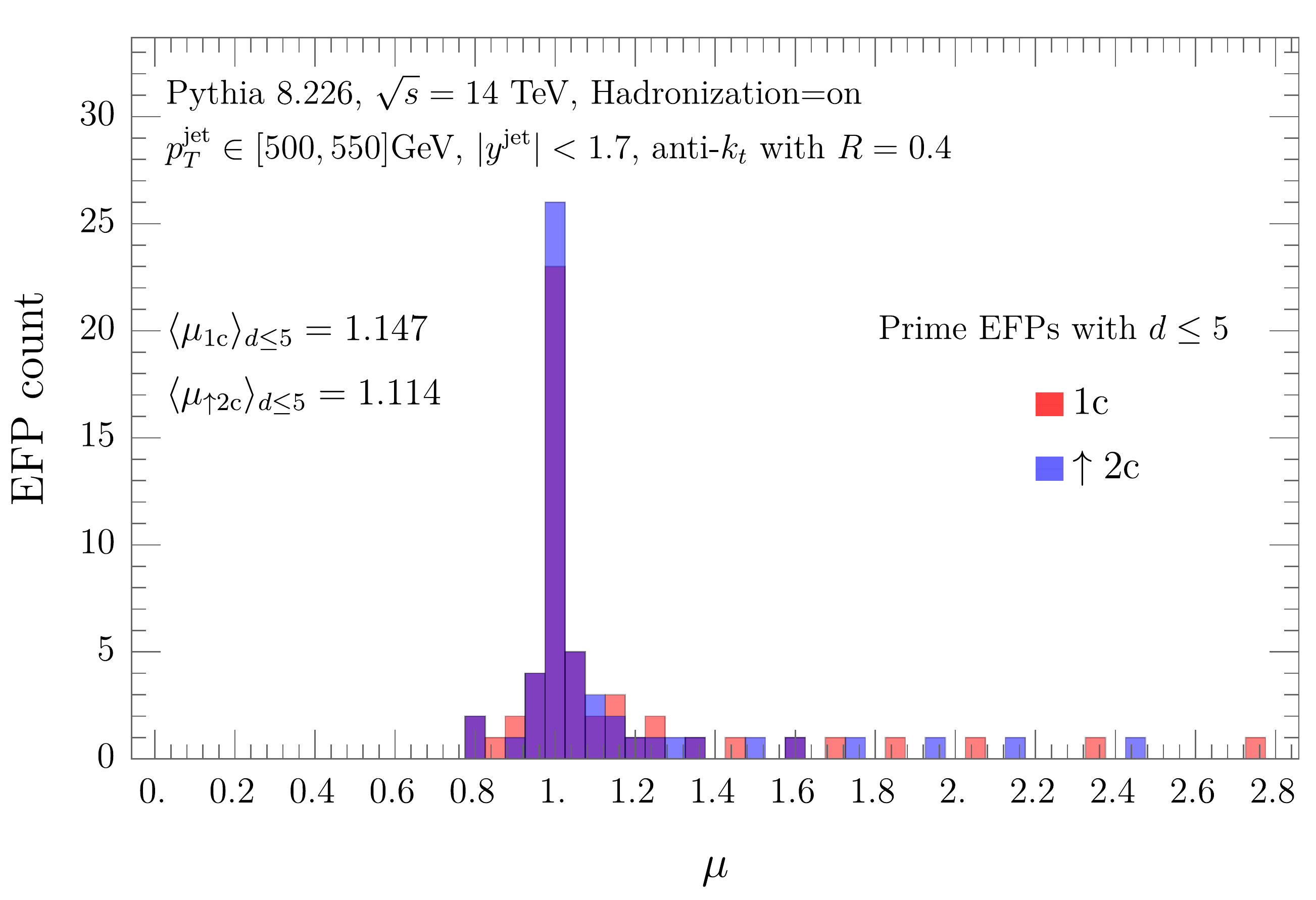}
     }
     \hfill 
     \subfloat[]{
     \includegraphics[width=0.48\textwidth]{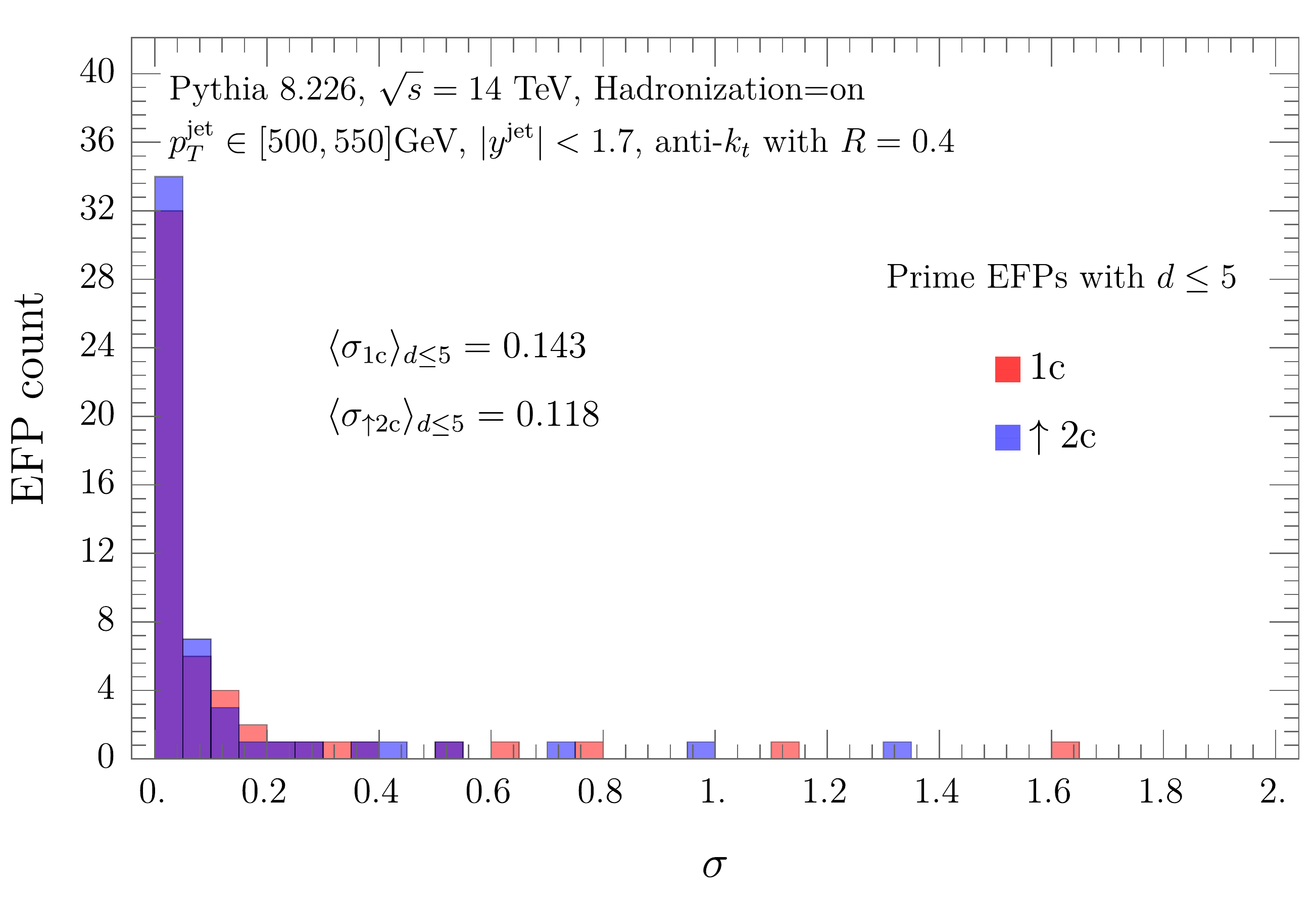}
     } \\ 
    \hfill
    \subfloat[]{\includegraphics[width=0.48\textwidth]{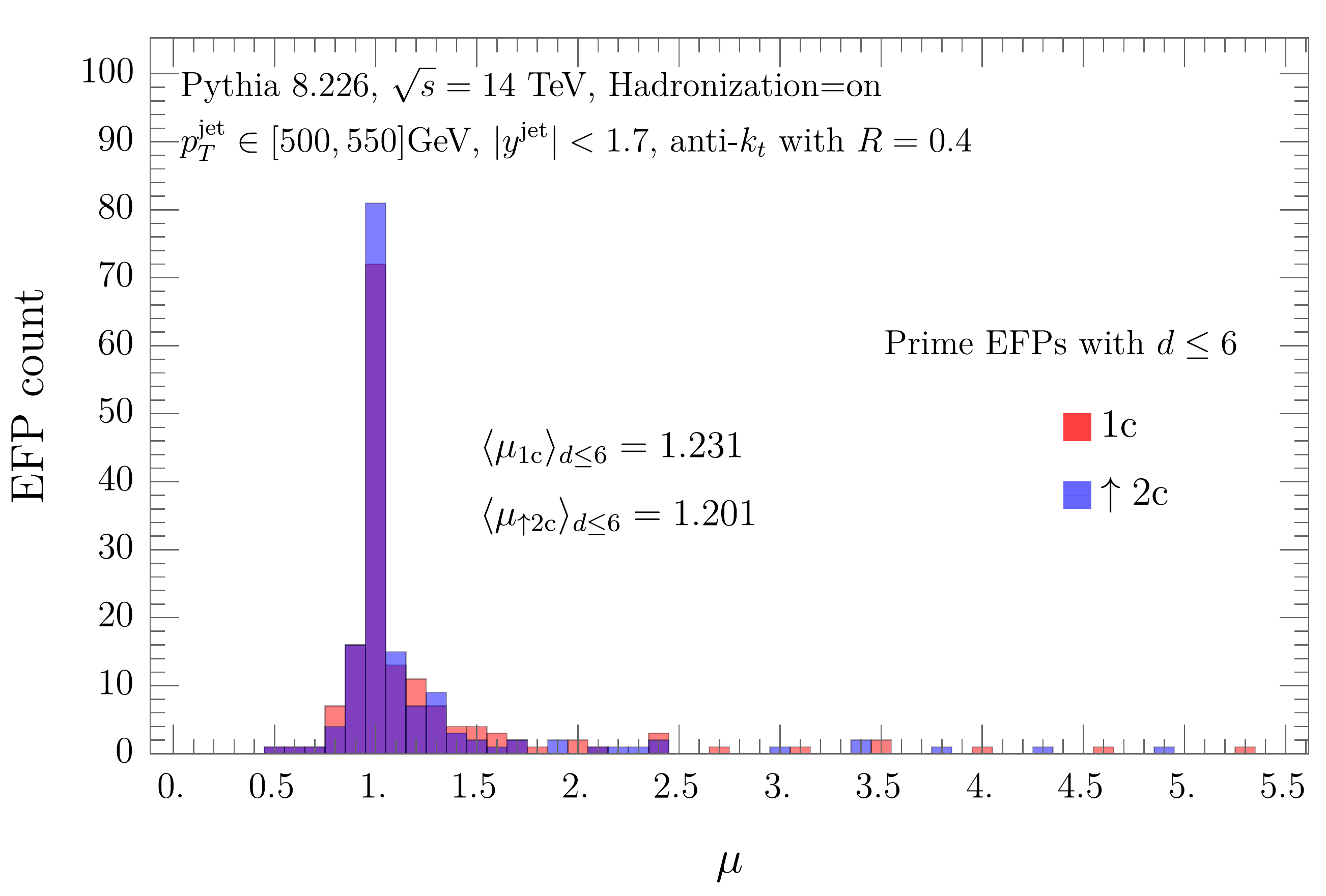}
    }
    \hfill 
     \subfloat[]{
     \includegraphics[width=0.48\textwidth]{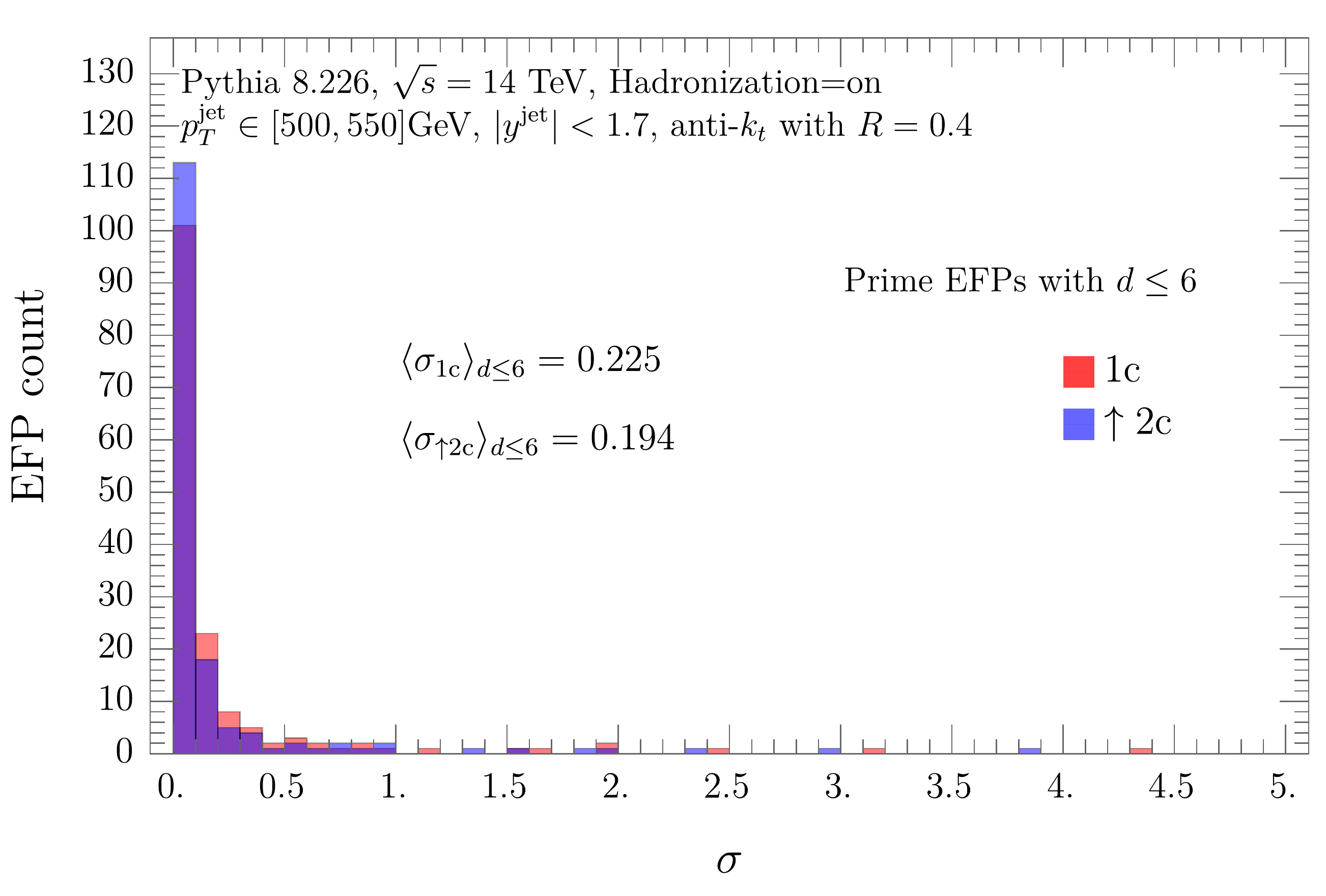}
     }
\caption{Histogram for the average $\mu$ (left) and standard deviation $\sigma$ (right) of the ratio between prime EFPs and their expression in terms of basis elements in the 1- (red) and 2-collinear (blue) approximation. Going from EFPs up to degree 5 (top) to degree 6 (bottom), $\mu$ and $\sigma$ become slightly worse because of a couple of outliers, which are star-like EFPs. \label{fig:corr_histogram}}
     \end{figure}

In \fig{corr_histogram}, we test the expression of all prime EFPs up to degree 6 in terms of the basis elements, in both the 1- and 2-collinear approximation.
     The results up to degree 5 and 6 are shown separately, allowing one to see an increase in the number of outliers at higher degree.
     We have identified these outliers as star-like EFPs and will have more to say about them in \sec{star}.
     As for the SO basis, $\mu$ is peaked around $1$ and $\sigma$ is peaked around 0.
     There is a noticeable improvement from including the 2-collinear approximation in both $\mu$ and $\sigma$.

   \begin{figure}[t]
          \subfloat[]{ \includegraphics[width=0.48\textwidth]{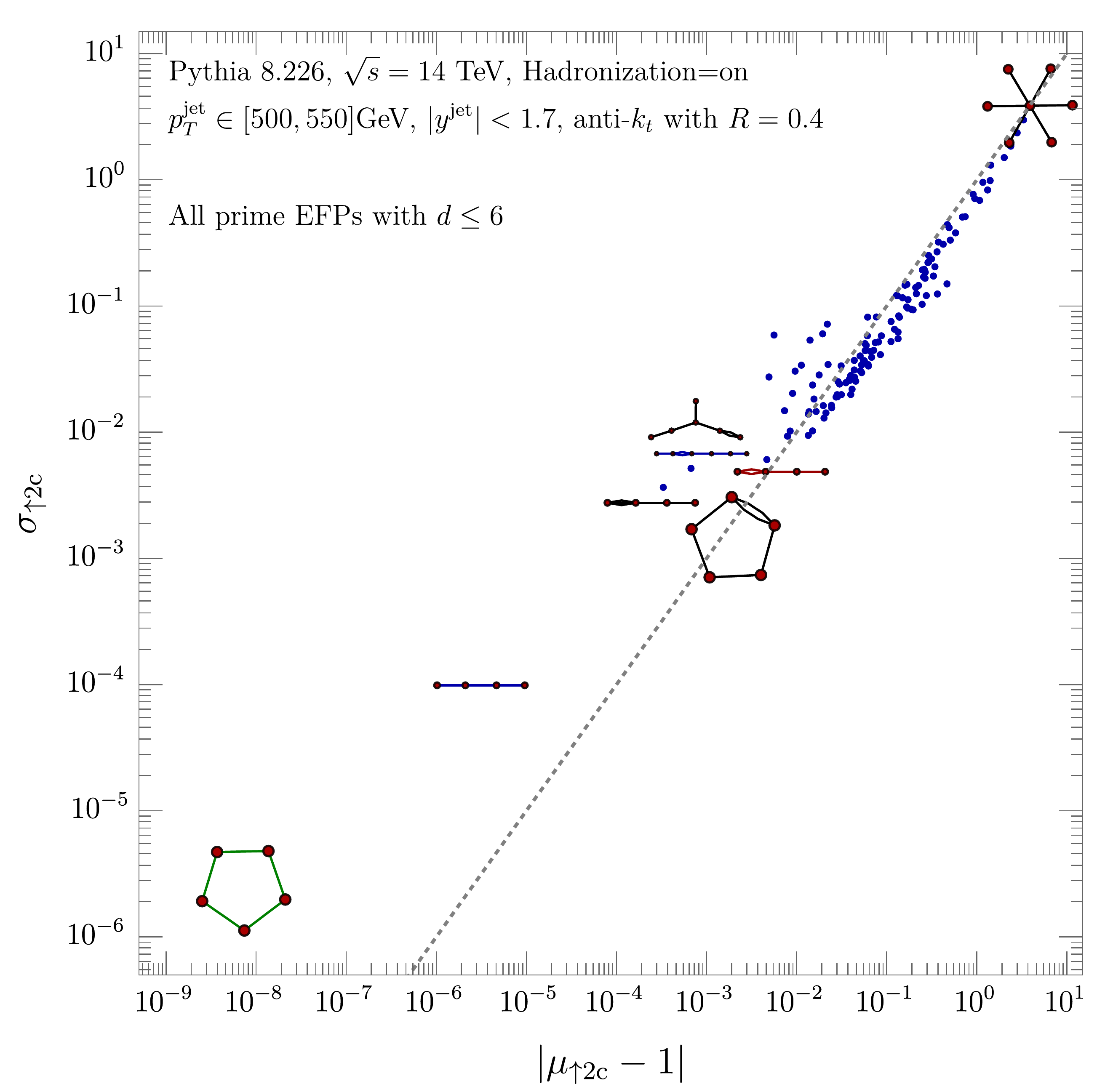}
          }
          \hfill 
     \subfloat[]{
     \includegraphics[width=0.48\textwidth]{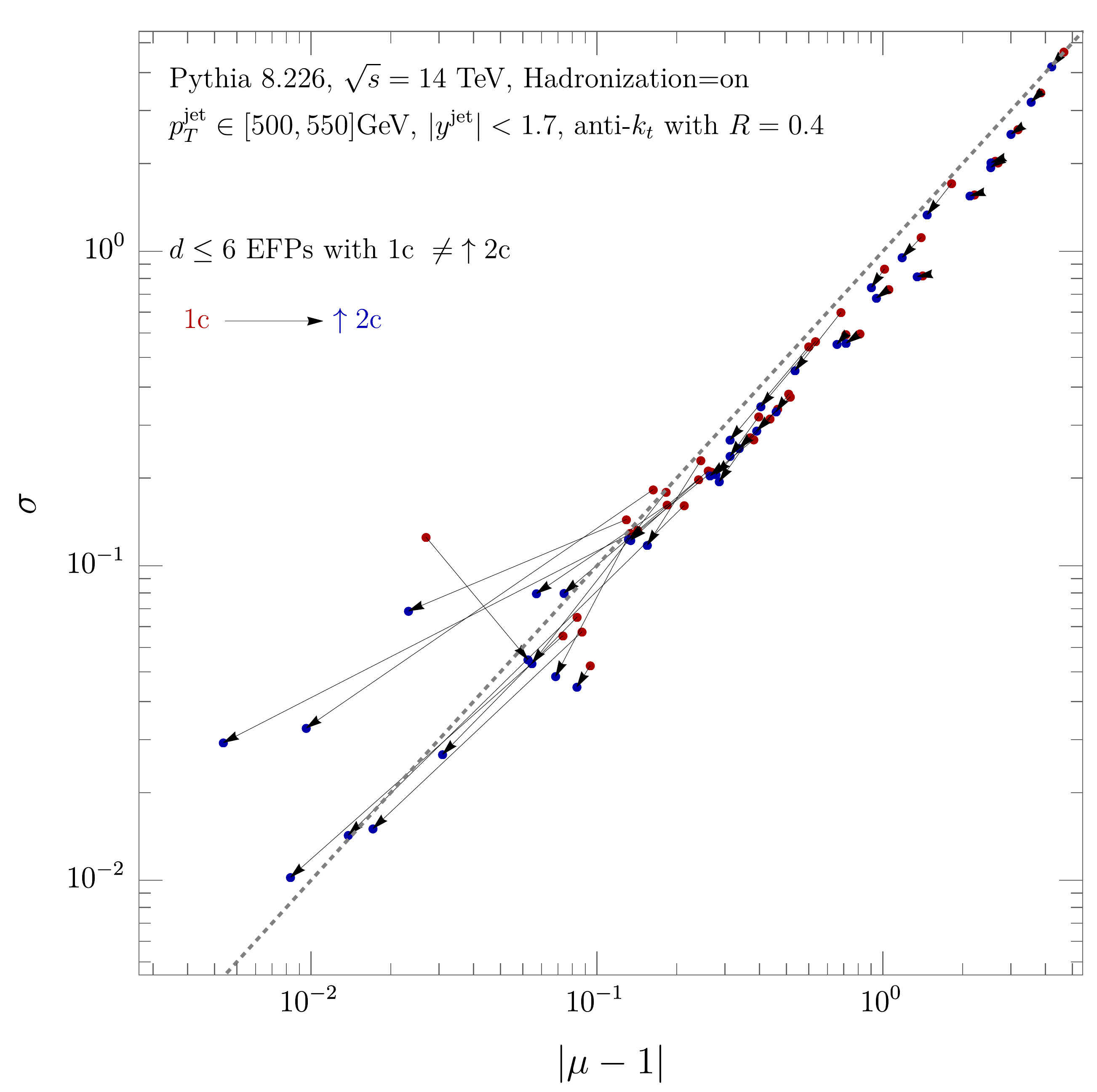}
     }
\caption{Scatter plot for $|\mu-1|$ and $\sigma$ of the  prime EFPs up to degree 6. Left: in the 2-collinear approximation. Right: the result in the 1- (red) and 2-collinear (blue) approximation, connected by an arrow.  We only show whose EFPs where the expression for 2-collinear is different; note the difference in scale.  The dashed line indicates $\sigma \simeq |\mu-1|$.
\label{fig:2d_scatter}}
     \end{figure}

Finally, in \fig{2d_scatter} we show the scatter plot with $|\mu-1|$ and $\sigma$ on the axes. 
In the left panel, results are shown for the 2-collinear approximation.
In this plot, most EFPs are represented by a dot, though for some, the actual graph is drawn.
As in the SO case, the EFPs fall along the line $|\mu - 1| = \sigma$.
In the right panel, EFPs are shown for which the expression in terms of basis elements differs at the 1-collinear and 2-collinear approximation, with an arrow indicating the improvement.

%-----------------------------------------------------------------------
\subsection{Linear regression for star graphs}
\label{sec:star}
%-----------------------------------------------------------------------

In \fig{SO_hist}, we saw that relationships obtained for the strongly-ordered basis performed poorly and could be improved by linear regression.
We can perform a similar study for star-like EFPs, for which the $n$-collinear expansion does not perform very well.

Before proceeding, we first check our fit procedure.
Performing regression on the 4-dot EFP, we reproduce the relationship in \eq{4dotsLL} up to expected power corrections:
\begin{align}
 \includegraphics[width=0.98\textwidth]{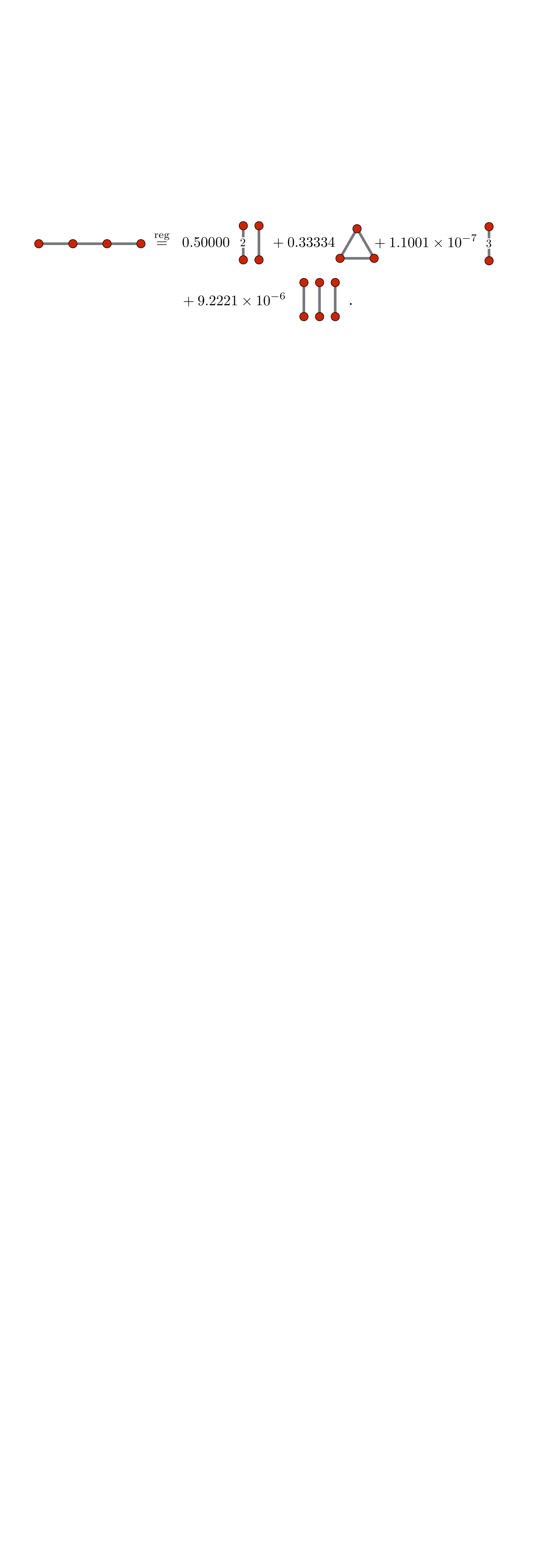}  \raisetag{2.0\baselineskip}  
\end{align}

\begin{figure}[!t]\centering
\subfloat[]{
  \includegraphics[width=.48\textwidth]{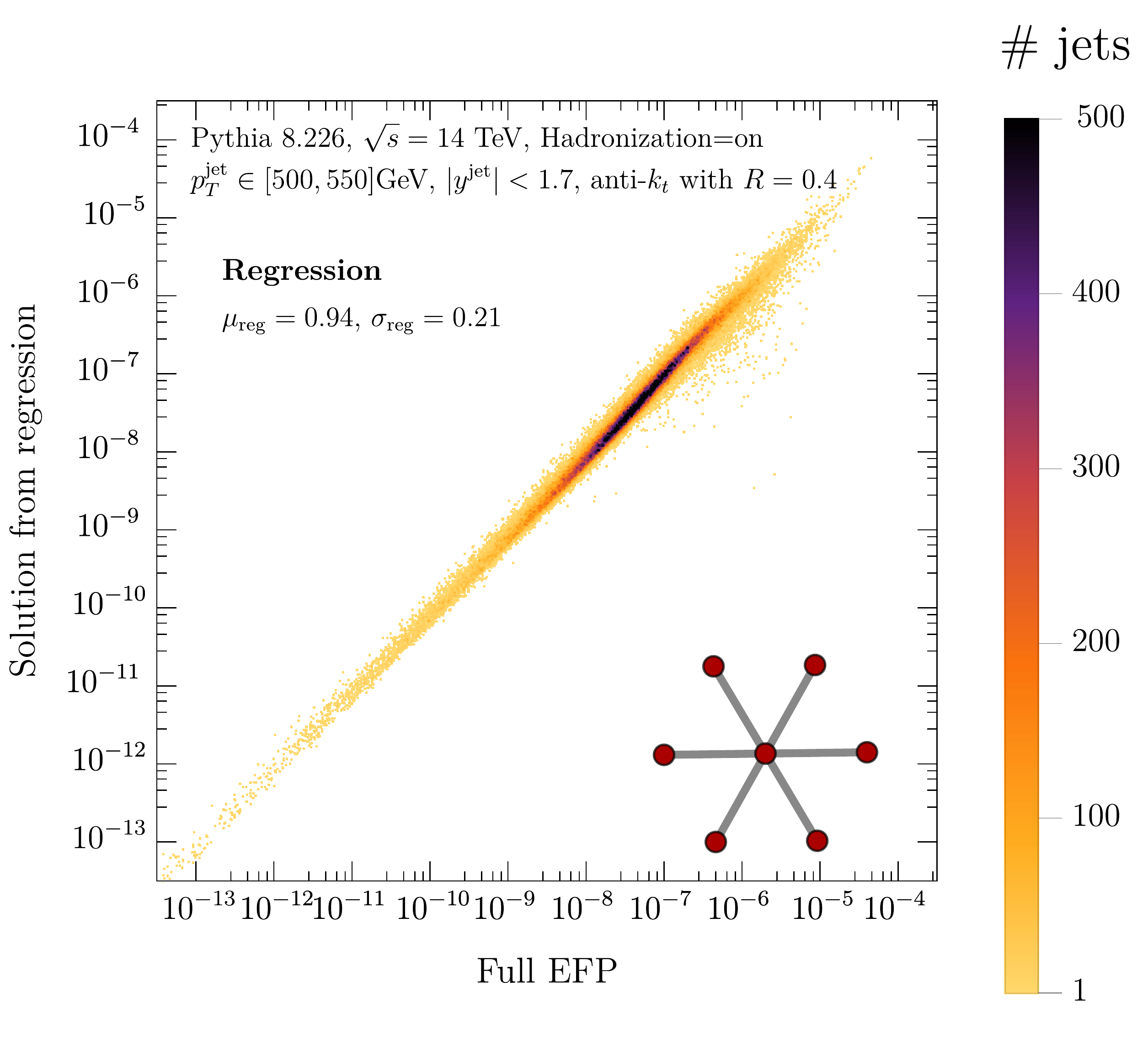}
  }
  \hfill
  \subfloat[]{
  \includegraphics[width=.48\textwidth]{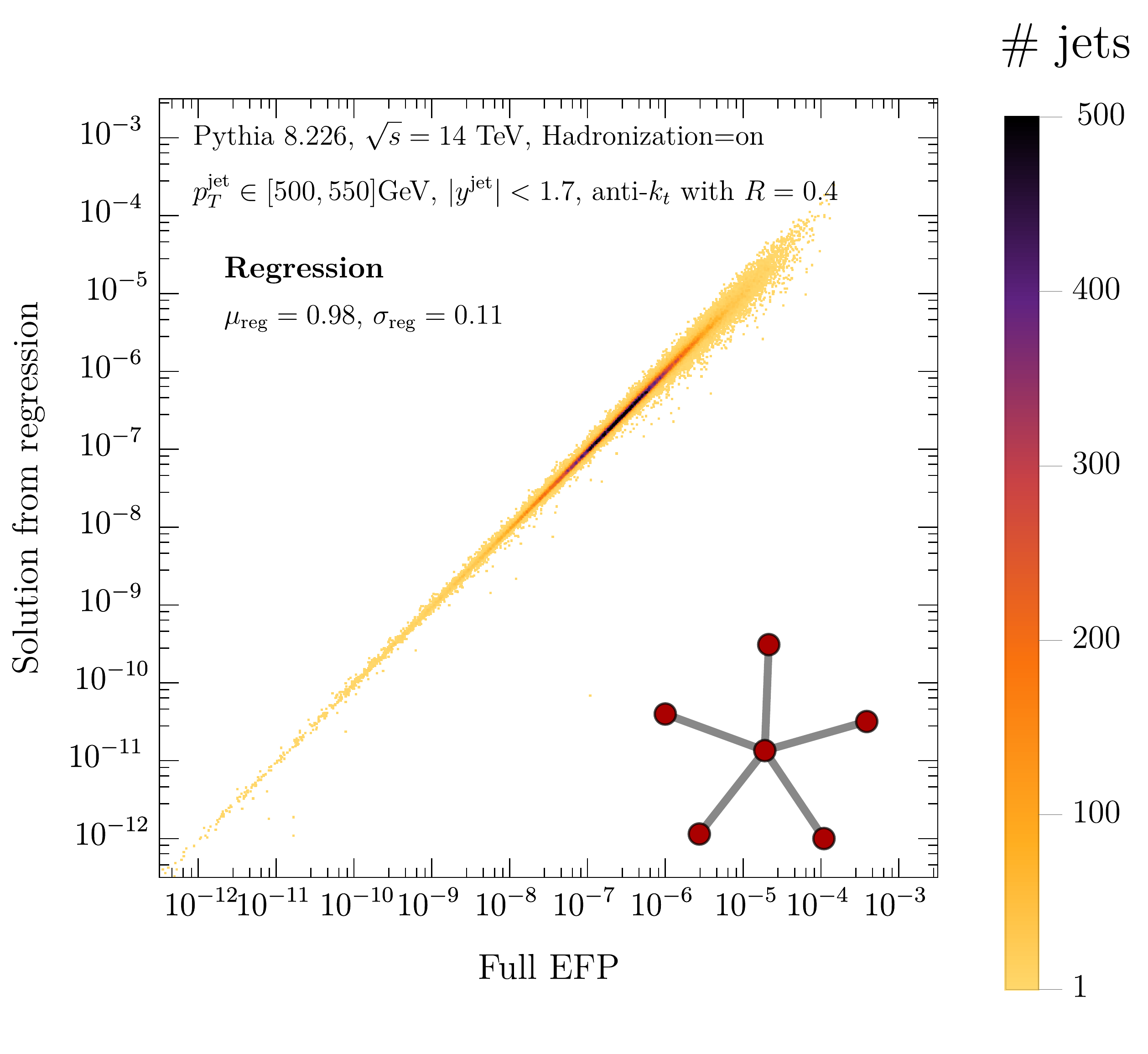}
  }
  \caption{Testing the relationship between the 6-pointed (left) and 5-pointed (right) star EFP and a corresponding linear fit of EFPs belonging to the 2-collinear basis. The average and standard deviation of the ratio is greatly improved over that for the relationship obtained using 2-collinear power counting, which was $(\mu,\sigma)_6= (4.89, 3.86)$ and $(\mu,\sigma)_5=(2.42,1.31)$, respectively. \label{fig:starfit}}
\end{figure}

For the star-like EFPs, the $n$-collinear expansion breaks down.
The EFP for which the collinear power counting relationship performs worst is a 6-pointed star, as indicated in \fig{2d_scatter}.
In the 2-collinear approximation, the 6-pointed star can be expressed as:
\begin{equation}
  \includegraphics[width=0.98\textwidth]{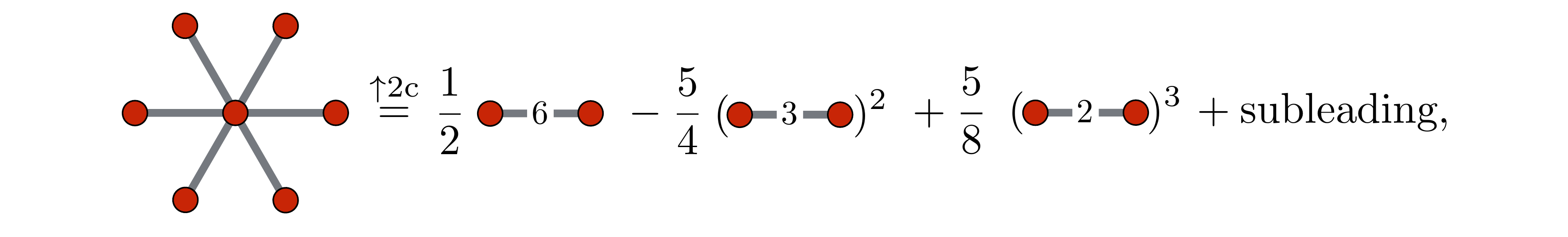}
\end{equation}
which is simply a polynomial of dumbbells.
At subleading power in $z$, though, one can assign a collinear-soft particle to the central node, which introduces strong dependence on the angle of this particle.
Such angular complexity is not captured by dumbbells, explaining the poor power counting performance.

Nevertheless, we find that star-like EFPs can be well approximated by a linear combination of basis elements.
Performing a linear regression (without a constant term) for the 6-pointed and 5-pointed stars in \fig{starfit}, we find $\mu$ much closer to 1 than from the 2-collinear power counting.
Inspected the regression solution, we found no discernible pattern.
In particular, we see no evidence that the regression solution looks like the power-counting relation plus corrections.
It is an interesting open question whether this regression relationship could have been derived from first principles.

%%%%%%%%%%%%%%%%%%%%%%%%%%%%%%%%%%%%%%%%%%%%%
\section{Logistic regression for quark/gluon jet tagging}
\label{sec:tagging}
%%%%%%%%%%%%%%%%%%%%%%%%%%%%%%%%%%%%%%%%%%%%%

For our last test of the power counting relations, we compare the performance of the reduced bases to that of all EFPs for the task of quark/gluon jet discrimination.
Distinguishing quark jets from gluon jets has a long history~\cite{Nilles:1980ys,Jones:1988ay,Fodor:1989ir,Jones:1990rz,Lonnblad:1990qp,Pumplin:1991kc}, with a recent revival of interest~\cite{Gallicchio:2011xq,Gallicchio:2012ez,Bhattacherjee:2015psa,FerreiradeLima:2016gcz,Gras:2017jty} including both analytic \cite{Larkoski:2014pca,Frye:2017yrw,Mo:2017gzp,Larkoski:2019nwj} and machine learning \cite{Komiske:2016rsd,Metodiev:2018ftz,Komiske:2018vkc,Cheng:2017rdo,Luo:2017ncs,Kasieczka:2018lwf} approaches.
Indeed, quark/gluon discrimination was one of the initial benchmark tests of EFPs \cite{Komiske:2017aww}.

For this study, we use the same dataset as in \sec{results}, which has parton-level quark and gluon labels from \Pythia.
The classification is accomplished via logistic regression, which receives as input either all EFPs or the EFPs in one of four power-counting bases:
\begin{itemize}
    \item strongly-ordered from \sec{SO_basis_present};
    \item 2-collinear from \sec{2c_basis_present};
    \item $z^2$-truncated from \app{energy-expansion}; 
    \item color-reduced (1-collinear) from \app{color-reduced}.
\end{itemize}
The classifier output for logistic regression takes the form:
\begin{equation}
    c(\Phi) = 
    \frac{1}{1 + e^{- \sum_G \, c_G \, {\rm EFP}_G(\Phi)}},
\end{equation}
where $\Phi$ represents the jet, $c_G$ are regression coefficients, ${\rm EFP}_G(\Phi)$ are the EFPs, and the sum is over the relevant graphs.
The classifier is trained to output 1 for quark jets and 0 for gluon jets.

To compare the performance of these classifiers, we examine their receiver operating characteristic (ROC) curves, which are obtained from the true-positive and false-positive rates as the decision threshold is varied.
We consider quark jets as signal and gluon jets as background, such that the true-positive rate corresponds to the quark jet efficiency and the false-positive to the gluon jet mistag rate.
To encapsulate the classifier performance into a single quantity, we take the area under the curve (AUC) of the ROC curve, with an AUC of 0.5 corresponding to a random classifier and an AUC of 1.0 to a perfect classifier. 

   \begin{figure}[p]
   \centering
     \subfloat[]{
     \includegraphics[width=0.4\textwidth]{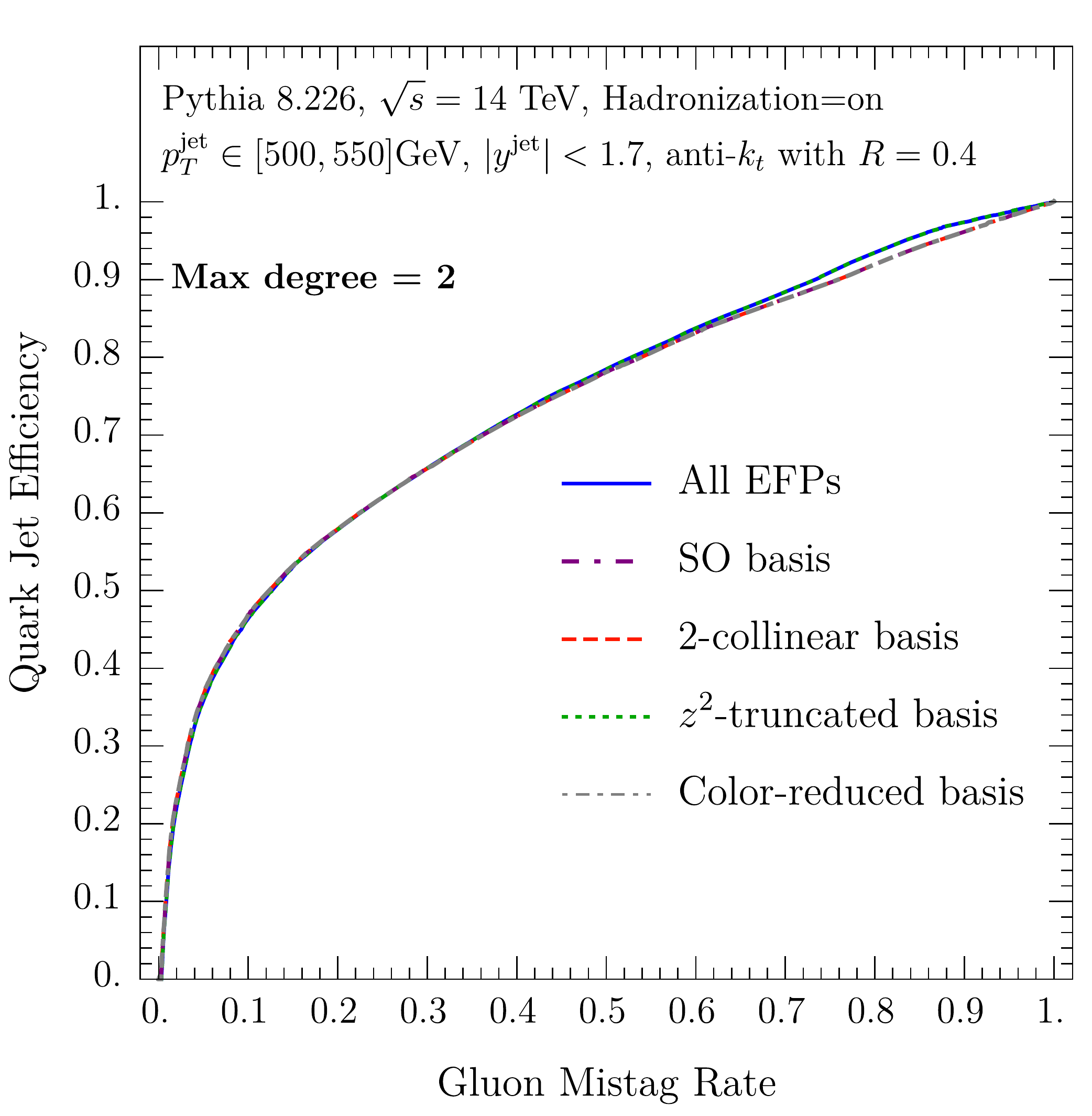}
     }
     $\qquad$
     \subfloat[]{
     \includegraphics[width=0.4\textwidth]{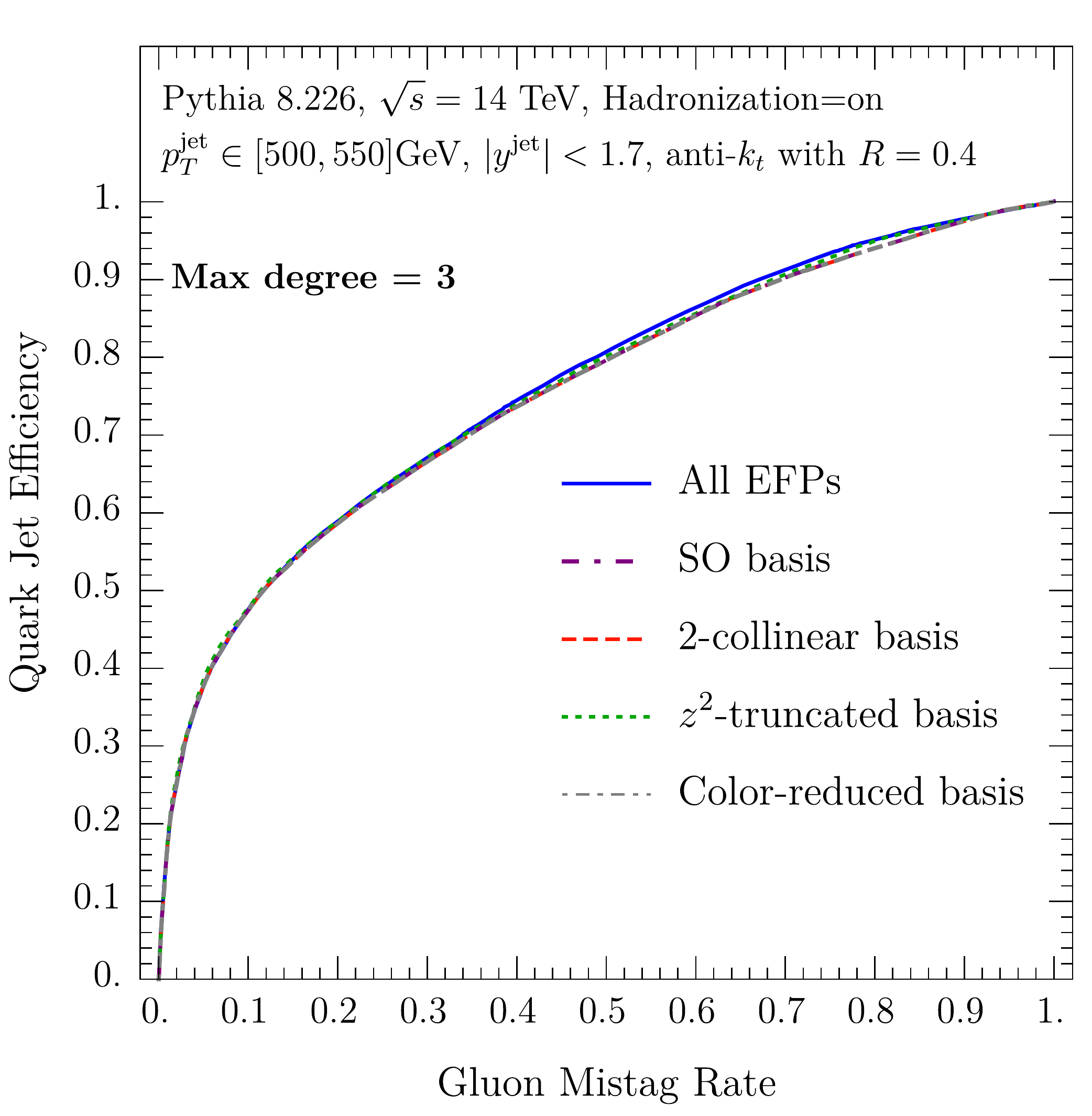}
     }
     \\
     \subfloat[]{
      \includegraphics[width=0.4\textwidth]{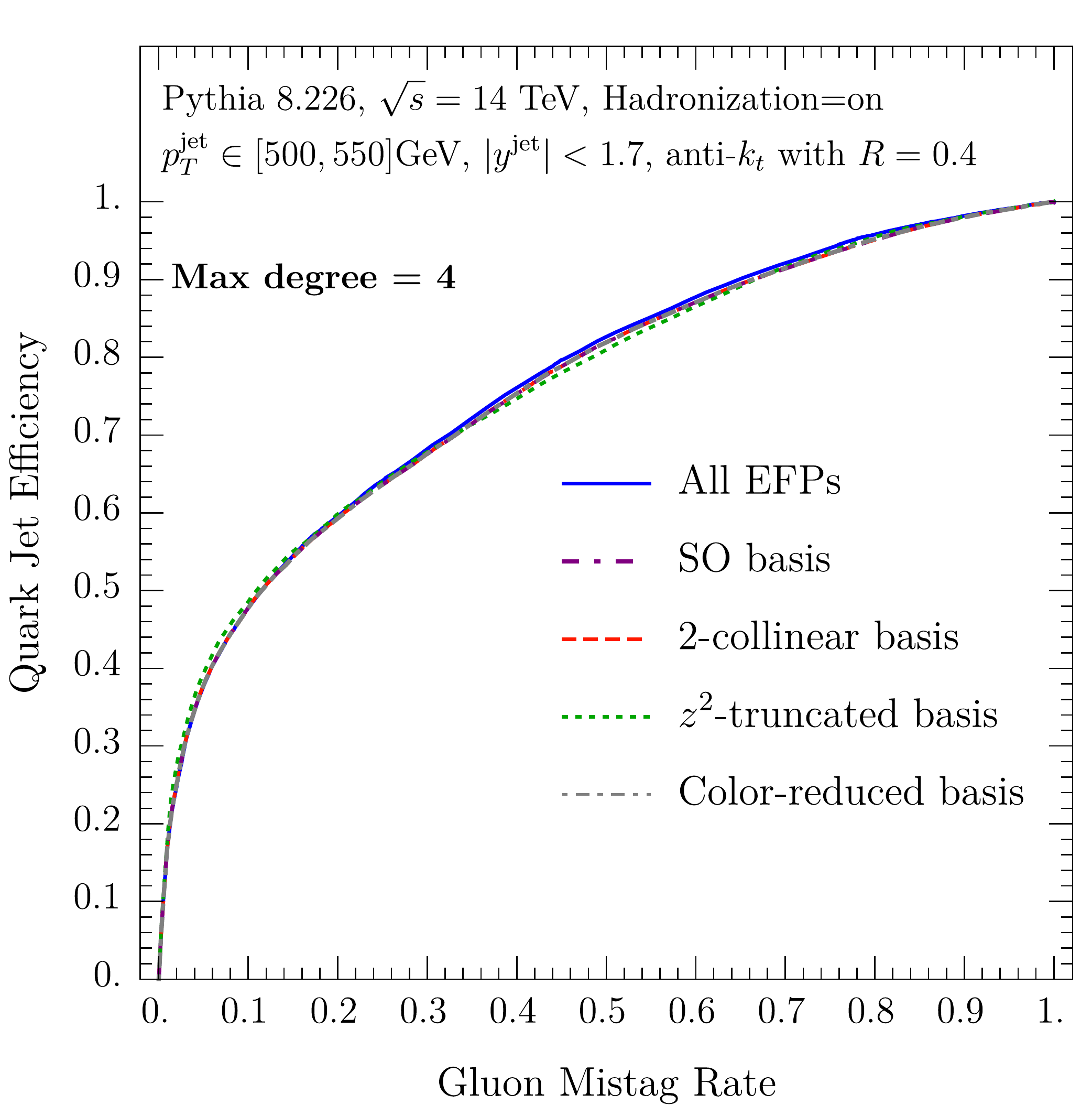}
      }
      $\qquad$
      \subfloat[]{
      \includegraphics[width=0.4\textwidth]{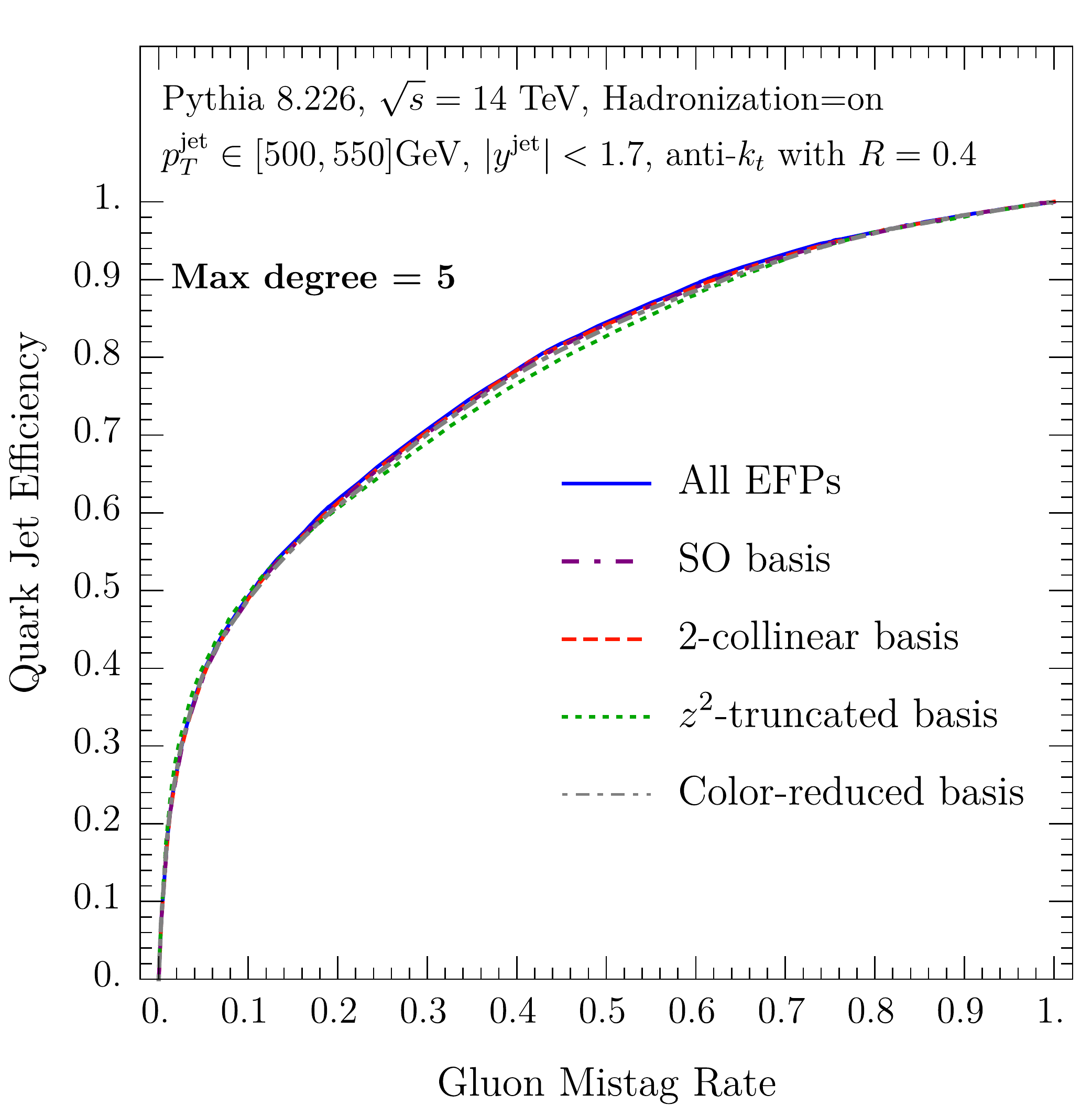}
      }
      \\
      \subfloat[]{
     \includegraphics[width=0.4\textwidth]{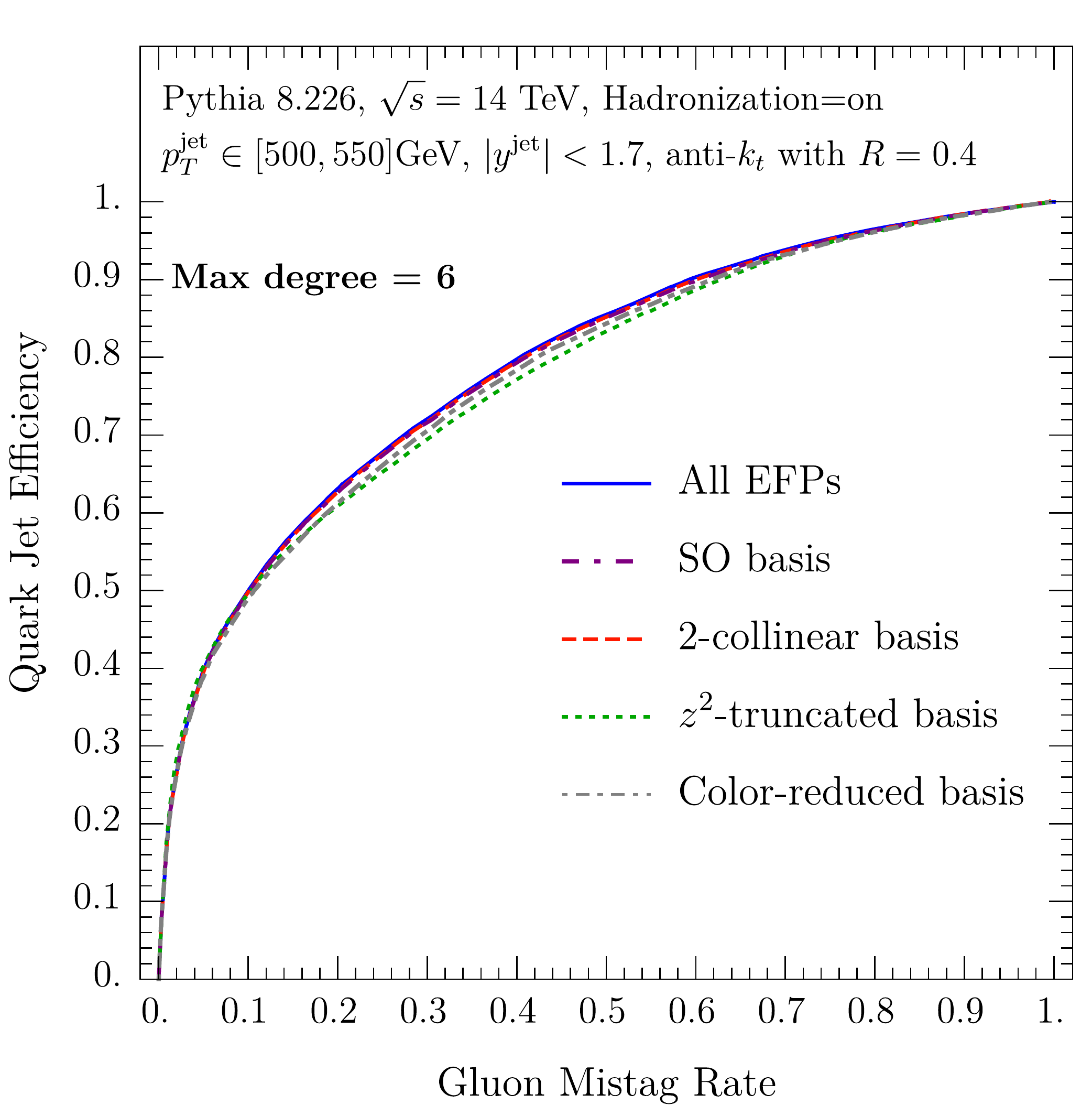}
     }
     $\qquad$
     \subfloat[]{
      \includegraphics[width=0.4\textwidth]{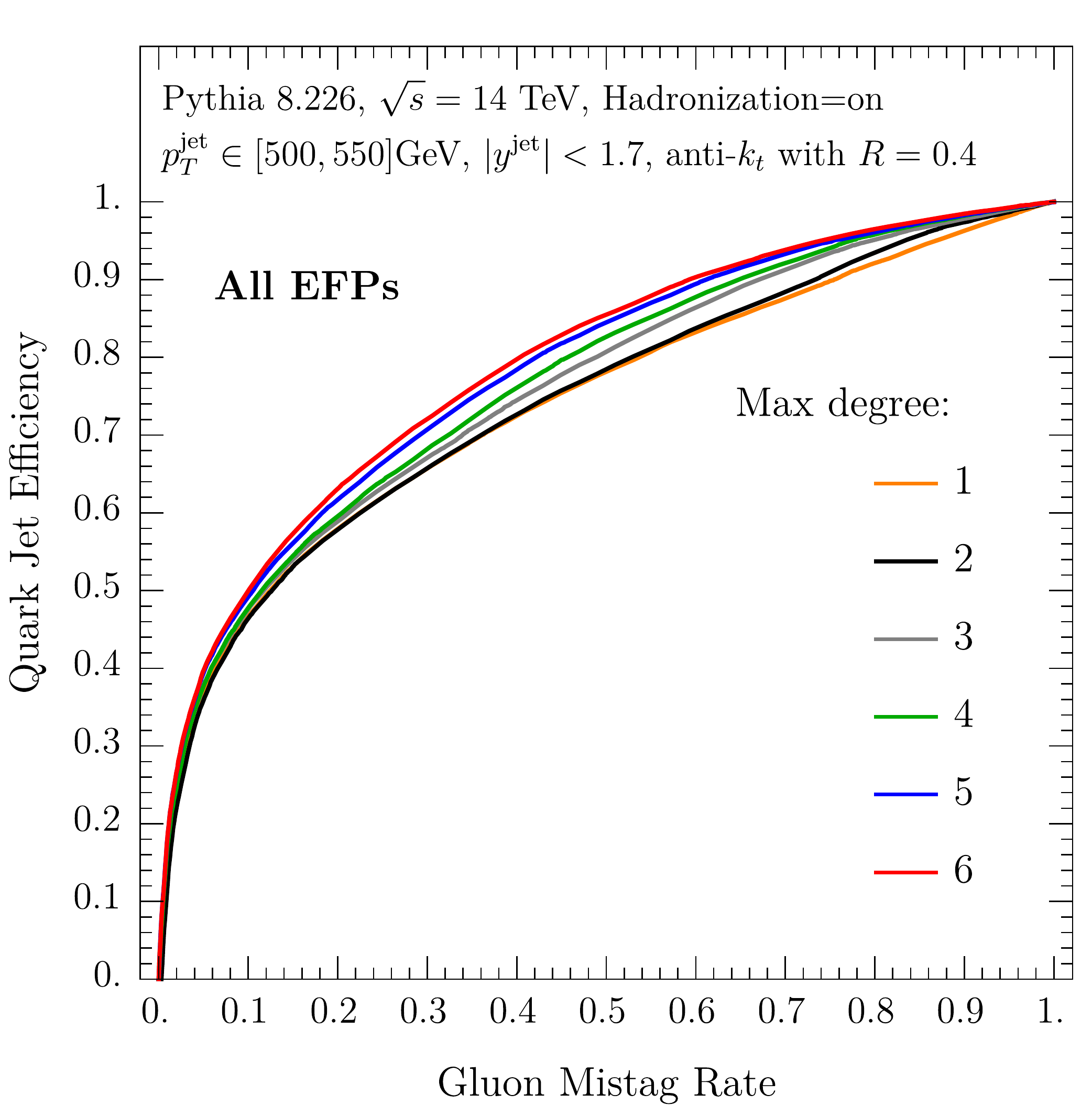}
      \label{fig:ROCs_f}
      }
\caption{ROC curves for quark/gluon discrimination, organized by maximum degree for the different inputs used for classification: all EFPs, strongly-ordered basis, 2-collinear basis, $z^2$-truncated basis, and color-reduced (1-collinear) basis. To help interpret the comparison, in  \fig{ROCs_f} we show the ROC curves for all EFPs by maximum degree. \label{fig:ROCs}}
     \end{figure}

The ROC curves for the different sets of inputs (all EFPs, strongly-ordered basis, 2-collinear basis, $z^2$-truncated, and color-reduced basis) are shown in \fig{ROCs}, where the different panels correspond to the maximum degree of the EFPs.
 The corresponding AUCs are shown in \tab{AUCs}, and the gluon mistag rate for a quark efficiency of 0.5 is shown in \tab{mistag-rate}.
At degree 1, there is only a single EFP and this is an element of all bases, which thus perform the same.
At degrees 2 and 3, the $z^2$-truncated basis has the best AUC of the reduced bases (performing as well as using all EFPs), as it keeps some power-suppressed terms.
It does not perform as well beyond degree 4, where the 2-collinear has the best AUC.

Perhaps surprisingly, the strongly-ordered basis has pretty much the same performance as the 2-collinear basis for degrees 4 and higher.
The poor correlation plots for the strongly ordered expansion in \fig{LL_corr} are not indicative of its tagging performance.
This was anticipated in \fig{SO_hist}, as we found much better (albeit training-set dependent) relations with regression.
The color-reduced basis also performs similarly to the 2-collinear case, except at degree 6.

\begin{table}[!t]
\centering
\begin{tabular}{c || c || c | c | c | c }
Max degree & All EFPs                   & SO                            &  2-collinear                &    $z^2$-truncated  &    Color-reduced   \\ \hline \hline
 1                 & $0.741 $   & \boldmath$0.741  $   & \boldmath$0.741  $   &    \boldmath$0.741  $   &    \boldmath $0.741 $   \\ \hline
 2                 & $0.745 $   & $0.741  $   & $0.741  $   &    \boldmath$0.745  $     &   $0.741 $          \\ \hline
 3                 & $0.761 $   & $0.755  $   & $0.755  $   &    \boldmath$0.759  $    &      $0.755 $          \\ \hline
 4                 & $0.770$   & $0.765  $   & \boldmath$0.766  $   &    \boldmath$0.766  $      &     $0.765 $         \\ \hline
 5                 & $0.784  $  & $0.781  $   & \boldmath$0.782  $   &    $0.776  $   &    $0.779 $       \\ \hline
 6                 &  $0.792$   & \boldmath$0.789  $   & \boldmath$0.789  $   &    $0.780  $    &  $0.782 $           \\ 
\end{tabular}

\caption{ AUCs for the ROC curves displayed in \fig{ROCs}. The uncertainty for all AUC values is $\pm 0.003$, obtained from the 95\% confidence interval coming from 10-fold cross validation. Bold-faced entries show the best performing approximations in each row.  \label{tab:AUCs}}
\end{table}

\begin{table}[!t]
\centering
\begin{tabular}{c || c || c | c | c | c }
Max degree & All EFPs                   & SO                            &  2-collinear                &    $z^2$-truncated  &    Color-reduced   \\ \hline \hline
 1                 &  $0.125$   &  \boldmath $0.125  $   &  \boldmath $0.125  $   &    \boldmath  $0.125  $   &   \boldmath $0.125 $   \\ \hline
 2                 &   $0.123$   &  \boldmath $0.123  $   & \boldmath $0.123  $   &    $0.127  $   &   \boldmath$0.123 $          \\ \hline
 3                 & $0.117$   & $0.117  $   & $0.117  $   &     \boldmath $0.114  $   &   $0.117 $          \\ \hline
 4                 & $0.115$   & $0.115  $   & $0.115  $   &     \boldmath $0.109  $   &   $0.115 $         \\ \hline
 5                 & $0.105$   & $0.107  $   & $0.106  $   &     \boldmath $0.103  $   &   $0.107 $       \\ \hline
 6                 & $0.100$   &  \boldmath $0.101  $   &  \boldmath  $0.101  $   &    $0.102  $   &   $0.106 $           \\ 
\end{tabular}
\caption{Gluon mistag rate at 0.5 quark efficiency. The uncertainty for all values is ±0.004, obtained from the 95\% confidence interval coming from 10-fold cross validation. Bold-faced entries show the best performing approximations in each row.
\label{tab:mistag-rate} }
\end{table}

Looking beyond AUCs at the structure of the ROC curves, we note that the strongly-ordered and 2-collinear bases have similar performance independent of the desired quark jet efficiency/gluon mistag rate.
On the other hand, the ROC curve for the $z^2$-truncated basis has a different shape.
In the high quark purity regime, it performs even better than using all EFPs, as also illustrated in \tab{mistag-rate} (compared to \tab{AUCs}).
As one goes towards higher quark efficiency, though, it performs worse.
It would be interesting to understand what $O(z^3)$ physics explains this behavior.

%%%%%%%%%%%%%%%%%%%%%%%%%%%%%%%%%%%%%%%%%%%%%
\section{Conclusions}
\label{sec:conc}
%%%%%%%%%%%%%%%%%%%%%%%%%%%%%%%%%%%%%%%%%%%%%

Energy flow polynomials (EFPs) form an (over)complete linear basis for jet substructure.
While one could simply use all EFPs as input for machine learning studies, computational cost limits how many EFPs can be used in practice.
By employing power counting, we find relations between EFPs that hold to a certain level of accuracy, and we can use these to substantially reduce the basis of EFPs, providing a more efficient choice of inputs.
Such reductions are also beneficial to streamline the interpretation of machine learning algorithms \cite{Faucett:2020vbu}.

We considered two power counting schemes in this body of this paper: strong ordering (SO) in the energy and angle of emissions; and an expansion involving $n$-collinear emissions and arbitrary collinear-soft emissions, with no further hierarchies.
We found that it was possible to choose a basis for the 1-collinear case such that only three new basis elements were needed up to degree 6 to obtain the 2-collinear basis.
Alternatively in \app{color-reduced}, we find that it is possible to obtain a color-reduced basis for the 1-collinear case, such that EFPs with chromatic number $c$ can be described using those with chormatic number $c-1$, reducing the computational overhead.
In \app{energy-expansion}, we consider keeping terms up to a certain power in the energy fraction ($z^n$ truncation).
One might expect that this would perform better at low degree (where additional basis elements are kept) than at high degree (where more are dropped), which was borne out in our quark/gluon discrimination study.

While the linear relationships obtained between EFPs in the SO expansion perform substantially worse than in the up-to-2-collinear case, interestingly, the two expansions have similar performance for quark/gluon discrimination, particularly at higher degree.
Thus, while the power counting relationships from the SO limit are not as clean, the same relevant information is apparently still present.
This echoes the argument in \refcite{Datta:2017rhs} that only a small number of observables are needed to map out $N$-body phase space.
Here, though, there is a crucial difference that we only performed simple logistic regression and not a fully non-linear machine learning study.
We leave a study of this curious result to future work.

We limited our study to the case of single-prong jets initiated by gluons or light quarks.
It will be interesting to extend this analysis to the case of jets produced from the hadronic decay of heavy resonances such as the Higgs boson or top quark.
Because we did not assume a relative hierarchy of angles in our $n$-collinear expansion, we expect that similar (and possibly the same) basis elements will appear for more complex decay topologies.

Power counting is sufficient for reducing the basis of EFPs, but further work is needed to make precise predictions for jet substructure.
The natural next step would be the calculation of cross sections that are \emph{simultaneously} differential in the basis EFPs.
There has been some work on multi-differential cross sections for angularities~\cite{Larkoski:2014tva,Procura:2014cba,Procura:2018zpn,Lustermans:2019gxu}, which correspond to dumbbells in the language of EFPs.
Following on the pioneering study of \refcite{Larkoski:2015kga}, it will be interesting to extend this analysis to higher-point EFPs.
Multi-differential EFP studies could provide insights into jet properties that are complementary to the parton showers approach.
We anticipate that power counting will continue to play an essential role in organizing and simplifying systematic studies of jet substructure.

\begin{acknowledgments}
We thank Patrick Komiske for collaboration in the initial stages of this research.
P.C.~thanks Miguel Jaques for machine learning related discussions. 
J.T. is supported by the U.S. Department of Energy (DOE) Office of High Energy Physics under Grant No.\ DE-SC0012567, and by the National Science Foundation under Cooperative Agreement PHY-2019786 (The NSF AI Institute for Artificial Intelligence and Fundamental Interactions, \url{http://iaifi.org/}).
W.W.~is supported by the D-ITP consortium, a program of NWO that is funded by the Dutch Ministry of Education, Culture and Science (OCW).
P.C. is supported by funding from the European Research Council (ERC) under the European Union’s
Horizon 2020 research and innovation programme (grant agreement No 101002090).

\end{acknowledgments}

\appendix

%%%%%%%%%%%%%%%%%%%%%%%%%%%%%%%%%%%%%%%%%%%%%
\section{$z^2$-truncated basis}
\label{app:energy-expansion}
%%%%%%%%%%%%%%%%%%%%%%%%%%%%%%%%%%%%%%%%%%%%%

In this appendix, we consider energy truncation as a strategy to organize the basis of EFPs.
We assume the same configuration of one hard emission and $M-1$ collinear-soft emissions as in the 1-collinear approximation in \eq{pc}.
Rather than keeping only leading terms and dropping all subleading terms, however, we use energy scaling to determine which elements to keep:
\begin{itemize}
\item \textbf{$z^n$-truncated basis:} We remove EFPs that scale as $\mathcal{O}(z^{n+1})$ and only use relationships between EFPs where the dropped terms are $\mathcal{O}(z^{n+1})$, where $z\ll 1$ is the collinear-soft momentum fraction.
\end{itemize}

The $z^n$-truncated basis seems natural when linearly combining EFPs, as it takes their relative $z$ scaling into account.
(Of course, this suppression could be compensated by the size of the coefficients.)
In the $z^n$-truncated bases, certain EFPs can be directly eliminated due to their scaling, but there are fewer relationships that can be used to reduce the basis.
Because the degree of the EFP limits the $z$ scaling,%
\footnote{EFPs with $N$ nodes and degree $d$ scale like $z^\kappa \theta^{d}$ with $\kappa \leq N-1 \leq d$, since always at least one of the nodes involves a collinear parton. Disconnected EFPs also satisfy $\kappa \leq d$, since they are simply products of connected EFPs.}
$z^2$-truncated bases will have more elements at low degree and fewer at high degree compared to the 2-collinear basis. 

        \begin{table}[t]\centering
      \includegraphics[width=0.98\textwidth]{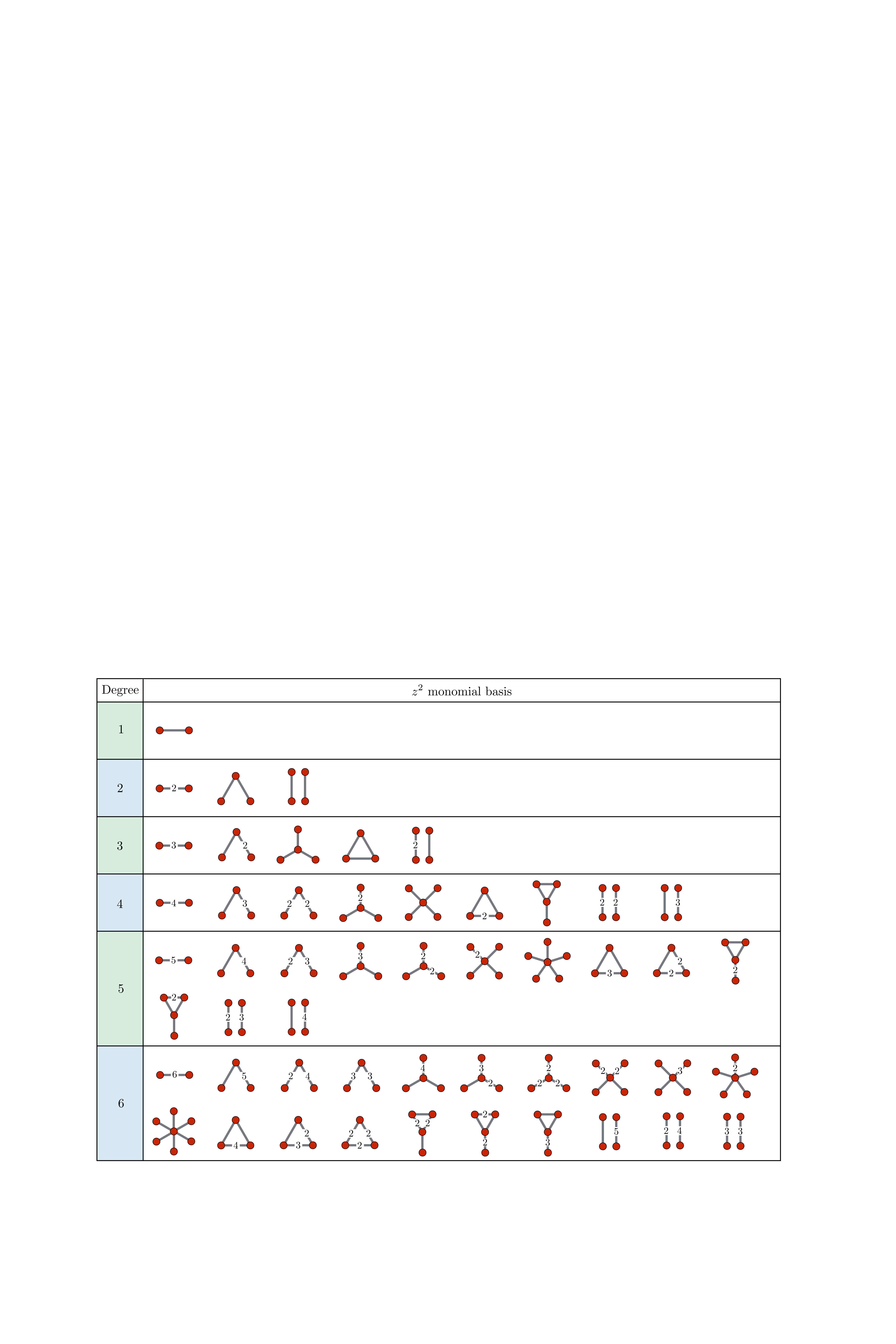}  
      \caption{
      The $z^2$-truncated basis of EFPs up to degree 6. A generic EFP up to degree 6 can be expressed as a \emph{linear} combination in terms of these bases elements.  \label{tab:z2_basis} }
     \end{table}

The $z$-truncated basis is rather simple: it consists of EFPs with two nodes connected by any number of lines, i.e. dumbbells.
Moving on to a more interesting case, we consider the $z^2$-truncated basis, for which the basis elements up to degree 6 are shown in \tab{z2_basis}.
Here we show composite EFPs, such that all EFPs can be expressed as a \emph{linear} combination of these ingredients.
This is different from \tabs{LLbasis}{EFP_basis_6} which only showed prime EFPs, and the EFP power counting relationships involved polynomials of the basis elements.

The reason for the different presentation in \tab{z2_basis} is that for $z^n$-truncation, it is in general difficult to determine by inspection which products are kept and whether they are related to other EFPs.
For $z^2$-truncation this is not too complicated yet: only products of two dumbbells are allowed. 
Because we can only use relationships between EFPs that hold up to corrections that are order $z^3$, new basis elements appear compared to the 2-collinear case in \tab{EFP_basis_6}, such as 3 dots in a row or star-like configurations.
On the other hand, other basis elements from \tab{EFP_basis_6} do not appear here because they themselves are order $z^3$, such as the fully-connected four dot graph or a martini glass with an elongated stem.

\section{1-collinear color-reduced basis}
\label{app:color-reduced}

    \begin{table}[t]\centering
      \includegraphics[width=0.98\textwidth]{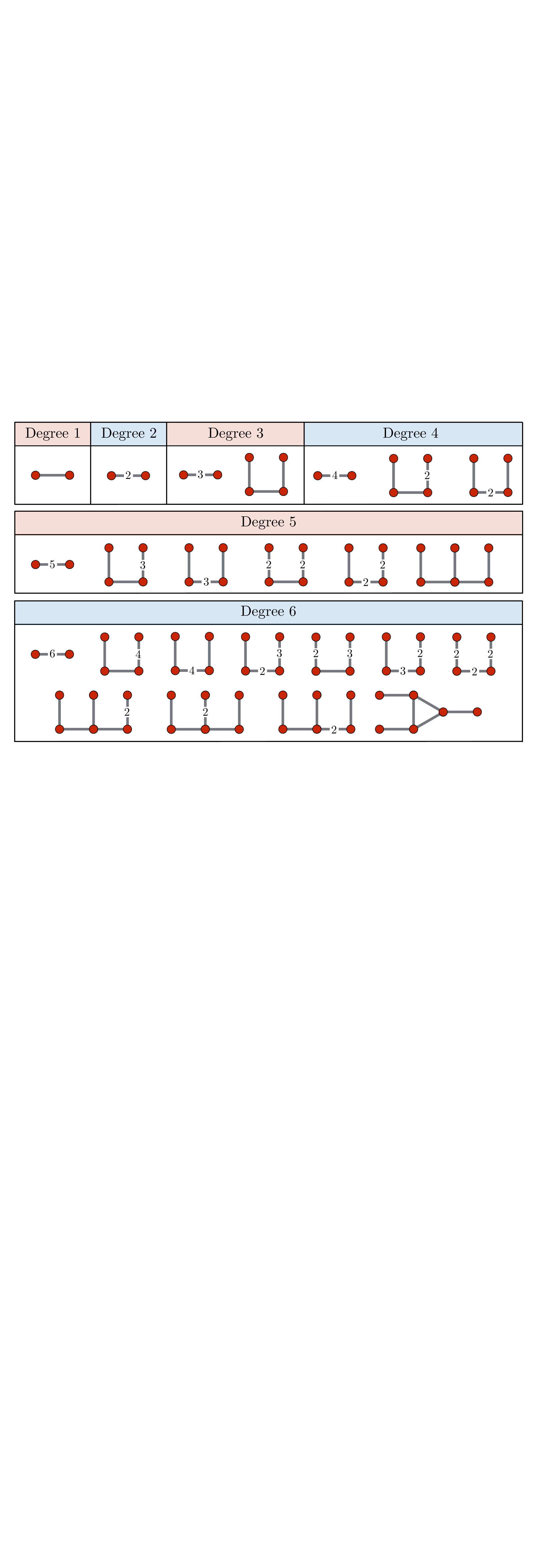}  
      \caption{Prime EFPs for the color-reduced basis of the 1-collinear expansion.  \label{tab:ColorReducedBasis}}
     \end{table}

As discussed in \sec{1c_basis_present}, we have freedom to choose a different basis set for the 1-collinear expansion which is more computational efficient.
The essential trick is shown in \eq{color-reduced}, where any vertex with a collinear parton can be cut open at 1-collinear order.
For a chromatic number $c$ graph, this trick allows it to be expressed in terms of chromatic number $c-1$ basis elements.

In \tab{ColorReducedBasis}, we present a color reduced basis that can be used in the 1-collinear expansion.
This basis has a reduced number of loop-like graphs, which offers a computational speedup in evaluating the EFP.
There are no loop-like graphs up to degree 5, and only one loop-like graph at degree 6.
Unfortunately, this color reduction does not persist to higher orders, and a non-minimal extension of \tab{ColorReducedBasis} is needed to lift the basis to 2-collinear accuracy.

For tagging and regression, this color-reduced 1-collinear basis exhibits somewhat worse performance than the default 1-collinear basis in \tab{EFP_basis_6}.
This is to be expected, since the default 1-collinear basis already contains the majority of the 2-collinear information.
It would be interesting to explore trade offs between performance and computational cost.
At degree 6, the fully-connected 4-point graph is the most costly EFP to compute, so one could consider a hybrid basis where this chromatic number 4 graph is cut open but most chromatic number 3 graphs are kept.

\bibliographystyle{JHEP}
\bibliography{bibliography}

\end{document}